\documentclass[11pt,nofootinbib]{article}
\pdfoutput=1
\usepackage{jheppub}
\usepackage{graphicx}
\usepackage{slashed}
\usepackage{subfig}
\usepackage[normalem]{ulem}

\newcommand\beq{\begin{equation}}
\newcommand\eeq{\end{equation}}

\newcommand\nn{\nonumber}

\newcommand\hc{\text{h.c.}}

\title{Probing Lepton Flavor Violation at the 13 TeV LHC}
\author[a]{Reinard Primulando,}
\author[b]{Patipan Uttayarat}

\affiliation[a]{Center for Theoretical Physics, Department of Physics, Parahyangan Catholic University, Jl. Ciumbuleuit 94, Bandung 40141, Indonesia}
\affiliation[b]{Department of Physics, Srinakharinwirot University, 114 Sukhumvit 23rd Rd., Wattana, Bangkok 10110, Thailand}

\emailAdd{rprimulando@unpar.ac.id}
\emailAdd{patipan@g.swu.ac.th}

\abstract{ 
We investigate the bounds on tau-mu lepton flavor violation (LFV). Our main focus is on the collider constrains on tau-mu LFV. We use the Type-III Two-Higgs-Doublet-Model (2HDM) as a set up for our study. While the LFV branching fraction of the 125 GeV is well constrained by current LHC searches, the heavier neutral states could have a large branching fraction to tau and muon. We estimate the LHC reach for the 13 TeV center of mass energy with 300 $\text{fb}^{-1}$ luminosity for a neutral boson decaying into a tau and a muon. We identify parts of the LFV parameter space where the searches for heavy scalar and pseudoscalar decaying into a tau and a muon are more sensitive than the similar search for the 125 GeV boson.
}

\begin{document}
\maketitle

\flushbottom
\section{Introduction}
The upgraded Large Hadron Collider (LHC) opens up the possibilities to explore a higher energy scale where new physics may lie. The lepton flavor volation (LFV) is an interesting possible new physics that might show up during this next run of the LHC. In Ref.~\cite{Harnik:2012pb,Banerjee:2016foh}, various LFV decay channels of the 125 GeV scalar $h$ were explored. The authors found that LHC constraints on  the decay $h\rightarrow \tau\mu$ can be superior to the bounds from low energy experiments such as $\tau\rightarrow \mu \gamma$ and $\tau \rightarrow 3 \mu$. This decay has been probed at the LHC run-1~\cite{Aad:2016blu, Khachatryan:2015kon, Aad:2015gha} and early run-2~\cite{CMS-PAS-HIG-16-005}. The ATLAS and CMS experiments constrained the LFV branching fraction to be BR$_{h\rightarrow \tau\mu}< 1.43\%$ and BR$_{h\rightarrow \tau\mu}<1.2\%$ respectively. Additionally there is a $2.4 \sigma$ hint of LFV branching fraction from run-1 CMS with BR$_{h\rightarrow \tau\mu} = 0.84_{-0.37}^{+0.39}$\%~\cite{Khachatryan:2015kon}, which is marginally compatible with the previous constraints.

The Two Higgs Doublet Model (2HDM) is one possible extension of the standard model (SM). In this model, the particle content of the SM is enlarged by an introduction of an additional scalar doublet. 
The extra doublet brings with it a host of interesting LHC phenomenology.
Firstly, there are additional new particles that might be observed at the LHC. Assuming the model is CP conserving, these new particles are neutral heavy scalar ($H$), neutral pseudoscalar ($A$) and a pair of charged scalars. 
Secondly, the new scalar doublet introduces new Yukawa couplings to the fermions which could give rise to LFV couplings.
The LFV couplings, in turn, lead to LFV decays of the neutral scalars and pseudoscalar\footnote{The charged scalar also has a LFV decay into $\ell \nu_{\ell'}$, where $\nu_{\ell'}$ is the neutrino with a different flavor than $\ell$. However since none of the LHC experiments can detect neutrino flavors, this LFV decay will be indistinguishable from the flavor conserving decay $\ell \nu_{\ell}$.}, in particular the decay into tau and muon. 
Hence this model can be a simple UV completion of LFV decay considered in Ref.~\cite{Harnik:2012pb,Banerjee:2016foh}. 
Many other models can incorporate the LFV decay, for example, the minimal supersymmetric Standard Model and its extension~\cite{DiazCruz:1999xe,DiazCruz:2008ry}, models with an electroweak triplet~\cite{Kakizaki:2003jk,Fukuyama:2009xk}, models with vector-like heavy lepton~\cite{Ishiwata:2013gma,Falkowski:2013jya}, composite Higgs model~\cite{Feruglio:2015gka}, and the Little Higgs models~\cite{Lami:2016mjf,Yang:2016hrh}.
As has already been mentioned, the LFV  $h\to\tau\mu$ has been searched for at the LHC and the branching ratio is tightly constrained. However, the corresponding LFV branching ratio of the heavy neutral particles can, in principle, be large.
Thus it is possible that the larger branching ratio compensates the smaller cross section of heavier scalars, making it possible to probe these decays at the LHC.
These LFV decay searches could explore more parameter space compared to the current $h \rightarrow \tau \mu$ search alone~\cite{Sher:2016rhh,Altmannshofer:2016zrn,Buschmann:2016uzg}.
While no specific LHC estimates have been given in refs.~\cite{Sher:2016rhh,Altmannshofer:2016zrn}, ref.~\cite{Buschmann:2016uzg} recasts the CMS $h\rightarrow\tau\mu$ search to incorporate the heavy CP-even scalar $H$. By considering an optimistic scenario, in which the heavy scalars couple to the top quark and is produced copiously through gluon-fusion, the authors of ref.~\cite{Buschmann:2016uzg} found that at the 8 TeV LHC the inclusion of $H\to\tau\mu$ search excludes more parameter space of 2HDM.

In this paper we examine the bounds on LFV decays at the 13 TeV LHC with a luminosity of 300 fb$^{-1}$. Moreover, we fully explore the parameter space of the 2HDM by including the pseudoscalar $A$ that has not been taken into account in the previous works. In addition, we consider various possibilities of heavy resonances couplings to SM particles in the context of type-III 2HDM. 
Hence we cover both the optimistic scenarios, where the heavy resonances production cross-section and their LFV branching fractions are large, and the pessimistic scenario (small production cross-sections and small LFV branching ratios) for discovering the LFV via the heavy resonance searches.

The paper is structured as follow. In Sec.~\ref{sec:2hdm} we review the Higgs phenomenology in 2HDM relevant for our LFV analysis. We then discuss the current constraints on LFV in the Higgs sector, including both the direct and indirect constraints, in Sec.~\ref{sec:LFV}. In Sec.~\ref{sec:LFVLHC} we study the collider constraints on the Higgs LFV decays. We also give the projected bound for the 13 TeV LHC with 300 $\text{fb}^{-1}$ luminosity. We then conclude in Sec.~\ref{sec:conclusion}.  

\section{Type-III 2HDM}
\label{sec:2hdm}
2HDM is one of the most studied extensions of the SM, for a review see Ref.~\cite{Branco:2011iw}. There are many realizations of the model in the literature. They can be characterized by the structure of Yukawa couplings. In Type-I, Type-II and Type-X 2HDM, each fermion type (up-type quarks, down-type quarks and leptons) is coupled to only one scalar doublet. Thus there is no flavor violating Yukawa couplings of the neutral scalar boson in this case~\cite{Glashow:1976nt,Paschos:1976ay}\footnote{Ref.~\cite{Crivellin:2015hha} shows that a small perturbation in the lepton Yukawa structure of the Type-X 2HDM could lead to an observable $h\to\tau\mu$ decay.}.  In type-III 2HMD, however, both of the scalar doublets couple to all the fermions. As a result, there is a possibility of flavor violation in the neutral scalars Yukawa couplings. Therefore, we will focus on the type-III 2HDM in this work.

\subsection{Conventions and notations}
In this section we set up our conventions and notations.  The two $SU(2)_L$ scalar doublets are taken to have hypercharge $Y=1/2$. With this convention, the electric charge generator is $Q=\frac{\sigma^3}{2}+Y$ where $\sigma^i$ are the Pauli matrices.  The scalar potential consistent with $SU(2)_L\times U(1)_Y$ is 
\begin{equation}
\begin{aligned}
	V &= M_{11}^2 (\Phi_1^{\dagger} \Phi_1^{\phantom{\dagger}}) + M_{22}^2 (\Phi_2^{\dagger} \Phi_2^{\phantom{\dagger}}) -  [M_{12}^2(\Phi_1^{\dagger} \Phi_2^{\phantom{\dagger}}) + \text{h.c.}]  \\
&+ \frac{1}{2} \lambda_1 (\Phi_1^{\dagger} \Phi_1^{\phantom{\dagger}})^2 + \frac{1}{2} \lambda_2 (\Phi_2^{\dagger} \Phi_2^{\phantom{\dagger}})^2 + \lambda_3 (\Phi_1^{\dagger} \Phi_1^{\phantom{\dagger}})(\Phi_2^{\dagger} \Phi_2^{\phantom{\dagger}}) + \lambda_4 (\Phi_1^{\dagger} \Phi_2^{\phantom{\dagger}})(\Phi_2^{\dagger} \Phi_1^{\phantom{\dagger}}) \\
&+ \{\frac{1}{2} \lambda_5 (\Phi_1^{\dagger} \Phi_2^{\phantom{\dagger}})^2 + [\lambda_6 (\Phi_1^{\dagger} \Phi_1^{\phantom{\dagger}}) + \lambda_7 (\Phi_2^{\dagger} \Phi_2^{\phantom{\dagger}})] (\Phi_1^{\dagger} \Phi_2^{\phantom{\dagger}}) + \text{h.c.} \}.
\end{aligned}
\end{equation}
In general, $M_{12}^2$, $\lambda_5$, $\lambda_6$ and $\lambda_7$ are complex while the rest of the parameters are real.
However, since in this work we are interested in the simplified scenario where CP is a good symmetry, we will take $M_{12}^2$, $\lambda_5$, $\lambda_6$ and $\lambda_7$ to be real. We leave the CP violating scenario for a future work.

In this paper we use the Higgs basis~\cite{Georgi:1978ri} where the vacuum expectation value (VEV) resides only in $\Phi_1$. In the Higgs basis, the fields $\Phi_1$ and $\Phi_2$ can be expanded as 
\begin{equation}
	\Phi_1 = \begin{pmatrix}G^+\\ \frac{1}{\sqrt{2}}\left(v+\phi_1+iG^0\right)\end{pmatrix},
	\qquad
	\Phi_2 = \begin{pmatrix}H^+\\ \frac{1}{\sqrt{2}}\left(\phi_2+iA\right)\end{pmatrix},
\end{equation}
where $v$ is the VEV, $G^\pm$ and $G^0$ are the would be Goldstone bosons, $H^\pm$ is the charged Higgs, $A$ is the neutral CP-odd Higgs, $\phi_1$ and $\phi_2$ are the neutral CP-even Higgs.  
Minimizing the potential leads to the relations
\begin{equation}
   	M_{11}^2 = -\frac{\lambda_1 v^2}{2} \text{\hspace{5mm} and \hspace{5mm}} M_{12}^2 = \frac{\lambda_6 v^2}{2}.
\end{equation}
The fields $H^\pm$ and $A$ are mass eigenstates with masses
\begin{equation}
	m_{H^\pm}^2 = \frac{1}{2} \left( 2 M_{22}^2 + \lambda_3 v^2 \right),\quad
	m_A^2 = m_{H^\pm}^2 + \frac{1}{2} \left( \lambda_4 - \lambda_5 \right) v^2.
\end{equation}
The fields $\phi_1$ and $\phi_2$ are in general not mass eigenstates. They are related to the mass eigenstates $h$ and $H$ by
\begin{equation}
	\begin{pmatrix}\phi_1\\ \phi_2\end{pmatrix}= \begin{pmatrix} \phantom{-}\cos\alpha & \sin\alpha \\ -\sin\alpha & \cos\alpha\end{pmatrix}
	\begin{pmatrix}h\\ H \end{pmatrix},
\end{equation}
where mixing angle $\alpha$ is given by
\begin{equation}
	\tan2\alpha = \frac{2\lambda_6v^2}{m_A^2-(\lambda_1-\lambda_5)v^2}.
\end{equation}
The scalar masses are
\begin{equation}
	m_{h,H}^2 = \frac{1}{2}\left( M_A^2 + (\lambda_1 + \lambda_5) v^2 \mp \sqrt{\left( M_A^2 - (\lambda_1 - \lambda_5) v^2\right)^2 + 4\lambda_6^2 v^4 } \right).
\end{equation}
We take the $h$ to be the 125 GeV scalar resonance discovered at the LHC.

The Yukawa sector in the Type-III 2HDM is given by 
\begin{equation}
\begin{split}
	\mathcal{L}_{yuk} &= -\frac{\sqrt{2}m^i_\ell}{v}\delta^{ij}\bar{L}_L^i\ell_R^j\Phi_1 - \sqrt{2}Y_\ell^{ij}\bar{L}_L^i\ell_R^j\Phi_2\\
	&\quad -\frac{\sqrt{2}m^i_U}{v}\delta^{ij}\bar{Q}_L^iu_R^j\tilde\Phi_1 - \sqrt{2}Y_U^{ij}\bar{Q}_L^iu_R^j\tilde\Phi_2\\
	&\quad -\frac{\sqrt{2}m^k_D}{v}V^{ik}\delta^{kj}\bar{Q}_L^id_R^j\Phi_1 - \sqrt{2}V^{ik}Y_D^{kj}\bar{Q}_L^id_R^j\Phi_2 + \hc,
\end{split}
\label{eq:yukbeforeewsb}
\end{equation}
where $m_f$ are fermion masses, $Y_f$ are the Yukawa coupling matrices, $V$ is the CMK matrix and the indices $i$, $j$, $k$ run over fermion families. 
The scalar doublet $\tilde\Phi$ is defined as $\tilde\Phi = i\sigma^2\Phi^\ast$. The fermion doublets are taken to be
\begin{equation}
	L_L = \begin{pmatrix}\nu_L \\ \ell_L\end{pmatrix},\qquad
	Q_L = \begin{pmatrix}u_L \\ V d_L\end{pmatrix}.
	\label{eq:fermiondoublet}
\end{equation}
Note that the fermion fields $\ell_{L(R)}$, $u_{L(R)}$and $d_{L(R)}$ in Eq.~\eqref{eq:yukbeforeewsb} and~\eqref{eq:fermiondoublet} are in the mass eigenbasis.
After electroweak symmetry breaking, the Yukawa couplings in the physical basis read 
\begin{equation}
\begin{split}
	\mathcal{L}_{yuk}  &\supset - y^{ij}_{f,h}\bar{f}_L^if_R^j h - y^{ij}_{f,H}\bar{f}_L^if_R^j H - y_{f,A}^{ij}\bar{f}_L^if_R^j A + \hc,  \\
\end{split}
\label{eq:yukawa}
\end{equation}
where $P_{R,L} = (1\pm\gamma^5)/2$ are the chiral projections, $f$ runs over fermion types ($\ell$, $U$ and $D$) and 
\begin{equation}
\begin{split}
	y_{f,h}^{ij} &= \frac{m_{f}^i}{v}\delta^{ij} \cos\alpha - Y_{f}^{ij} \sin\alpha,  \\
	y_{f,H}^{ij} &= \frac{m_{f}^i}{v}\delta^{ij} \sin\alpha + Y_{f}^{ij} \cos\alpha,  \\
	y_{U,A}^{ij} &= -iY_U^{ij}, \quad y_{\ell(D),A}^{ij}  = iY_{\ell(D)}^{ij}.
\end{split}
\label{eq:reducedcoupling}
\end{equation}
Notice that the flavor violating Higgs Yukawa couplings are encoded in the Yukawa matrices $Y_\ell$, $Y_U$ and $Y_D$.

The Yukawa matrices $Y_f$'s affect the production cross-sections and decay rates of the Higgs bosons.  Since we're interested in the lepton flavor violating (LFV) decay of the neutral scalar $\phi \to \tau^\pm \mu^\mp$, we will restrict our attention to the neutral scalars sector.

\subsection{Production cross-sections of neutral Higgs bosons}
The production cross-sections of the neutral Higgs bosons, $\phi$, in each channel can be most conveniently described in terms of the would be SM Higgs boson cross-sections.  
They are~\cite{Branco:2011iw}
\begin{equation}
\begin{split}
	\frac{\sigma^{\phi}_{gF}}{\sigma^{SM}_{gF}} &\simeq \frac{\left| \sum\limits_{q,i} \frac{v}{m^i_q} \Big[\text{Re}(y_{q,\phi}^{ii})A_{1/2}^H(\tau_{q^i}) + \text{Im}(y_{q,\phi}^{ii})A_{1/2}^A(\tau_{q^i})\Big] \right|^2 }{\left| A_{1/2}^H(\tau_t)\right|^2},\\
	\frac{\sigma^{\phi}_{t\bar t h}}{\sigma^{SM}_{t\bar th}} &\simeq \left|\frac{v}{m_t}y_{U,\phi}^{tt}\right|^2, \quad
	\frac{\sigma^{\phi}_{b\bar b h}}{\sigma^{SM}_{b\bar bh}} \simeq \left|\frac{v}{m_b}y_{D,\phi}^{bb}\right|^2,\\
	\frac{\sigma^{\phi}_{VBF}}{\sigma^{SM}_{VBF}} &= \frac{\sigma^{h}_{Vh}}{\sigma^{SM}_{Vh}} = \delta_\phi^2,
\end{split}
\label{eq:xsection}
\end{equation}
where $q = U,D$, $\delta_h=\cos\alpha$, $\delta_H=\sin\alpha$, $\delta_A=0$, $i$ runs over generation index, $\tau_{x} = 4m_x^2/m_\phi^2$ and the loop functions $A_{1/2}^H(\tau)$ and $A_{1/2}^A(\tau)$ are given in Eq.~\eqref{eq:loopfunctions}. In our analysis below, we use the would be SM Higgs boson cross-sections provided by the LHC Higgs Cross Section Working Group~\cite{Heinemeyer:2013tqa}.

\begin{figure}
        \centering
       \includegraphics[width= 0.5\textwidth]{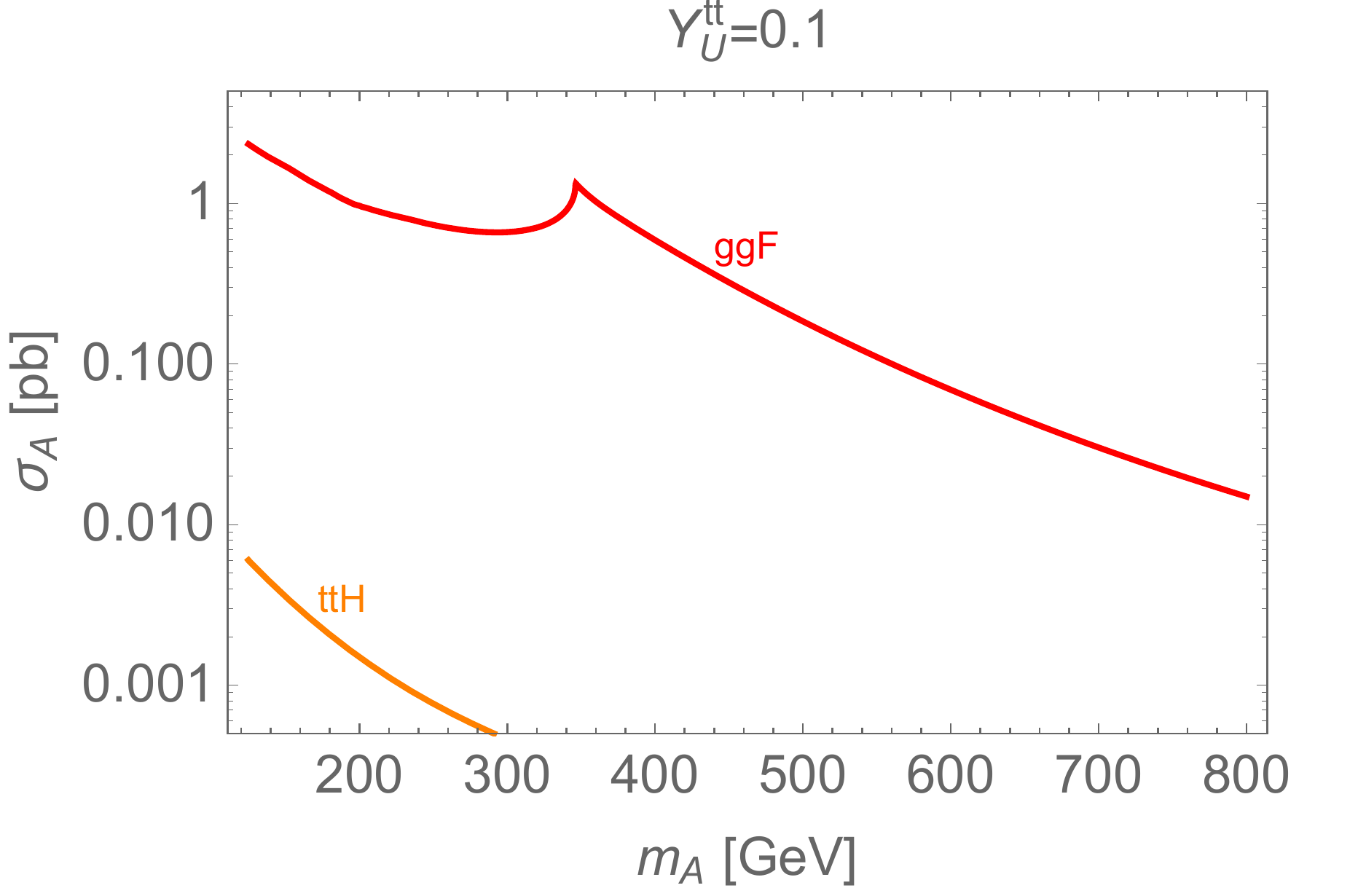}
       \includegraphics[width= 0.49\textwidth]{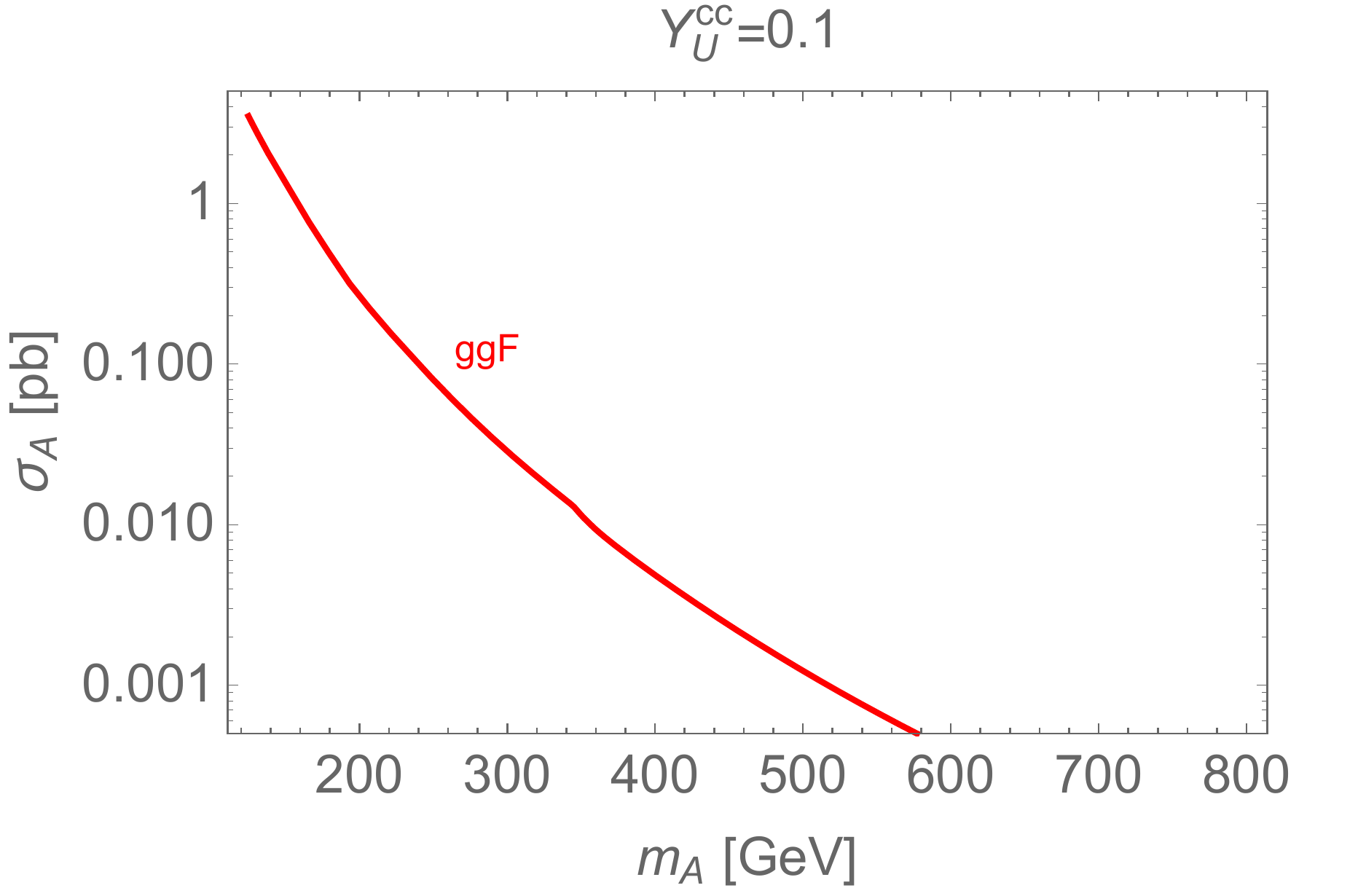}
        \caption{The production cross-section of the pseudoscalar $A$ as a function of its mass. The gluon-fusion channel is shown in red and the $t\bar tA$ channel is shown in orange. On the left pane $Y_U^{tt}=0.1$ while on the right pane $Y_U^{cc}=0.1$. Other Yukawa couplings are taken to be 0.}
         \label{fig:xsecA}
\end{figure}
\begin{figure}[h]
        \centering
       \includegraphics[width= 0.49\textwidth]{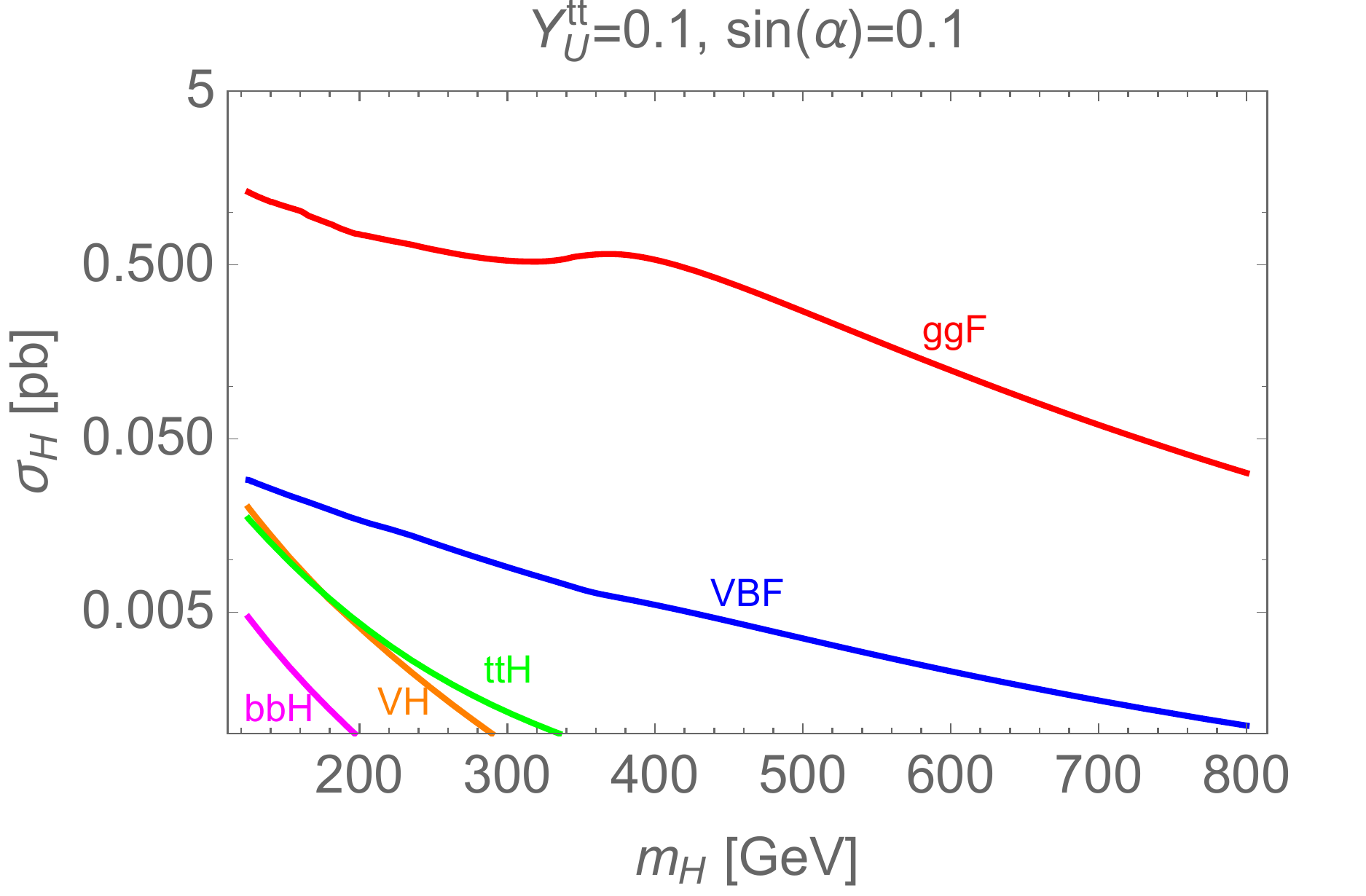}
       \includegraphics[width= 0.49\textwidth]{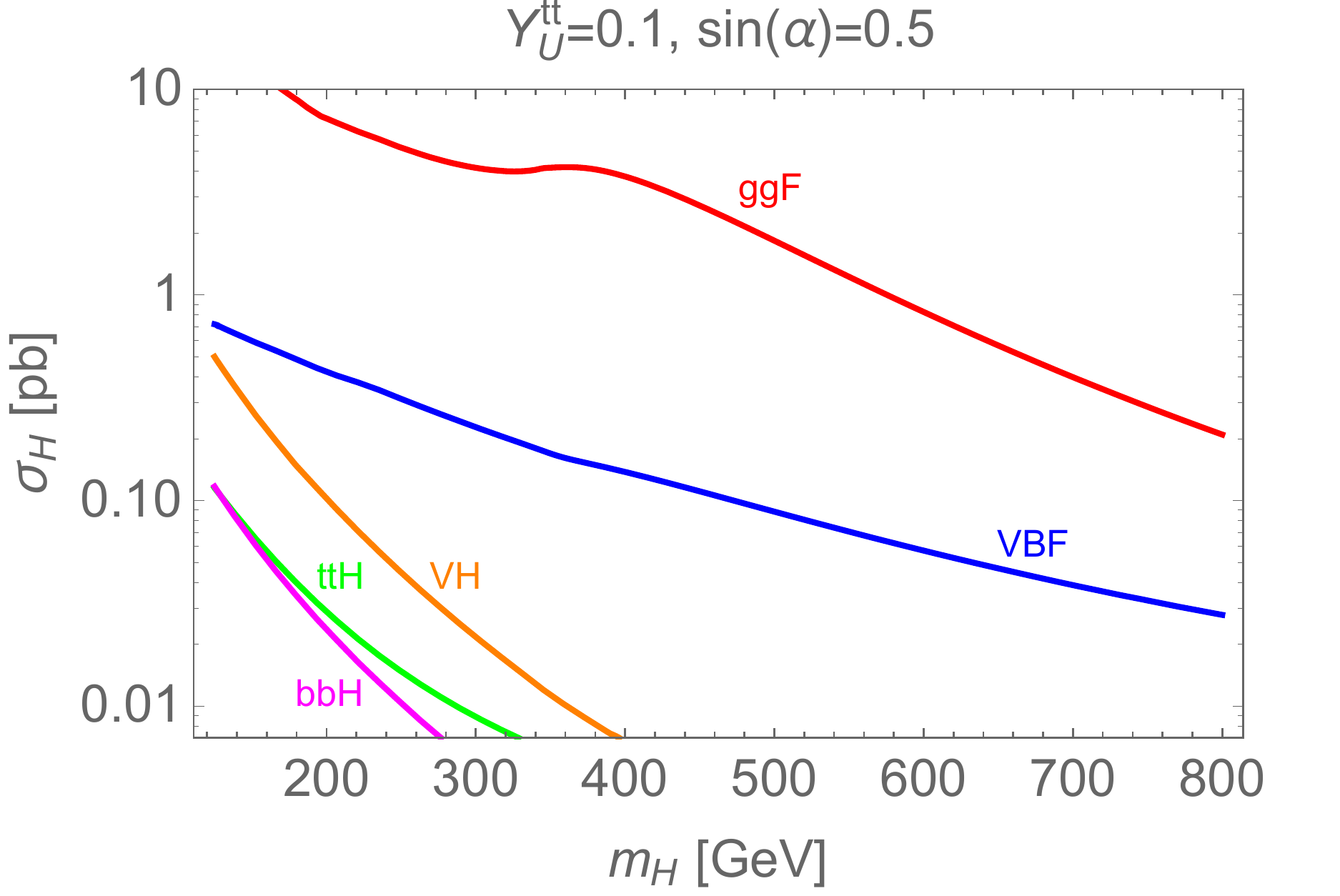}\\[0.1cm]
       \includegraphics[width= 0.49\textwidth]{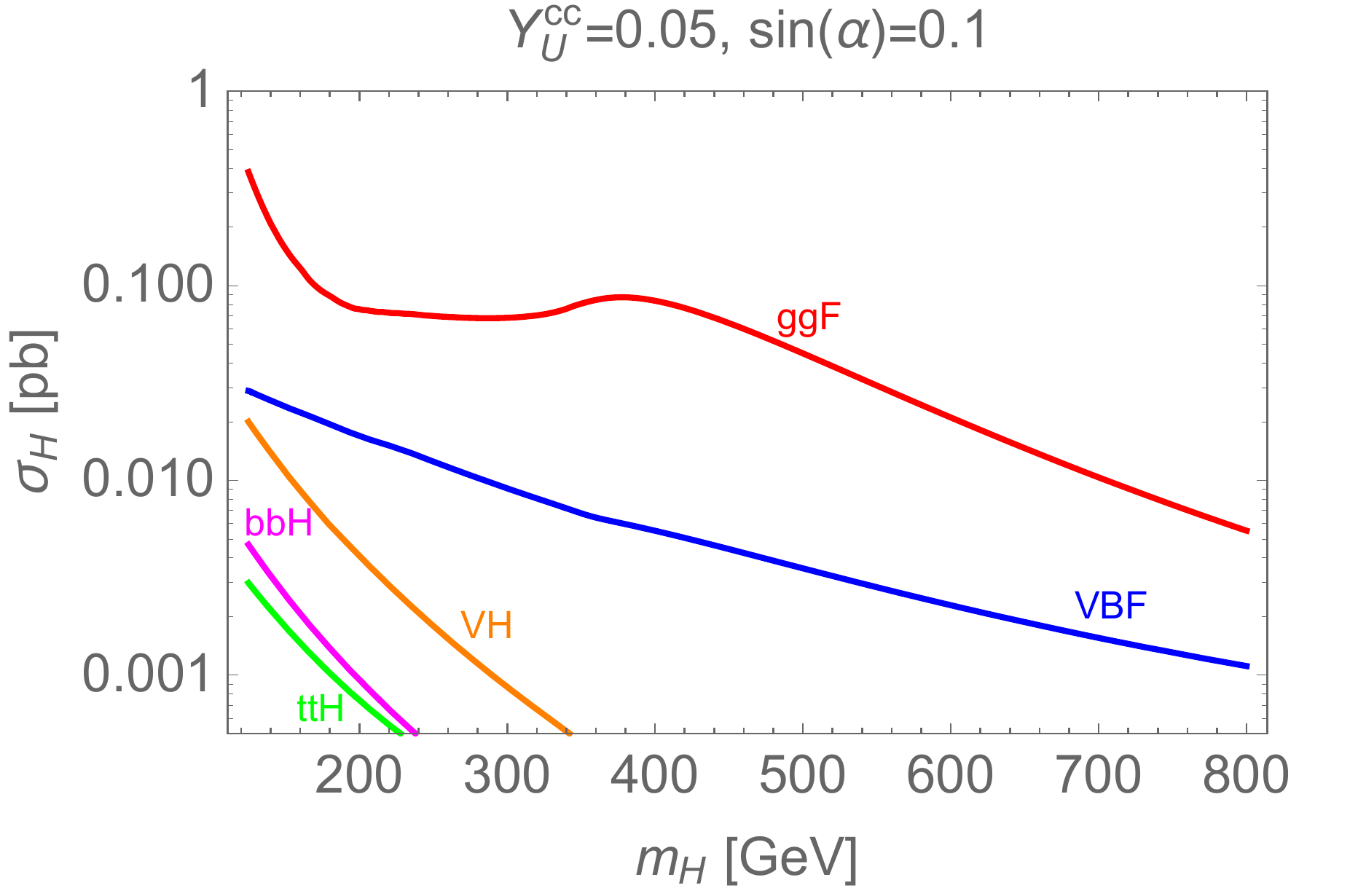}
       \includegraphics[width= 0.49\textwidth]{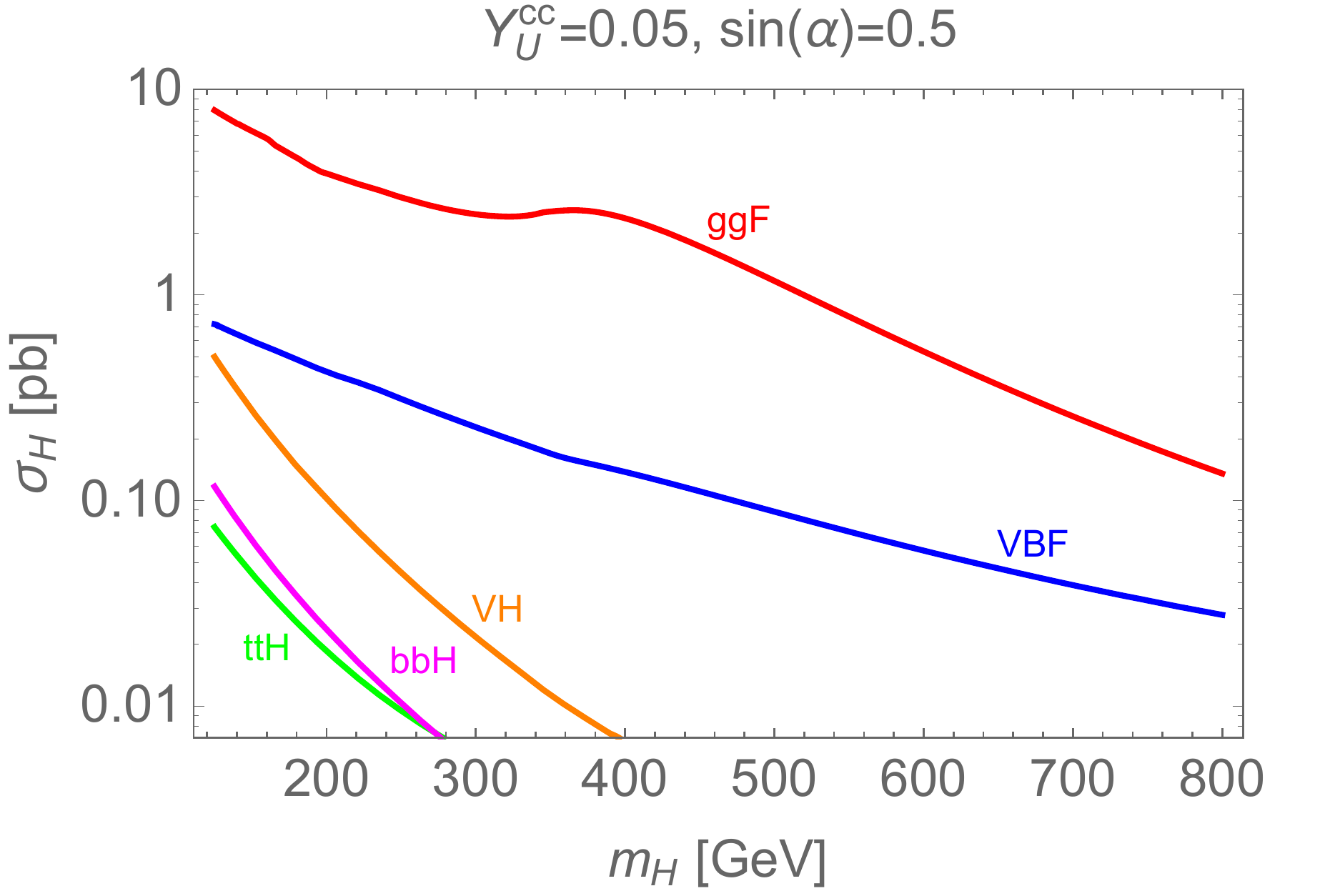}
        \caption{The production cross-section of the neutral scalar $H$ for the 13 TeV LHC as a function of its mass. The production channels are color coded as follow: gluon-fusion (red), $t\bar tH$ (orange), $b\bar bH$ (magenta), VBF (blue) and VH (green). On the top row $Y_U^{tt}=0.1$ while on the bottom row $Y_U^{cc}=0.05$. On the left pane $\sin\alpha=0.1$ and on the right pane $\sin\alpha=0.5$. Other Yukawa couplings are taken to be 0.}
         \label{fig:xsecH}
\end{figure}

The pseudoscalar $A$ can only be produced via the Yukawa interactions. Thus its production depends most sensitively on Yukawa coupling matrix $Y_U$ and $Y_D$. Fig.~\ref{fig:xsecA} shows the production cross-section of $A$, $\sigma_A$, at the 13 TeV LHC as a function of its mass, $m_A$, for the cases $Y_U^{tt}=0.1$ (relevant for the analysis in Sec.~\ref{sec:large}) and $Y_U^{cc}=0.1$ (relevant for the analysis in Sec.~\ref{sec:smallYcc}). 

For the neutral scalar $H$, its production cross-section also depends on the mixing angle $\alpha$. Fig.~\ref{fig:xsecH} shows the production cross-section of $H$, $\sigma_H$, at the 13 TeV LHC as a function of its mass, $m_H$, for the cases $Y_U^{tt}=0.1$ (top row, relevant for the analysis in Sec.~\ref{sec:large}) and $Y_U^{cc}=0.05$ (bottom row, relevant for the analysis in Sec.~\ref{sec:smallYcc}) with the mixing angle $\sin\alpha=0.1$ and 0.5.

\subsection{Decays of neutral Higgs bosons}
The decays of the neutral Higgs bosons, $\phi$, can be expressed in terms of the would be SM Higgs decays.  They are
\begin{equation}
\begin{split}
	\frac{\Gamma^\phi_{VV}}{\Gamma^{SM}_{VV}} &= \delta_\phi^2,\quad
	\frac{\Gamma^\phi_{t\bar t}}{\Gamma^{SM}_{t\bar t}} = \left|\frac{v}{m_t}y_{U,\phi}^{tt}\right|^2,\quad
	\frac{\Gamma^\phi_{b\bar b}}{\Gamma^{SM}_{b\bar b}} = \left|\frac{v}{m_b}y_{D,\phi}^{bb}\right|^2,\quad
	\frac{\Gamma^\phi_{\tau\bar \tau}}{\Gamma^{SM}_{\tau\bar \tau}} = \left|\frac{v}{m_\tau}y_{\ell,\phi}^{\tau\tau}\right|^2,\\
	\frac{\Gamma^\phi_{gg}}{\Gamma^{SM}_{gg}} &\simeq \frac{\left| \sum\limits_{q,i} \frac{v}{m^i_q} \Big[\text{Re}(y_{q,\phi}^{ii})A_{1/2}^H(\tau_{q^i}) + \text{Im}(y_{q,\phi}^{ii})A_{1/2}^A(\tau_{q^i})\Big] \right|^2 }{\left| A_{1/2}^H(\tau_t)\right|^2},\\
	\frac{\Gamma^\phi_{\gamma\gamma}}{\Gamma^{SM}_{\gamma\gamma}} &\simeq \frac{\left| \sum\limits_{f,i} N_cQ^2_{f^i}\frac{v}{m^i_f} \Big[\text{Re}(y_{f,\phi}^{ii})A_{1/2}^H(\tau_{f^i}) + \text{Im}(y_{f,\phi}^{ii})A_{1/2}^A(\tau_{q^i})\Big] + \delta_\phi A_1^H(\tau_w)\right|^2 }{\left| \frac43A_{1/2}^H(\tau_t) +A_1^H(\tau_w)\right|^2},
\end{split}
\end{equation}
where $\delta_h=\cos\alpha$, $\delta_H=\sin\alpha$, $\delta_A=0$, $N_c$ is the number of color, $Q_f$ is the electric charge of fermion $f$ and the loop functions $A_{1}^H(\tau)$ is defined in Eq.~\eqref{eq:loopfunctions}. 
For the $\phi\to\gamma\gamma$ decay, there is also a contribution from the charged Higgs loop. However, this contribution is small thus we will drop it from our analysis.
In our analysis below, we use the would be SM Higgs boson branching ratios provided by the LHC Higgs Cross Section Working Group~\cite{Heinemeyer:2013tqa}.

The off-diagonal elements of the Yukawa coupling matrices in Eq.~\eqref{eq:yukawa} lead to flavor violating decays of the neutral Higgs bosons. 
The partial decay width into final states $ f^i \bar f^j$ can be written as
\begin{equation}
\begin{split}
	\frac{\Gamma_\phi^{f^i\bar f^j}}{m_\phi} &= \frac{1}{8\pi\, m_\phi^2}\left[p_{f^i}\cdot p_{f^j}\left(\left|y_{f,\phi}^{ij}\right|^2+\left|y_{f,\phi}^{ji}\right|^2\right)- m_{f^i}m_{f^k}\left(y_{f,\phi}^{ij}y_{f,\phi}^{ji} +y_{f,\phi}^{ij\ast}y_{f,\phi}^{ji\ast}\right)\right]\\
	&\qquad\times\frac{\sqrt{\left(m_\phi^2 - (m_{f^i}+m_{f^j})^2\right)  \left(m_\phi^2 - (m_{f^i}-m_{f^j})^2\right)}}{m_\phi^2}\\
	&\simeq \frac{1}{16\pi}\left(\left|y_{f,\phi}^{ij}\right|^2+\left|y_{f,\phi}^{ji}\right|^2\right),
\end{split}
\label{eq:partialwidth}
\end{equation}
where in the last line we make the approximation $m_\phi \gg m_f$. 
Since our main interest in this work is on the LFV decays involving tau and muon, therefore the only nonzero off-diagonal elements of $Y$'s that we consider are $Y_{\ell}^{\tau\mu}$ and $Y_{\ell}^{\mu\tau}$.
\begin{figure}
        \centering
       \includegraphics[width= 0.49\textwidth]{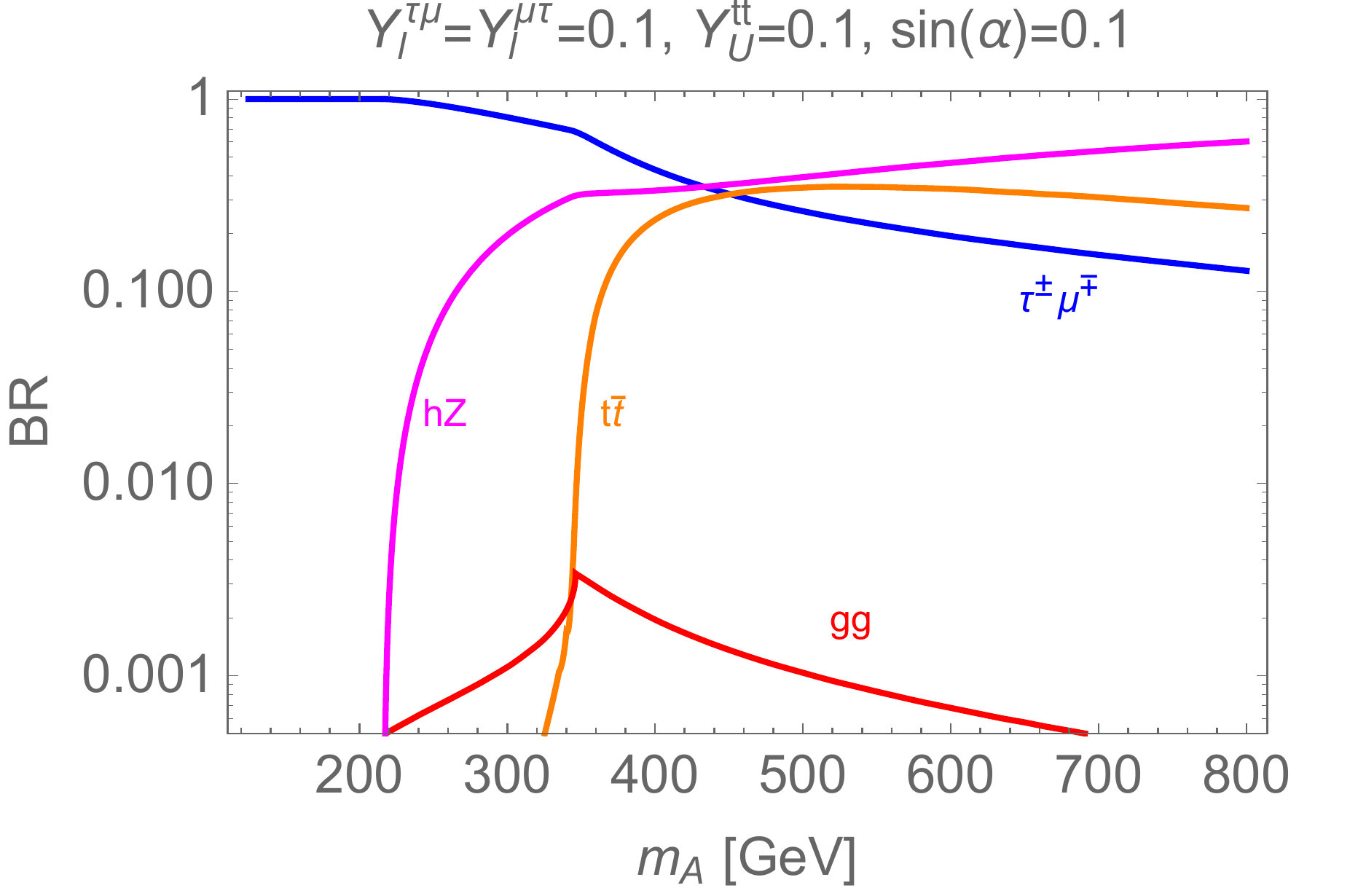}
       \includegraphics[width= 0.49\textwidth]{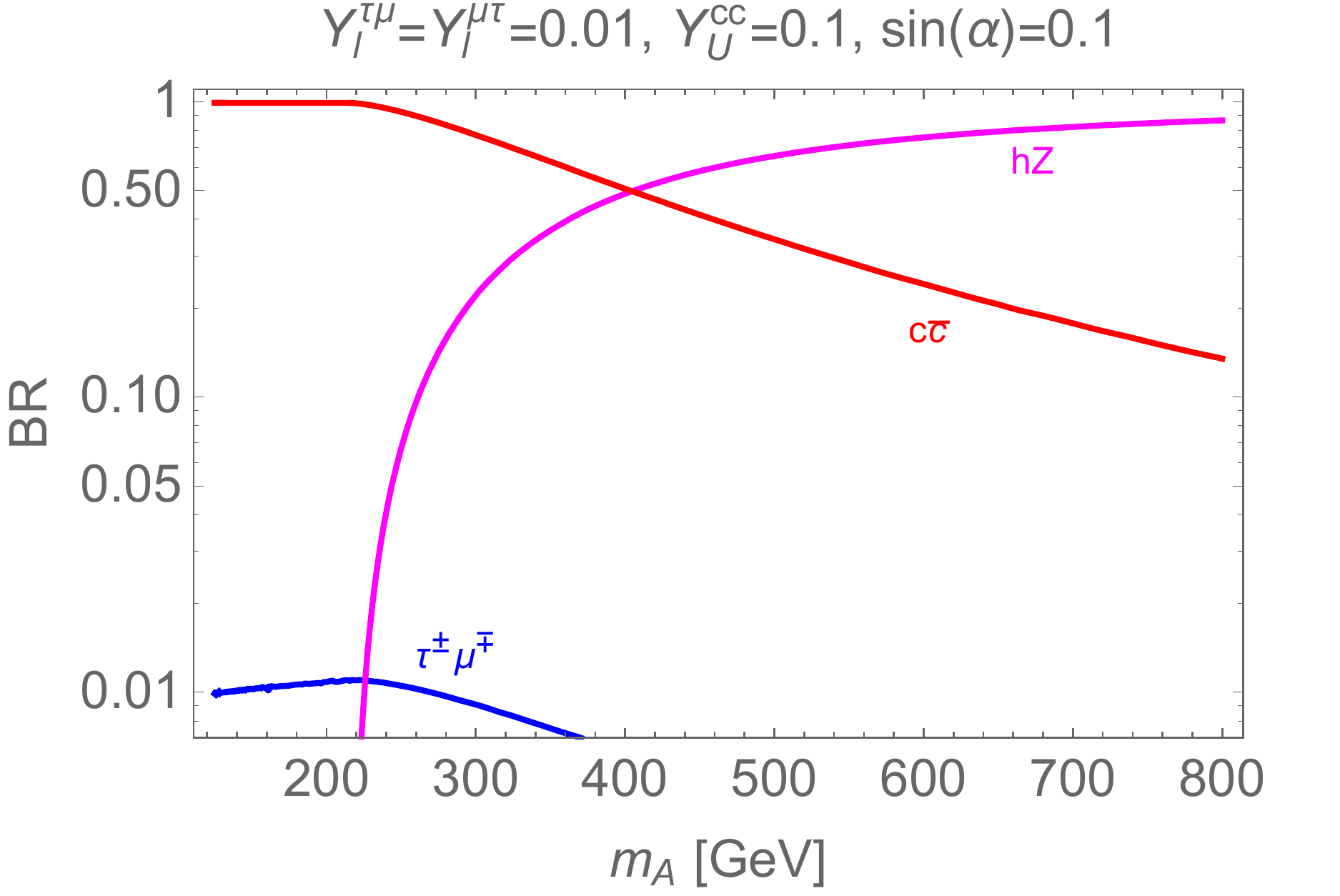}
        \caption{The branching ratios of the pseudoscalar $A$ as a function of its mass. 
        In both plots, $Y_\ell^{\tau\mu}=Y_\ell^{\mu\tau}=0.1$ and $\sin\alpha=0.1$. 
        The $A\to\tau\mu$ channel is shown in blue, $t\bar t$ channel in orange, $hZ$ channel in magenta and $gg\,(c\bar c)$ channel in red. On the left pane, $Y_U^{tt}=0.1$ while on the right pane $Y_U^{cc}=0.1$.
        Other Yukawa couplings are taken to be 0.}
         \label{fig:brA}
\end{figure}

In addition to the decay channels listed in Eq.~\eqref{eq:partialwidth}, the pseudoscalar $A$ can decay to a scalar and a $Z$ boson.
The partial decay width for $A\to hZ$ is given by
\begin{equation}
	\frac{\Gamma^A_{h Z}}{m_A} = \frac{\sin^2\alpha}{16\pi}\frac{m_A^2}{v^2}\left[\left(1-\frac{(m_h+m_Z)^2}{m_A^2}\right)\left(1-\frac{(m_h-m_Z)^2}{m_A^2}\right)\right]^{3/2}.
\end{equation}
The $\Gamma^A_{HZ}$ can be obtained by making a replacement $\sin\alpha\to\cos\alpha$ and $m_h\to m_H$. If the mass $m_H\simeq m_A$, as is the case when an approximate $SO(3)$ symmetry is imposed on the scalar sector~\cite{Dev:2014yca}, the decay channel $A\to HZ$ is closed.
Fig.~\ref{fig:brA} shows the branching ratio of $A$ as a function of its mass for the cases $Y_U^{tt}=0.1$ (relevant for the analysis in Sec.~\ref{sec:large}) and $Y_U^{cc}=0.1$ (relevant for the analysis in Sec.~\ref{sec:smallYcc}). 

\begin{figure}
       \centering
       \includegraphics[width= 0.49\textwidth]{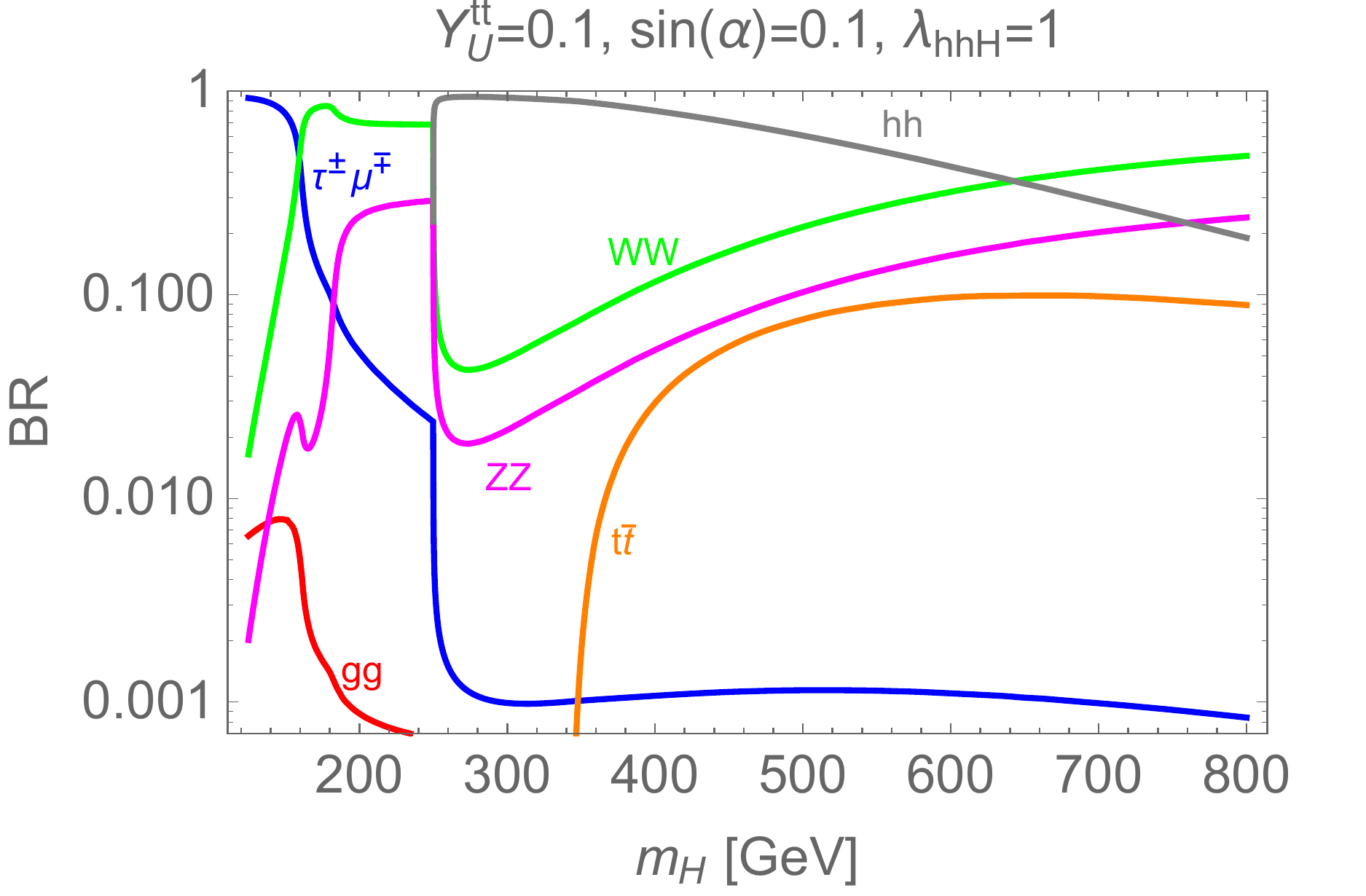}
       \includegraphics[width= 0.49\textwidth]{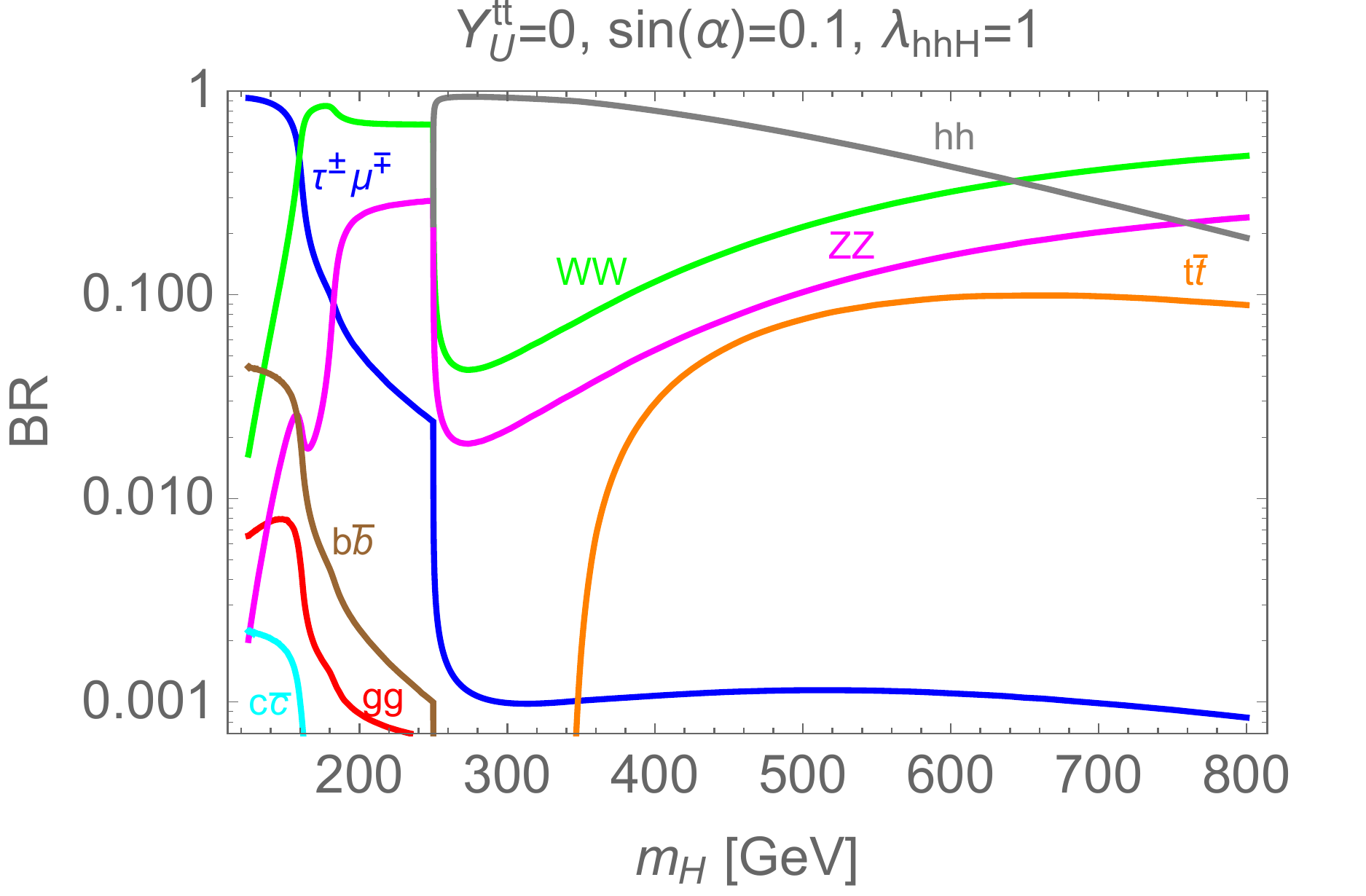}\\[0.1cm]
       \includegraphics[width= 0.49\textwidth]{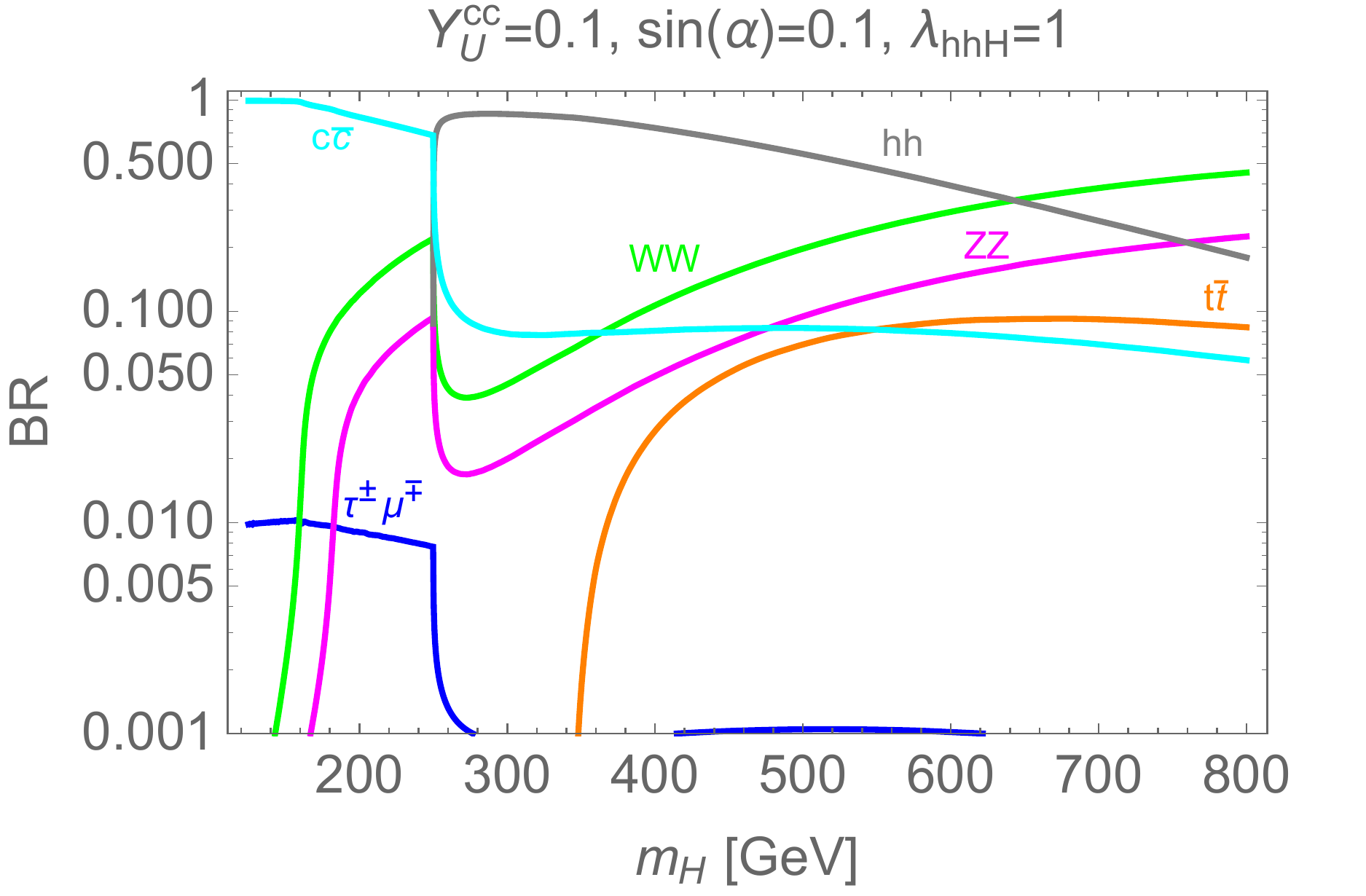}
       \includegraphics[width= 0.49\textwidth]{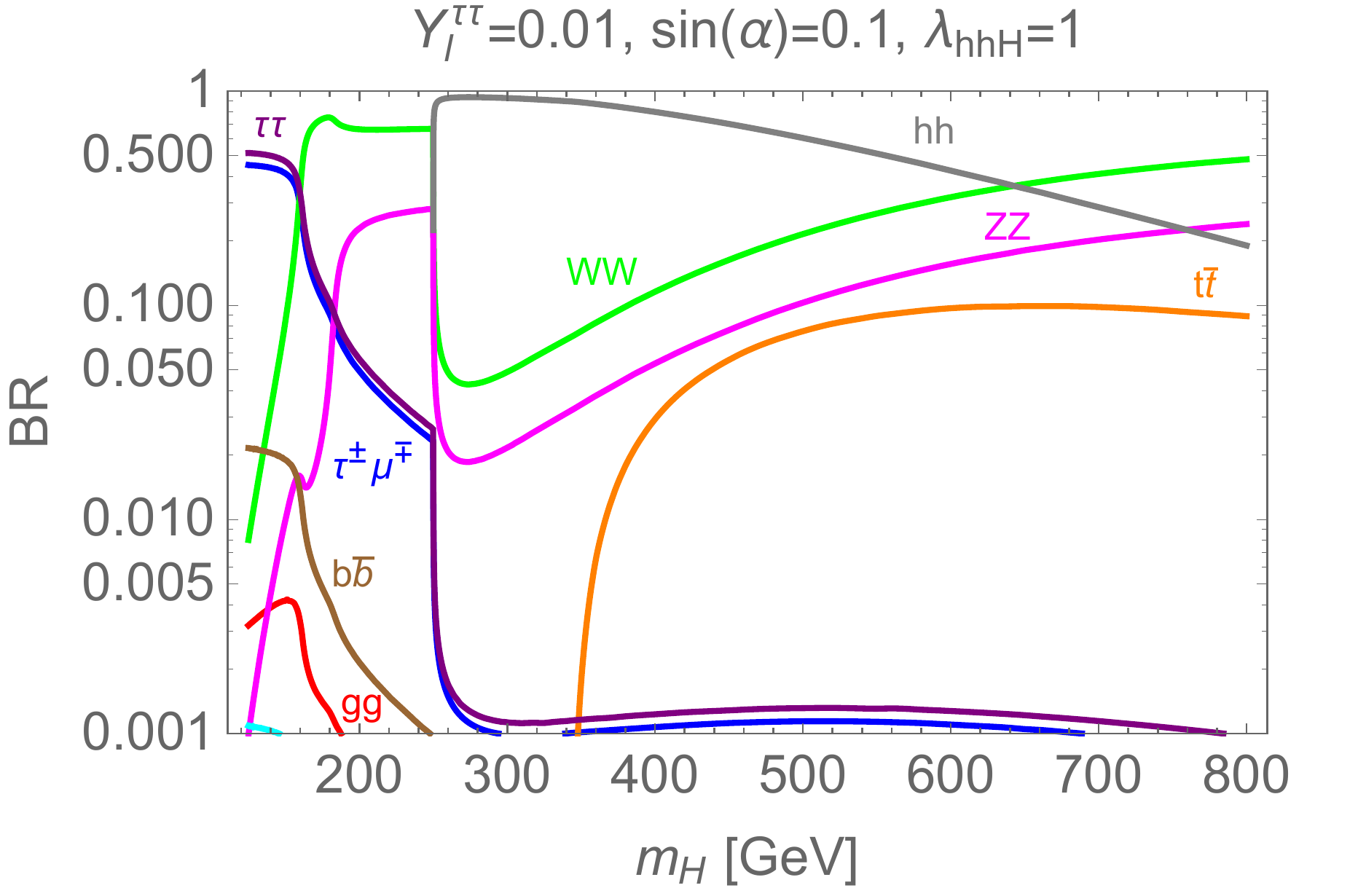}
       \caption{The branching ratios of the heavy neutral scalar $H$ as a function of its mass. The decay into $\tau^\pm\mu^\mp$ is shown in blue, the $t\bar tA$ in orange, the digluon in red, the $W^\pm W^\mp$ in green, the $ZZ$ in magenta, the $hh$ in gray, the $b\bar b$ in brown, the $c\bar c$ in cyan and the $\tau^+\tau^-$ in purple. In all plots, $Y_\ell^{\tau\mu}=Y_\ell^{\mu\tau}=0.01$, $\sin\alpha=0.1$ and $\lambda_{hhH}=1$. On the upper left, $Y_U^{tt} = 0.1$, on the upper right $Y_U^{tt}=0$, on the bottom left $Y_U^{cc}=0.1$ and on the bottom right $Y_\ell^{\tau\tau}=0.01$. Other Yukawa couplings are taken to be 0.}
         \label{fig:brH}
\end{figure}

The heavy neutral scalar $H$ can also decay into a pair of lighter Higgs bosons, $\phi=h, A$ and $H^+H^-$ if it is kinematically open.  We can parametrize this decay channel by introducing the coupling $\lambda_{\phi\phi H}$ such that $\mathcal{L}\supset \frac12 \lambda_{\phi\phi H}vH\phi^2$ for neutral resonance $\phi$. For the charged Higgs, the coupling is defined without a 1/2 factor. Then, the partial decay width $H\to\phi\phi$ ($\phi=h,A$) is
\begin{equation}
	\Gamma_{H}^{\phi\phi} = \frac{\lambda_{\phi\phi H}^{2}}{32\pi}\frac{v^2}{m_H}\sqrt{1-\frac{4m_\phi^2}{m_H^2}}.
\end{equation}
For the case of $H\to H^+H^-$, there is an extra factor of 2 dues to $H^+$ and $H^-$ being distinct particles.
We note the coupling $\lambda_{\phi\phi H}$ depends strongly on the scalar potential. 
Fig.~\ref{fig:brH} show the branching ratios of $H$ as a function of its mass for 4 different benchmark scenarios. Each benchmark is relevant for our analysis in Sec.~\ref{sec:result}. 
In all these plots, we assume the decay $H\to AA$ and $H\to H^+H^-$ are kinematically closed.

\section{Lepton flavor violation in the Higgs sector}
\label{sec:LFV}
The off-diagonal elements in the Yukawa matrice $Y_f$'s induce flavor violating decays of the Higgs bosons. In this paper, for simplicity, we will assume that flavor violations reside only in the lepton sector.  Moreover, we will focus our attention on the $\tau-\mu$ LFV. Thus the only non-zero off-diagonal entries of $Y_\ell$ that we consider are $Y_\ell^{\tau\mu}$ and $Y_\ell^{\mu\tau}$.

The couplings $Y_\ell^{\tau\mu}$ and $Y_\ell^{\mu\tau}$ can be probed directly at the LHC or indirectly via low energy precision measurements.  Here we summarize the current constraints on $Y_\ell^{\tau\mu}$ and $Y_\ell^{\mu\tau}$.
\subsection{Direct constraints on $Y_\ell^{\tau\mu}$ and $Y_\ell^{\mu\tau}$}

\begin{figure}
        \centering
       \includegraphics[width= 0.49\textwidth]{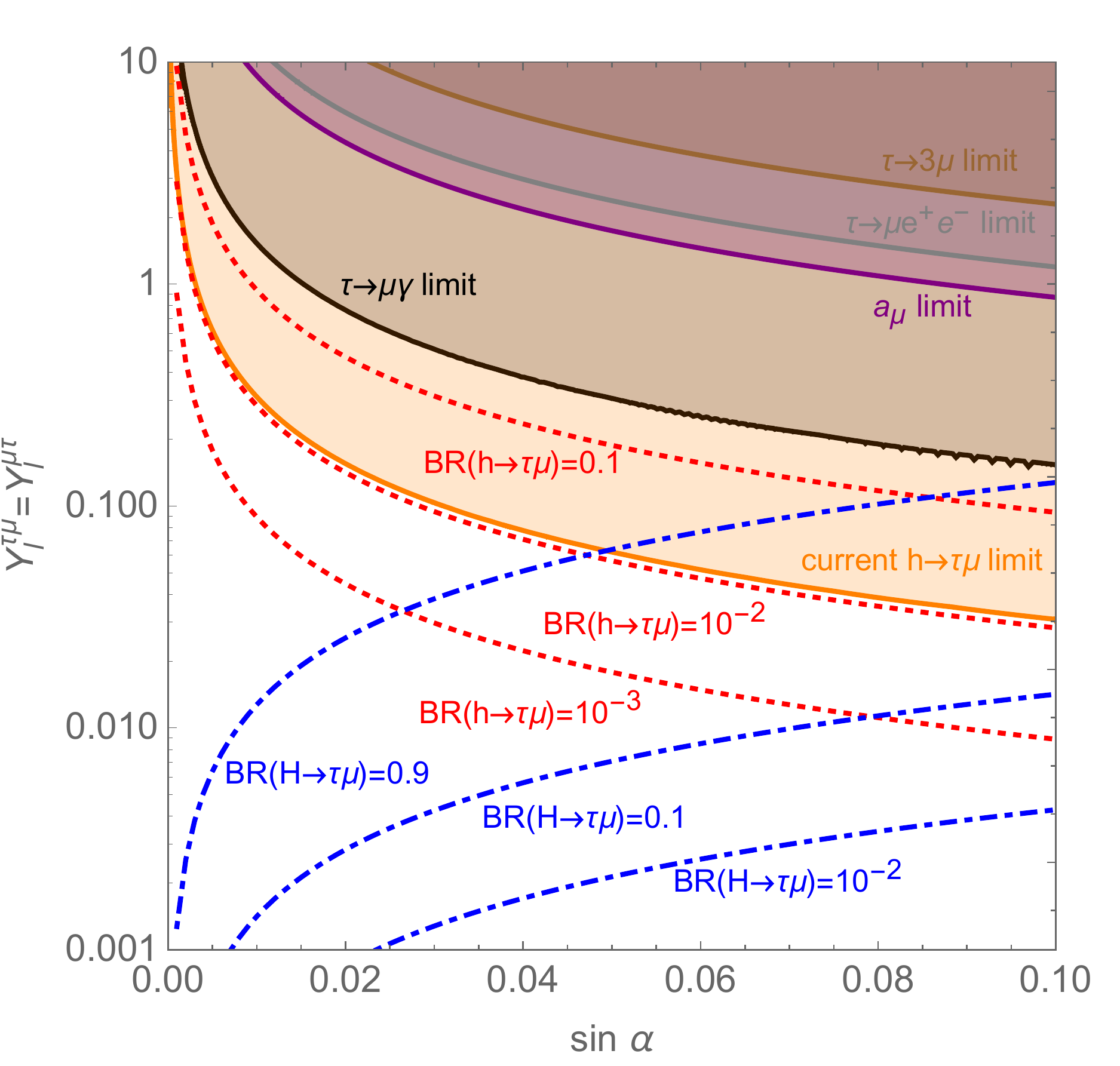}
       \includegraphics[width= 0.49\textwidth]{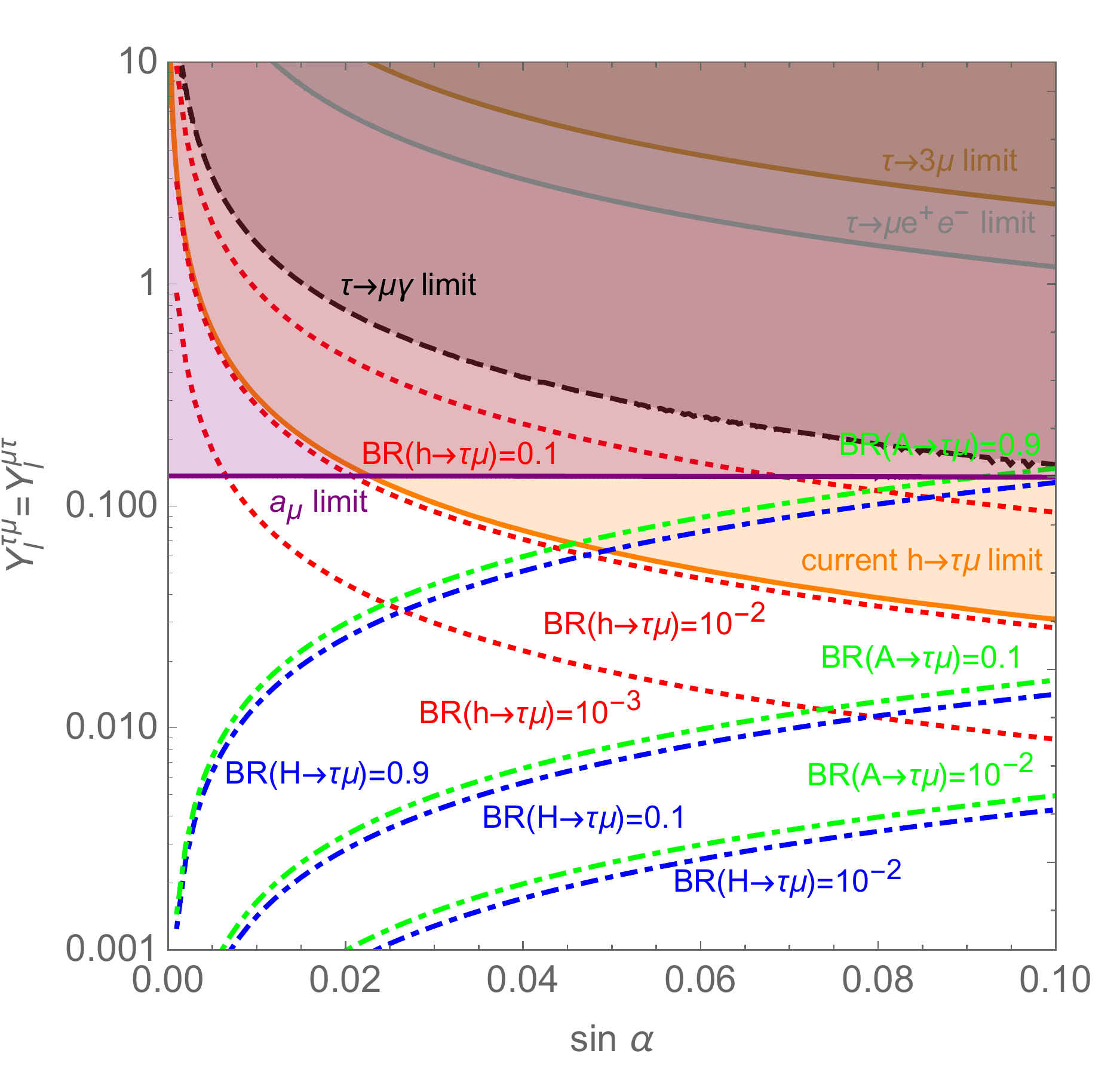}
        \caption{The contours of constant branching ratio for $h\to\tau\mu$ (dotted red lines) and $H\to\tau\mu$ (dashed blue lines). The orange shaded region is excluded by CMS search for $h\to\tau\mu$ decays. The black shaded region is excluded by the $\tau\to\mu\gamma$ search. The purple shaded region is excluded by the muon magnetic dipole moment measurement. The gray shaded region and the brown shaded region are excluded by the $\tau\to3\mu$ and $\tau\to\mu e^+e^-$ search respectively.
        Note in making these plots the other Yukawa couplings $Y_f^{ij}$ are taken to be vanishing and $m_H=200$ GeV. On the left pane, $m_A=m_{H^+}=200$ GeV, thus $Br(A\to\tau\mu)=1$. On the right pane, $m_A=m_{H^+}=300$ GeV.}
         \label{fig:ytaumuvssinalpha}
\end{figure}

The coupling $Y_\ell^{\tau\mu}$ and $Y_\ell^{\mu\tau}$ lead to flavor violating decays $\phi\to\tau\mu$ for the neutral Higgs bosons.  In particular, for the 125 GeV resonance, $h$, the partial width for such a decay is
\begin{equation}
\begin{aligned}
	\Gamma_h^{\tau^\pm\mu^\mp} &= \frac{m_h}{16\pi}\left(\left|y_{\ell,h}^{\tau\mu}\right|^2 + \left|y_{\ell,h}^{\mu\tau}\right|^2\right)\\
	&= \frac{m_h \sin^2\alpha}{16\pi}\left(\left|Y_{\ell}^{\tau\mu}\right|^2 + \left|Y_{\ell}^{\mu\tau}\right|^2\right).
\end{aligned}
\label{eq:h2taumu}
\end{equation}
This decay mode has been searched for at the LHC by the CMS collaboration. Currently, CMS has placed an upper bound on the branching ratio at $Br(h\to\tau\mu)<1.20\%$~\cite{CMS-PAS-HIG-16-005}. This bound translates to the bound on the mixing angle $\alpha$ and the coupling $Y_\ell^{\tau\mu}$ and $Y_\ell^{\mu\tau}$ as shown by the solid orange line in Fig.~\ref{fig:ytaumuvssinalpha}. Notice that the branching ratio $h\to\tau\mu$ is tightly constrained by the CMS limit.  However, the branching ratio $H\to\tau\mu$ (or $A\to\tau\mu$) can still be large.
\subsection{Indirect constraints on $Y_\ell^{\tau\mu}$ and $Y_\ell^{\mu\tau}$}
Extensive analyses of flavor constraints on the Yukawa structure of the Type-III 2HDM have been carried out in Ref.~\cite{Crivellin:2013wna}. Here we'll focus on the flavor observables relevant for tau-muon LFV.
\subsubsection{$\tau \to \mu\gamma$}
The flavor violating couplings $Y_\ell^{\tau\mu}$ and $Y_\ell^{\mu\tau}$ lead to a rare decay $\tau\to\mu\gamma$. Their contributions can be computed by first matching to the effective operators~\cite{Harnik:2012pb} 
\begin{equation}
	\mathcal{L}_{\tau\to\mu\gamma} = \frac{e m_\tau}{8\pi^2}c_L \bar{\mu}\sigma^{\alpha\beta}P_L\tau F_{\alpha\beta} + \frac{e m_\tau}{8\pi^2}c_R \bar{\mu}\sigma^{\alpha\beta}P_R\tau F_{\alpha\beta} + \hc,
	\label{eq:dipole}
\end{equation}
where $F_{\alpha\beta}$ is the $U(1)_{EM}$ field strength tensor. In terms of the Wilson coefficients $c_L$ and $c_R$, the decay rate for $\tau\to\mu\gamma$ can be written as
\begin{equation}
	\Gamma_{\tau\to\mu\gamma} = \frac{\alpha m_\tau^5}{64\pi^4}\left(|c_L|^2 + |c_R|^2\right).
\end{equation}
The experimental bound on the branching ratio is Br$(\tau\to\mu\gamma)<4.4\times10^{-8}$~\cite{Olive:2016xmw}.

\begin{figure}
        \centering
       \includegraphics[width= 0.3\textwidth]{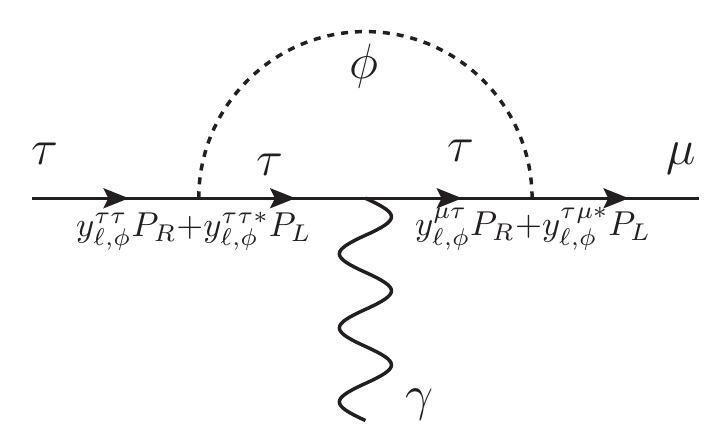}
       \includegraphics[width= 0.3\textwidth]{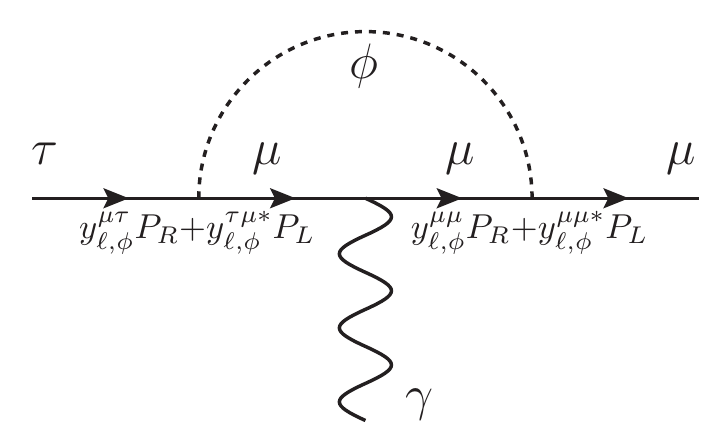}
       \includegraphics[width= 0.3\textwidth]{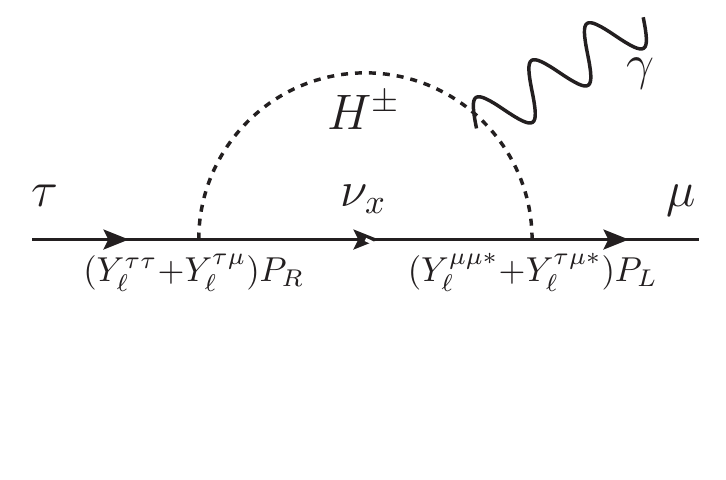}
        \caption{The one-loop diagrams contributing to $\tau\to\mu\gamma$ decays induced by the Higgs bosons with flavor violating Yukawa couplings $Y_\ell^{\tau\mu}$ and $Y_\ell^{\mu\tau}$.}
         \label{fig:tau2mugamma1}
\end{figure}

The Wilson coefficients $c_{L,R}$ get contributions from both one-loop and two-loop diagrams. At the one-loop level, the contributions arise from diagrams shown in Fig.~\ref{fig:tau2mugamma1}. Their  contributions are~\cite{Efrati:2016uuy}
\begin{align}
	c_L &=  \sum_{\phi=h,H} \frac{y_{\ell,\phi}^{\tau\tau}y_{\ell,\phi}^{\tau\mu\ast}}{12m_\phi^2}\left(-4 + 3\ln\frac{m_\phi^2}{m_\tau^2}\right) - \frac{Y_{\ell}^{\tau\tau}Y_{\ell}^{\tau\mu\ast}}{24m_A^2}\left(-5 + 3\ln\frac{m_A^2}{m_\tau^2}\right)  -\frac{Y_{\ell}^{\tau\tau}Y_{\ell}^{\tau\mu\ast}}{12m_{H^+}^2},\\
	c_R &=  \sum_{\phi=h,H} \frac{y_{\ell,\phi}^{\tau\tau\ast}y_{\ell,\phi}^{\mu\tau}}{12m_\phi^2}\left(-4 + 3\ln\frac{m_\phi^2}{m_\tau^2}\right) - \frac{Y_{\ell}^{\tau\tau\ast}Y_{\ell}^{\mu\tau}}{24m_A^2}\left(-5 + 3\ln\frac{m_A^2}{m_\tau^2}\right) -\frac{Y_{\ell}^{\tau\tau\ast}Y_{\ell}^{\mu\tau}}{12m_{H^+}^2},
\end{align}
where we have assumed $|y_{\ell,\phi}^{\tau\tau}|\gg |y_{\ell,\phi}^{\mu\mu}|$. Notice that $c_L$ and $c_R$ are related by $Y_\ell^{ij}\leftrightarrow Y_\ell^{ji\ast}$. 
The coefficients $c_L$ and $c_R$ also get corrections from two-loop processes which can be as large as the one-loop contributions. The expressions for the two-loop contributions, $\Delta c_L$ and $\Delta c_R$, are given in App.~\ref{app:tautomugamma}. In the case that $Y_\ell^{\tau\mu}=Y_\ell^{\mu\tau}$ are the only non-vanishing Yukawa couplings, the bound from $\tau\to\mu\gamma$ is shown in the solid black line in the left pane of Fig.~\ref{fig:ytaumuvssinalpha}.

\subsubsection{Muon magnetic dipole moment}
\begin{figure}
        \centering
       \includegraphics[width= 0.4\textwidth]{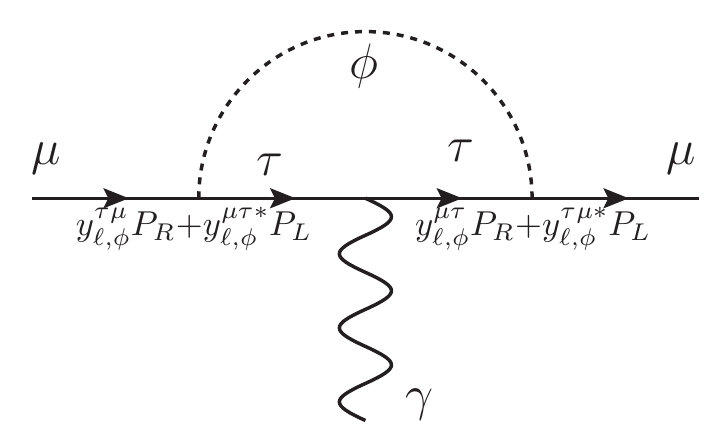}
        \caption{The $Y_\ell^{\tau\mu}$ and $Y_\ell^{\mu\tau}$ contribution to the muon magnetic dipole moment.}
         \label{fig:muon_amu}
\end{figure}
The $Y_\ell^{\tau\mu}$ and $Y_\ell^{\mu\tau}$ coupling also contribute to a magnetic dipole moment for the muon, $a_\mu\equiv (g_\mu-2)/2$, as shown in Fig.~\ref{fig:muon_amu}. Their contribution can be easily translated from the result in Ref.~\cite{Harnik:2012pb} 
\begin{equation}
\begin{aligned}
	\delta a_\mu &\simeq \frac{m_\mu m_\tau}{16\pi^2}\sum_\phi \frac{\text{Re}(y_{\ell,\phi}^{\mu\tau}y_{\ell,\phi}^{\tau\mu}) }{m_{\phi}^2}\left(2\ln\frac{m_{\phi}^2}{m_\tau^2}-3\right),
\end{aligned}
\end{equation}
where $\phi=h$, $H$ and $A$.
Note in the above expression we have dropped terms suppressed by $m_\mu/m_\tau$ and $m_\tau/m_{\phi}$. 

The discrepancy between SM prediction and the measured value is~\cite{Bennett:2006fi,Olive:2016xmw}
\begin{equation}
	\Delta a_\mu \equiv a_\mu^{exp} - a_\mu^{SM} = (2.87 \pm 0.80)\times 10^{-9}.
\end{equation}
This can be used to bound the LFV contribution to $a_\mu$, $\delta a_\mu \le 4.47\times10^{-9}$ at 95\% C.L..
From the form of the Yukawa couplings in Eq.~\eqref{eq:reducedcoupling}, one can see that the pseudoscalar contribution and the scalar contributions have opposite sign. Thus in the case where $m_H$ and $m_A$ are nearly degenerate and a small mixing angle, their contributions to $\delta a_\mu$ cancel each other. Therefore the bound from the muon magnetic dipole moment is expected to be weak, see for example the left pane of Fig.~\ref{fig:ytaumuvssinalpha}. 
On the other hand, if $m_H$ and $m_A$ are different, the bound from $a_\mu$ could be strong as can be seen from the right pane of Fig.~\ref{fig:ytaumuvssinalpha}.

\subsubsection{Muon electric dipole moment}
In effective theory, the muon electric dipole moment is described by
\begin{equation}
	\mathcal{L}_{EDM} = -\frac{i}{2}d_\mu\left(\bar\mu \sigma^{\alpha\beta}\gamma^5\mu\right)F_{\alpha\beta},
\end{equation}
where $F_{\alpha\beta}$ is the $U(1)_{EM}$ field strength tensor.  The coefficient $d_\mu$ is~\cite{Harnik:2012pb}
\begin{equation}
\begin{aligned}
	d_\mu &\simeq -\frac{em_\tau}{32\pi^2}\sum_\phi \frac{\text{Im}(Y_{\ell, \phi}^{\mu\tau}Y_{\ell,\phi}^{\tau\mu}) }{m_{\phi}^2}\left(2\ln\frac{m_{\phi}^2}{m_\tau^2}-3\right),
\end{aligned}
\end{equation}
where $\phi=h$, $H$ and $A$ and we have dropped the term suppressed by $m_\mu/m_\tau$ and $m_\tau/m_{\phi}$.
Since in this work we are interested in the simple case where $Y_\ell^{\tau\mu}$ and $Y_\ell^{\mu\tau}$ are real, the muon electric dipole moment constraint does not apply to our scenario.

\subsubsection{$\tau\to3\mu$ and $\tau\to\mu e^+e^-$} \label{sec:taumull}
\begin{figure}
        \centering
       \includegraphics[width= 0.4\textwidth]{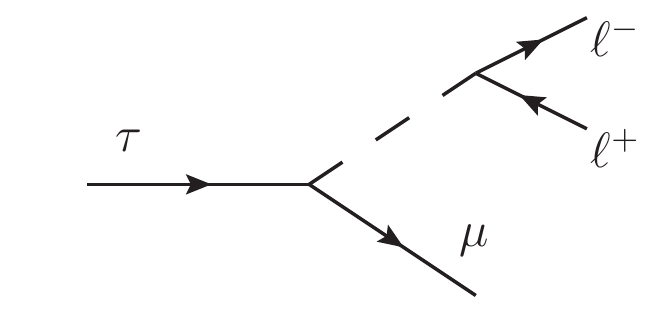}
       \includegraphics[width= 0.4\textwidth]{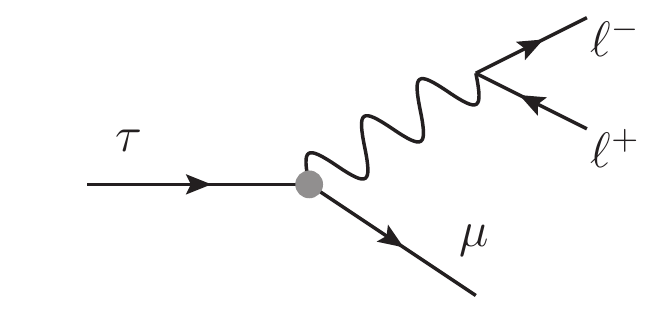}
        \caption{The diagrams contributing to $\tau\to\mu\ell^+\ell^-$ decays. The diagram on the right arises from the effective dipole operators in Eq.~\eqref{eq:dipole}.}
         \label{fig:tau3mu}
\end{figure}

The flavor violating couplings $Y_\ell^{\tau\mu}$ and $Y_\ell^{\mu\tau}$ also lead to a decay $\tau \to 3\mu$ as well as $\tau\to\mu e^+e^-$ as shown in Fig.~\ref{fig:tau3mu}. 
Their contributions can be easily computed by matching onto effective operators. The relevant effective operators for $\tau\to3\mu$ are the dipole operators in Eq.~\eqref{eq:dipole} and the 4-fermion operators
\begin{equation}
	\mathcal{L}_{4f} \supset c^\mu_{xy} \,\bar\mu P_x\tau \,\bar\mu P_y\mu + c^e_{xy} \,\bar\mu P_x\tau \,\bar e P_ye,
\end{equation} 
where $\{x,y\} = \{L,R\}$ and we assume LFV resides only in the tau-mu couplings. For the case of a real Yukawa matrix $Y_\ell$ and dropping terms suppressed by $m_\mu/v$, we get
\begin{equation}
\begin{aligned}
	c^\mu_{LL} &= Y^{\tau\mu}_\ell Y^{\mu\mu}_\ell\left(\frac{\sin^2\alpha}{m_h^2}+\frac{\cos^2\alpha}{m_H^2} - \frac{1}{m_A^2}\right),\\
	c^\mu_{RR} &= Y^{\mu\tau}_\ell Y^{\mu\mu}_\ell\left(\frac{\sin^2\alpha}{m_h^2}+\frac{\cos^2\alpha}{m_H^2} - \frac{1}{m_A^2}\right),
\end{aligned}
\,
\begin{aligned}
	c^\mu_{RL} &= Y^{\tau\mu}_\ell Y^{\mu\mu}_\ell\left(\frac{\sin^2\alpha}{m_h^2}+\frac{\cos^2\alpha}{m_H^2} + \frac{1}{m_A^2}\right),\\
	c^\mu_{LR} &= Y^{\mu\tau}_\ell Y^{\mu\mu}_\ell\left(\frac{\sin^2\alpha}{m_h^2}+\frac{\cos^2\alpha}{m_H^2} + \frac{1}{m_A^2}\right).
\end{aligned}
\label{eq:cLL}
\end{equation}
The expressions for $c^e_{xy}$ can be obtained by a replacement $Y_\ell^{\mu\mu}\to Y_\ell^{ee}$. 
In terms of these Wilson coefficients, the doubly differential partial width for $\tau^-(p)\to\mu^-(p_1)\mu^+(p_2)\mu^-(p_3)$ is~\cite{Celis:2014asa}
\begin{align}
	\frac{d^2\Gamma_{\tau\to3\mu} }{dm_{13}^2\,dm_{23}^2} &= \frac{1}{1024\pi^3m_\tau^3}\left\{\frac{4\alpha^2m_\tau^2}{\pi^2m_{23}^2(m_{13}^2+m_{23}^2-m_\tau^2)}\left[-2m_\tau^2(2m_{13}^4+4m_{13}^2m_{23}^2+m_{23}^4)\right.\right. \nn\\
	&\qquad\left. +2m_{13}^2(m_{13}^4+3m_{13}^2m_{23}^2+3m_{23}^4) + m_\tau^4(3m_{13}^2+2m_{23}^2)-m_\tau^6\right] \left(|c_L|^2+|c_R|^2\right)\nn\\
	&\qquad + \left(m_{23}^2(m_\tau^2-m_{23}^2) + (m_{13}^2+m_{23}^2)(m_\tau^2-m_{13}^2-m_{23}^2)\right)\left(|c^\mu_{LR}|^2 + |c^\mu_{RL}|^2\right)\nn\\
	&\qquad + m_{13}^2(m_\tau^2-m_{13}^2)\left(|c^\mu_{LL}|^2+|c^\mu_{RR}|^2\right) \bigg\},
\end{align}
where $m_{ij}^2=(p_i+p_j)^2$. Note in the above expression we set $m_\mu=0$. Similarly, the doubly differential partial decay width  for $\tau^-(p)\to\mu^-(p_1)e^+(p_2)e^-(p_3)$ is
\begin{align}
	\frac{d^2\Gamma_{\tau\to\mu e^+e^-} }{dm_{13}^2\,dm_{23}^2} &= \frac{1}{1024\pi^3m_\tau^3}\left\{\frac{4\alpha^2m_\tau^2}{\pi^2m_{23}^2(m_{13}^2+m_{23}^2-m_\tau^2)}\left[-2m_\tau^2(2m_{13}^4+4m_{13}^2m_{23}^2+m_{23}^4)\right.\right. \nn\\
	&\qquad\left. +2m_{13}^2(m_{13}^4+3m_{13}^2m_{23}^2+3m_{23}^4) + m_\tau^4(3m_{13}^2+2m_{23}^2)-m_\tau^6\right] \left(|c_L|^2+|c_R|^2\right)\nn\\
	&\qquad + m_{23}^2(m_\tau^2-m_{23}^2)\left(|c^e_{LL}|^2+|c^e_{RR}|^2 + |c^e_{LR}|^2 + |c^e_{RL}|^2\right) \bigg\}.
\end{align}
Numerically, they are
\begin{align}
	\Gamma_{\tau\to3\mu} &= 4.46\times10^{-8}\left(|c_L|^2+|c_R|^2\right) + 4.37\times10^{-5}\left(|c^\mu_{LL}|^2+|c^\mu_{RR}|^2\right) \nn\\
	&\qquad + 8.65\times10^{-5}\left(|c^\mu_{RL}|^2+|c^\mu_{LR}|^2\right)\text{GeV}^5\\
	\Gamma_{\tau\to\mu e^+e^-} &= 1.41\times10^{-7}\left(|c_L|^2+|c_R|^2\right)\nn\\
	&\qquad + 4.50\times10^{-5}\left(|c^e_{LL}|^2+|c^e_{RR}|^2 + |c^e_{LR}|^2 + |c^e_{RL}|^2\right) \text{GeV}^5.
\end{align}
As can be seen from the Wilson coefficients in Eq.~\eqref{eq:cLL}, $\Gamma_{\tau\to3\mu}$ and $\Gamma_{\mu e^+e^-}$ depends on the coupling $Y_\ell^{\mu\mu}$ and $Y_\ell^{ee}$. These couplings are constrained by the direct $h\to\mu\mu$~\cite{ATLAS:2016zzs} and $h\to ee$~\cite{Altmannshofer:2015qra} search with $\sin\alpha\,Y_\ell^{\mu\mu}\lesssim 2.13\times10^{-3}$ and $\sin\alpha\,Y_\ell^{ee}\lesssim 1.76\times10^{-3}$. 

Current experimental bounds on the $\tau\to3\mu$ and $\tau\to\mu e^+e^-$ decays are~\cite{Olive:2016xmw} 
\begin{equation}
	\text{Br}(\tau\to3\mu) = 2.1\times10^{-8},\qquad
	\text{Br}(\tau\to\mu e^+e^-) = 1.8\times10^{-8}.
\end{equation}
In the limit that $Y_\ell^{ee} = Y_\ell^{\mu\mu} = 0$, these bounds translate to the limit on the LFV couplings $Y_\ell^{\tau\mu}$ and $Y_\ell^{\mu\tau}$ as shown in Fig.~\ref{fig:ytaumuvssinalpha}.

\subsubsection{$\tau \rightarrow \mu M$} \label{subsection:taumum}
\begin{figure}
        \centering
        \subfloat[Bounds on $Y_U^{uu}$ from $\tau \rightarrow \mu \pi$]{\includegraphics[width= 0.4\textwidth]{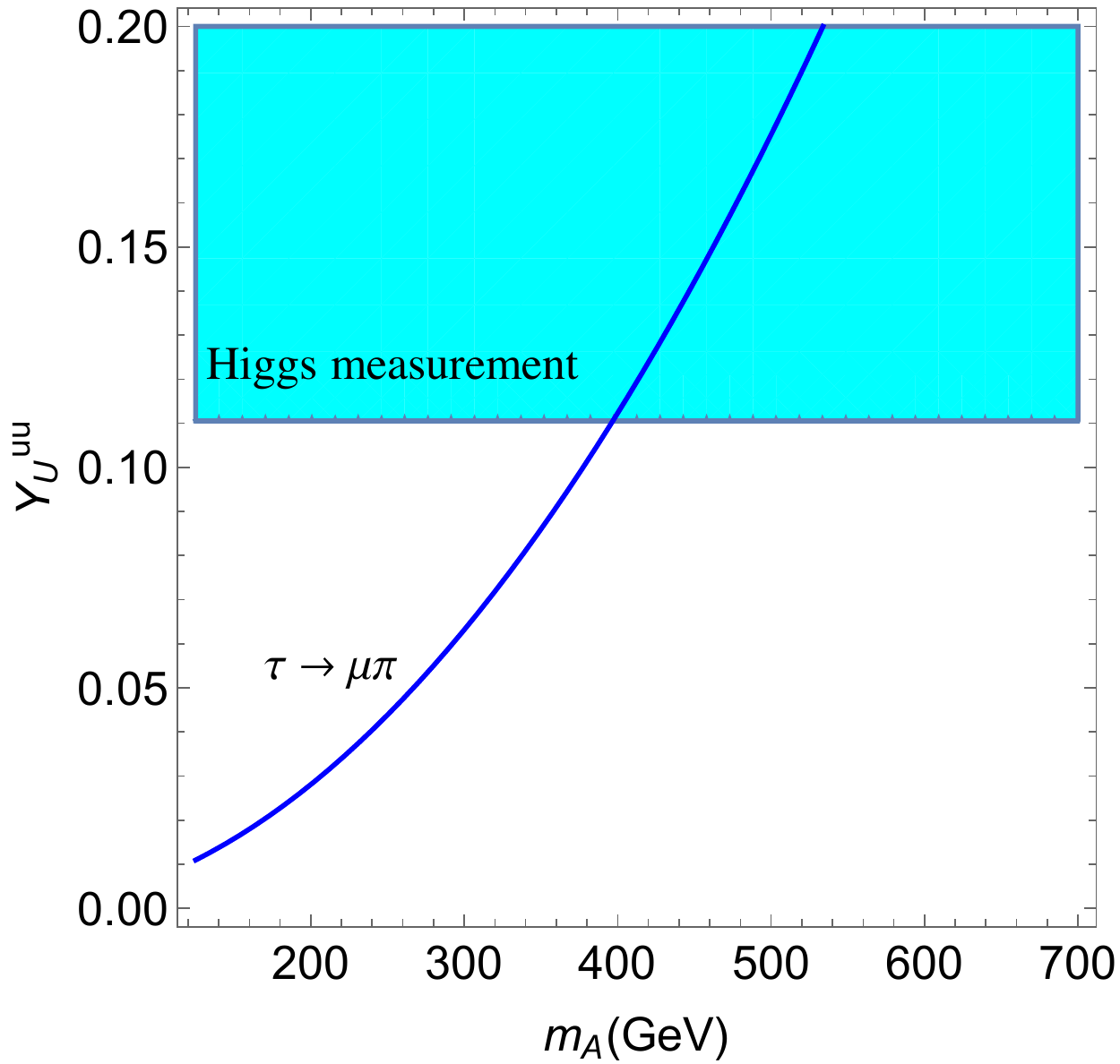}}
        \subfloat[Bounds on $Y_D^{dd}$ from $\tau \rightarrow \mu \pi$]{\includegraphics[width= 0.4\textwidth]{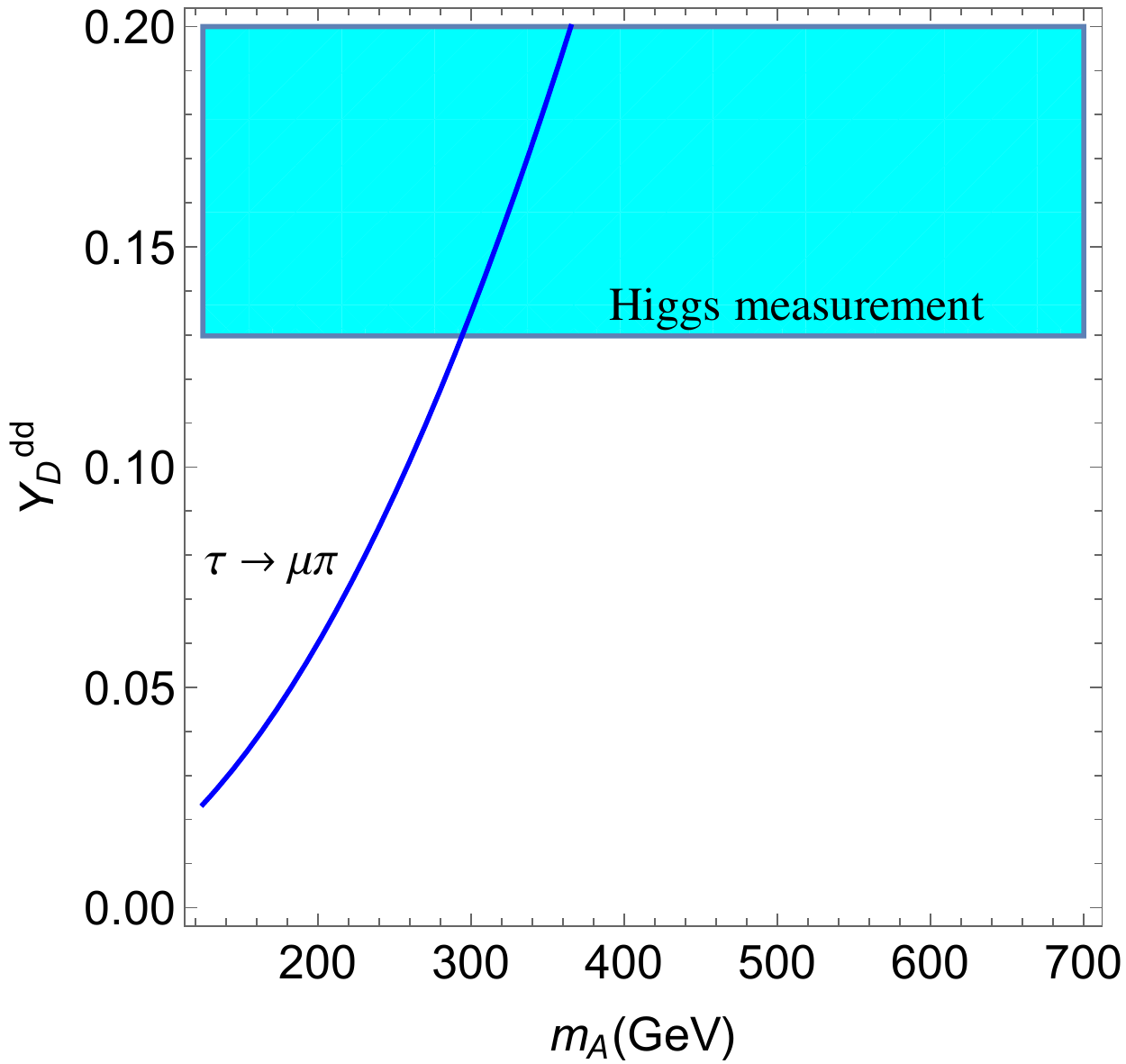}} \\
        \subfloat[Bounds on $Y_D^{ss}$ from $\tau \rightarrow \mu \eta$]{\includegraphics[width= 0.4\textwidth]{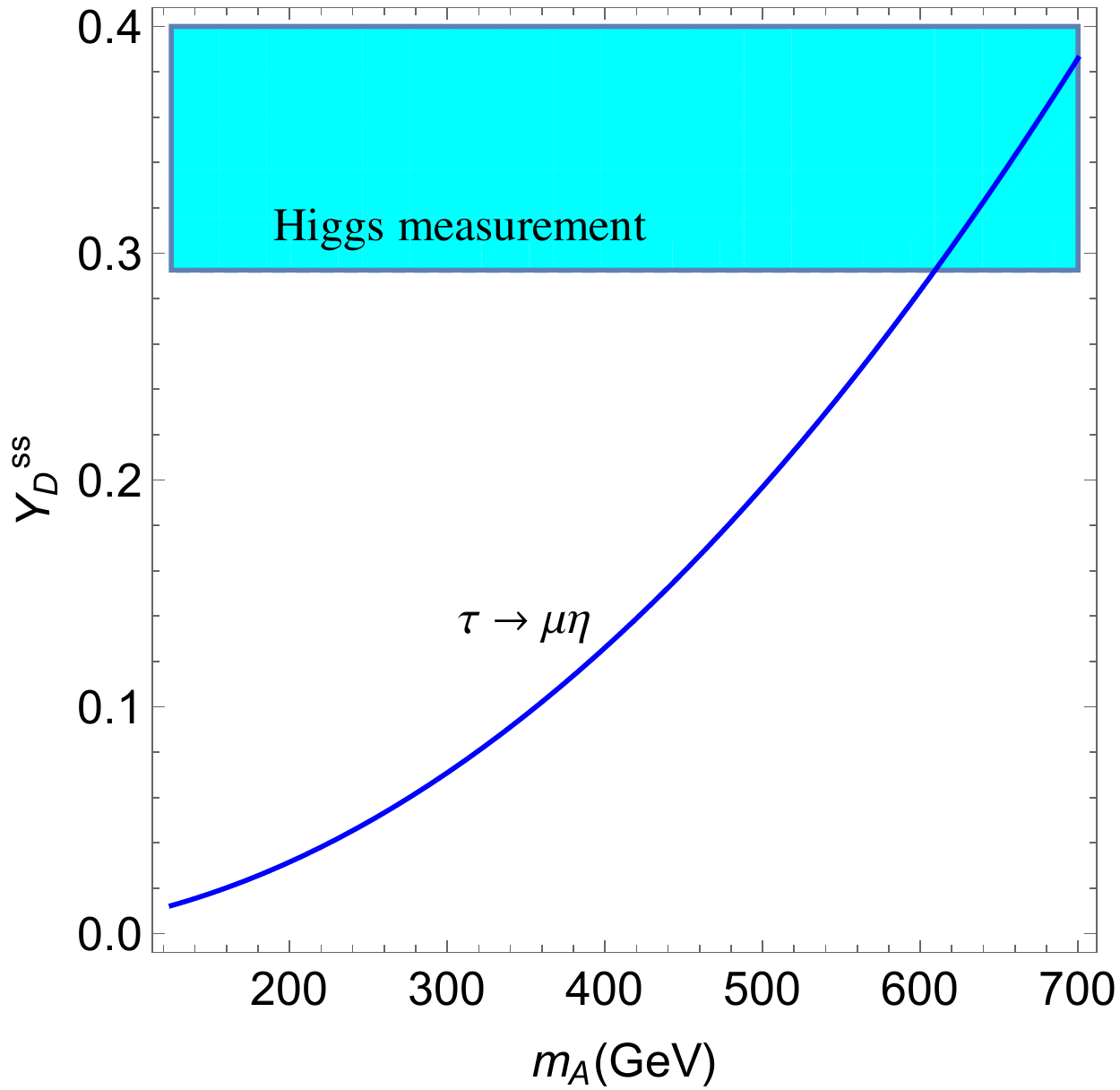}}        
        \caption{The solid blues line in these plots show the bounds for $Y_U^{uu}$, $Y_D^{dd}$ and $Y_D^{ss}$ from $\tau \rightarrow \mu P$ as a function of $m_A$ for $Y_\ell^{\tau\mu} = 0.01$. Here $P$ represents the pseudoscalars mesons: $\pi$ or $\eta$. The region above the lines are excluded by the searches. The shaded regions are excluded from the kinematics of light Higgs productions for $\sin\alpha = 0.1$. In calculating these bounds, we take only one of the $Y$'s to be nonzero, while the others are zero.}
         \label{fig:taumum}
\end{figure}

In this subsection we discuss low energy constraints from the decay of $\tau \rightarrow \mu M$, where $M$ represents a light meson. Tau decays into a muon and a pseudoscalar meson ($\pi$, $\eta$, $\eta'$) constrains the coupling of the pseudoscalar, $A$, to light quarks, in addition to the LFV coupling ($Y_\ell^{\tau\mu}$ and $Y_\ell^{\mu\tau}$). We follow Ref. \cite{Celis:2014asa} to calculate the branching fraction of  LFV decays $\tau \rightarrow \mu M$. The expression for the decay width are given in Appendix \ref{app:taumum}. 
The decay of $\tau \rightarrow \mu \pi$ constrains the coupling the pseudoscalar to the up and down quarks. We can compare these bounds against the bounds obtained from the kinematics of the light Higgs ($h$) productions. These are $y_{U,h}^{uu} < 0.011$ and  $y_{D,h}^{dd} < 0.013$ \cite{Soreq:2016rae}. The coupling of the pseudoscalar to the strange quarks are best constrained by the decay $\tau \rightarrow \mu \eta.$\footnote{We found that $\tau\rightarrow\mu\eta$ constraints the coupling $Y_D^{ss}$ better than the bounds from the decay $\tau\rightarrow\mu\eta'$.} This bound can also be compared against bounds from the LHC light Higgs precision measurement: $y_{D,h}^{ss} < 0.029$ \cite{Yu:2016rvv}. Figure \ref{fig:taumum} shows the bounds on $Y_{Q'}^{qq}$ from the LFV $\tau \rightarrow \mu \pi$ and $\tau \rightarrow \mu \eta$ decays for  $Y_\ell^{\tau\mu} = 0.01$. In the figure, we also show the bounds from the light Higgs precision measurements for $\sin\alpha = 0.1$.

\begin{figure}
        \centering
        \subfloat[Bounds on $Y_U^{uu}$ from $\tau \rightarrow \mu \pi^+\pi^-$]{\includegraphics[width= 0.4\textwidth]{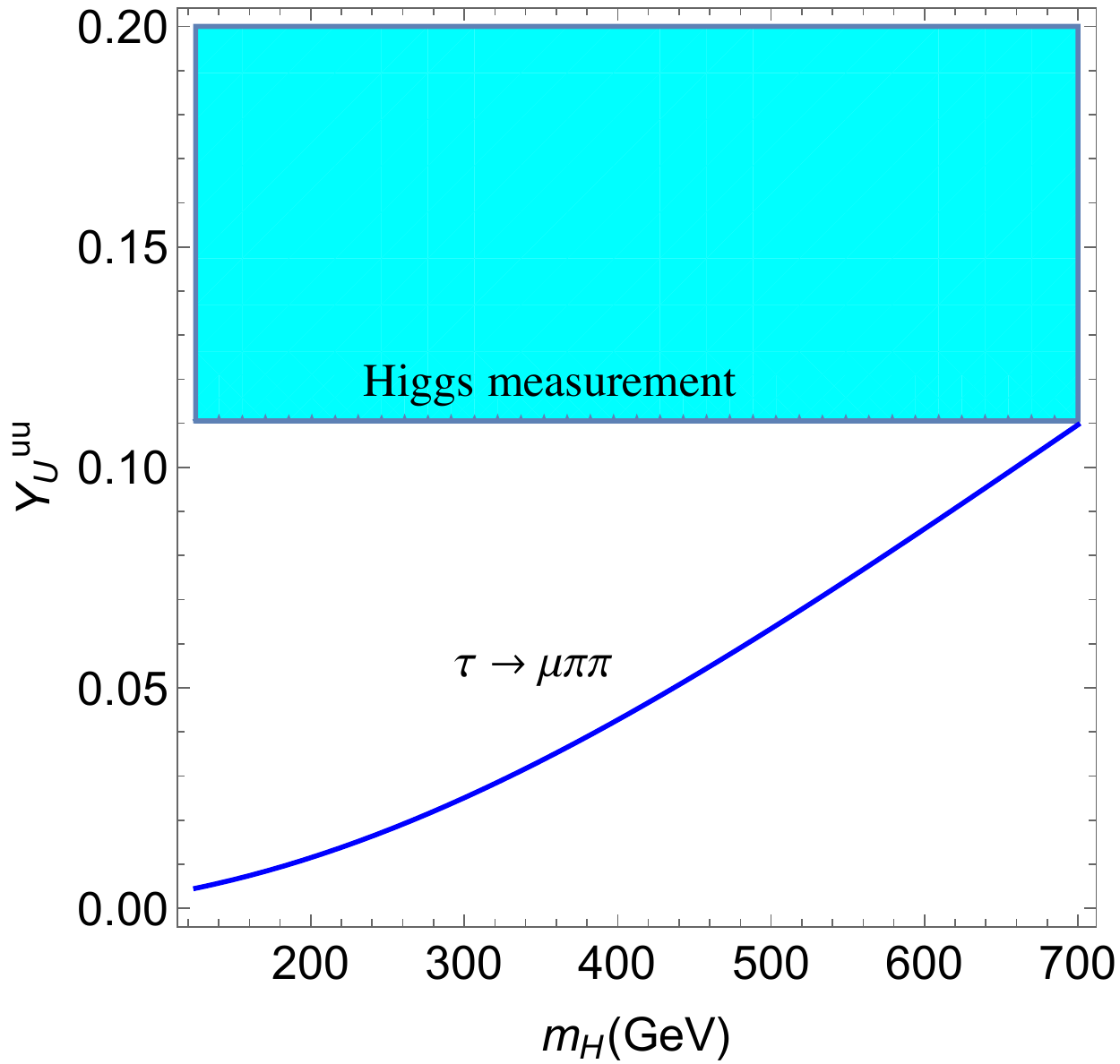}}
        \subfloat[Bounds on $Y_D^{dd}$ from $\tau \rightarrow \mu \pi^+\pi^-$]{\includegraphics[width= 0.4\textwidth]{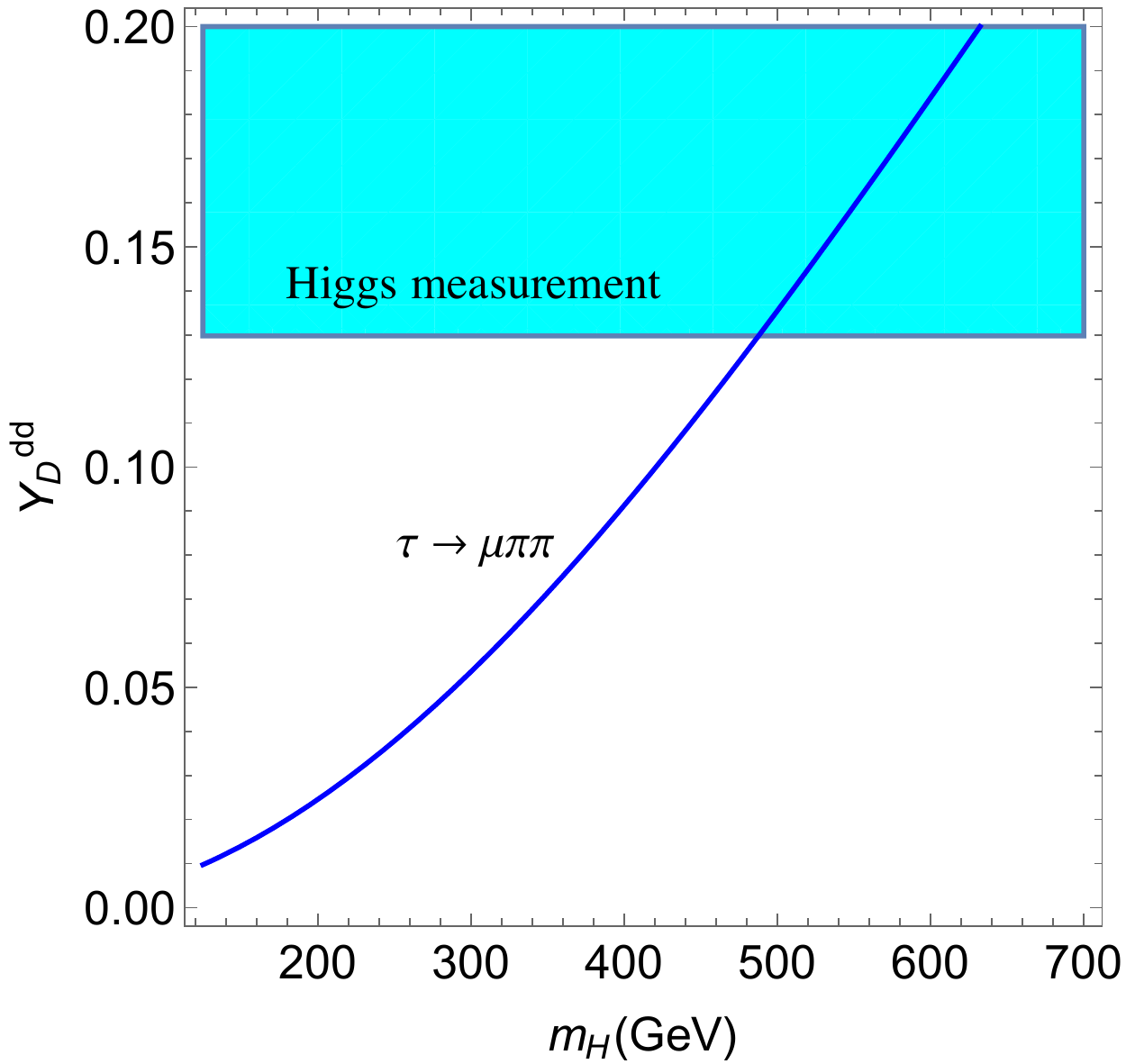}} \\
        \subfloat[Bounds on $Y_D^{ss}$ from $\tau \rightarrow \mu \pi^+\pi^-$]{\includegraphics[width= 0.4\textwidth]{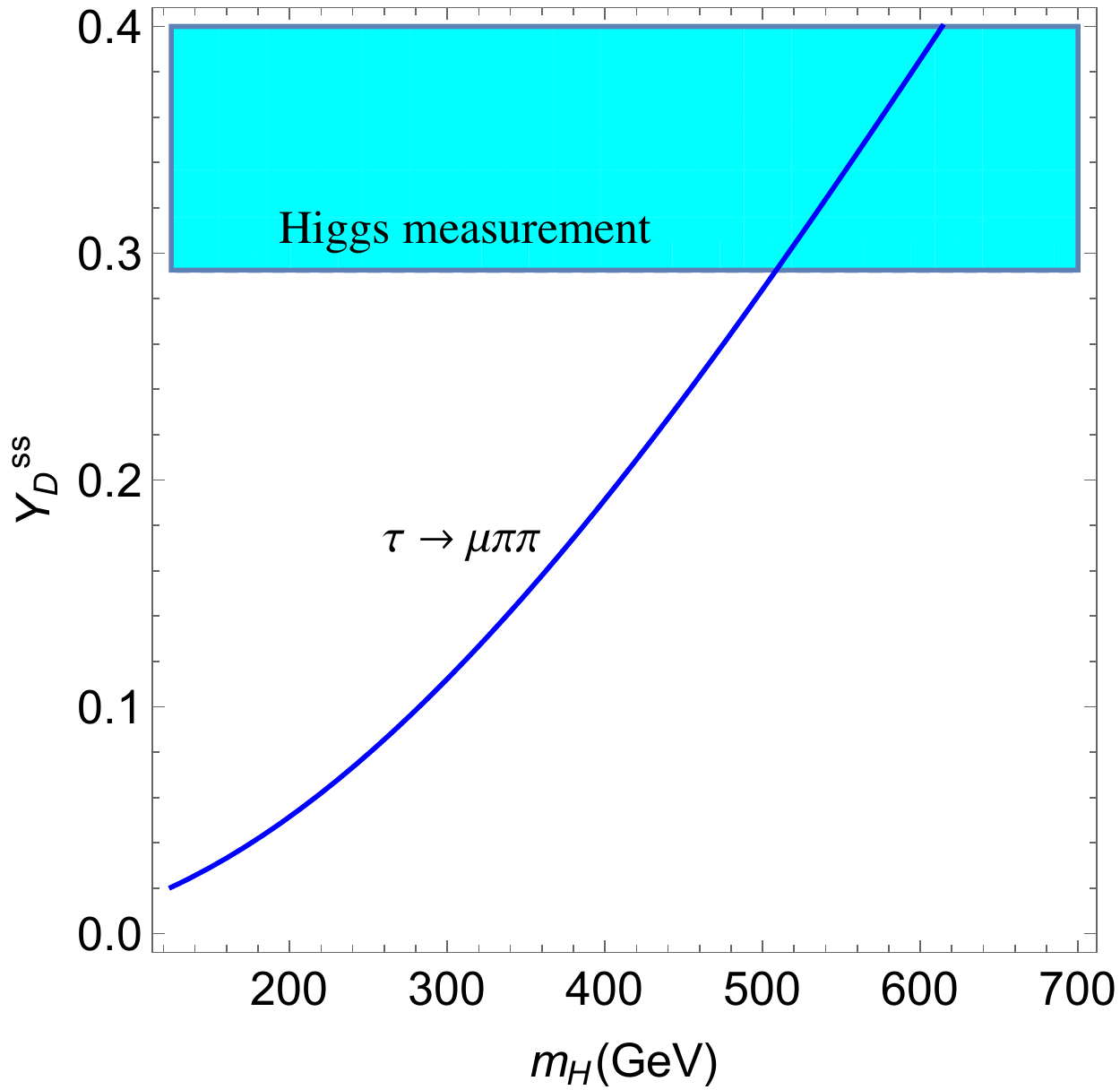}}        
        \caption{The solid blue lines in these plots show the bounds for $Y_U^{uu}$, $Y_D^{dd}$ and $Y_D^{dd}$ from $\tau \rightarrow \mu \pi^+\pi^-$ as a function of $m_H$ for $Y_\ell^{\tau\mu} = 0.01$ and  $\sin\alpha = 0.1$. The region above the lines are excluded by the searches. The shaded regions are excluded from the light Higgs precision measurements. In calculating these bounds, we take only one of the $Y$'s to be nonzero, while the others are zero.}
         \label{fig:taumupipi}
\end{figure}

The light and heavy scalar coupling to light quarks can be probed using the decay $\tau \rightarrow \mu \pi^+\pi^-$ and $\tau \rightarrow \mu \rho$. We follow Ref. \cite{Celis:2014asa} to calculate the partial decay width of the tau. We have found that the bounds from $\tau \rightarrow \mu \pi^+\pi^-$ are slightly stronger than the one from $\tau \rightarrow \mu \rho$. Hence we only show the $\tau \rightarrow \mu \pi^+\pi^-$ results in this paper. In addition to the scalar couplings to the light quarks, the bounds also depends on $\sin\alpha$, $Y_\ell^{\tau\mu}$ and $m_H$. Figure \ref{fig:taumupipi} shows the bounds for $Y_\ell^{\tau\mu} = 0.01$ and $\sin\alpha = 0.1$. The constraints from the light Higgs precision measurements are also shown in the plots.

\section{LHC constraint on LFV}
\label{sec:LFVLHC}
\subsection{Analysis}

\begin{table}
   \centering
   
   \begin{tabular}{c|c|c|c} 
	& 0-jet & 1-jet & 2-jet \\
   \hline
	$p^\mu_T$ & $> 50$ GeV & $> 45$ GeV & $> 25$ GeV \\   
	$p^e_T$ & $> 10$ GeV & $> 10$ GeV & $> 10$ GeV \\
	$M^e_T$ & $> 65$ GeV & $> 65$ GeV & $> 25$ GeV \\  
	$M^\mu_T$ & $> 50$ GeV & $> 40$ GeV & $> 15$ GeV \\ 
	$\Delta\phi_{\vec p_T^e - \vec E^\text{miss}_T}$ & $< 0.5$ & $< 0.5$ & $< 0.3$ \\  
	$\Delta\phi_{\vec p_T^e - \vec p_T^\mu}$ & $> 2.7$ & $> 1.0$ & $-$ \\      
   \end{tabular}
   \caption{ Selection criteria for each categories jet categories. The transverse mass $M^\ell_T$ is defined as $M^\ell_T = \sqrt{2 |\vec p^{\,\ell}_T| |\vec E_T^\text{miss}| \left(1 - \cos\Delta\phi_{\vec p_T^{\,\ell} - \vec E^\text{miss}_T}\right)}$.}
   \label{Tab:cuts}
\end{table}

In this section we discuss the LHC search for the LFV decay of the scalar and pseudoscalar resonances. We follow closely the analysis of CMS run-1~\cite{Khachatryan:2015kon}. In the CMS analysis, two decay channels of tau are discussed: tau decays into electron and neutrinos (denoted by $\mu\tau_e$ channel) and the hadronically decay tau (denoted by $\mu\tau_h$ channel). The background for the $\mu\tau_h$ channel is dominated by $j(W \rightarrow \mu \nu)$, which the jet is misidentified as $\tau_h$. Since we cannot simulate the jet misidentifications accurately, we will not consider the $\mu\tau_h$ channel in our work. This exclusion weakens our expected bound, hence our estimated bound is a conservative one. The $\mu\tau_e$ channel requires exactly one muon with $p_T > 25$ GeV and $|\eta| < 2.1$, and one opposite charge electron with $p_T > 10$ GeV and $|\eta| < 2.3$. The events are categorized according to the number of jets in the event. The jets are required to have $p_T > 30$ GeV and $|\eta| < 4.7$. The cuts for each jet categories are defined in Table~\ref{Tab:cuts}. In order to distinguish the signal from the background, the collinear approximation is used. The collinear mass of the system is defined as
\begin{equation}
m_\text{col} = \frac{m_{e\mu}}{\sqrt{x}}, 
\end{equation} 
where $m_{e\mu}$ is the invariant mass of the electron and the muon and 
\begin{equation}
x = \frac{|\vec p_T^{\,e}|}{|\vec p_T^{\,e}| + \vec E^\text{miss}_T \cdot \vec p_T^{\,e}}.
\end{equation}

\begin{figure}
        \centering
        \subfloat[$Z\rightarrow \tau \tau$]{\includegraphics[width= 0.5\textwidth]{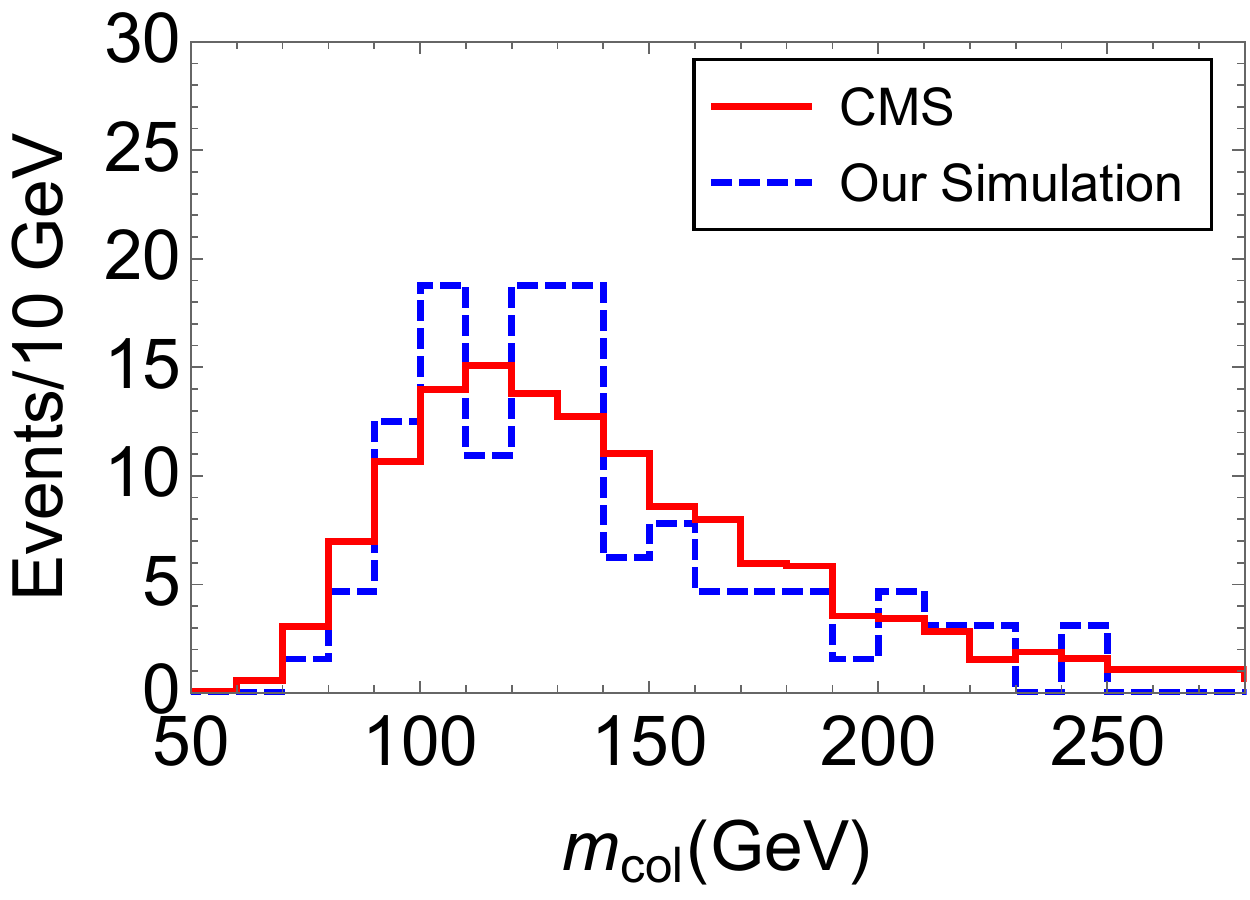}}
        \subfloat[$WW$ and $ZZ$]{\includegraphics[width= 0.5\textwidth]{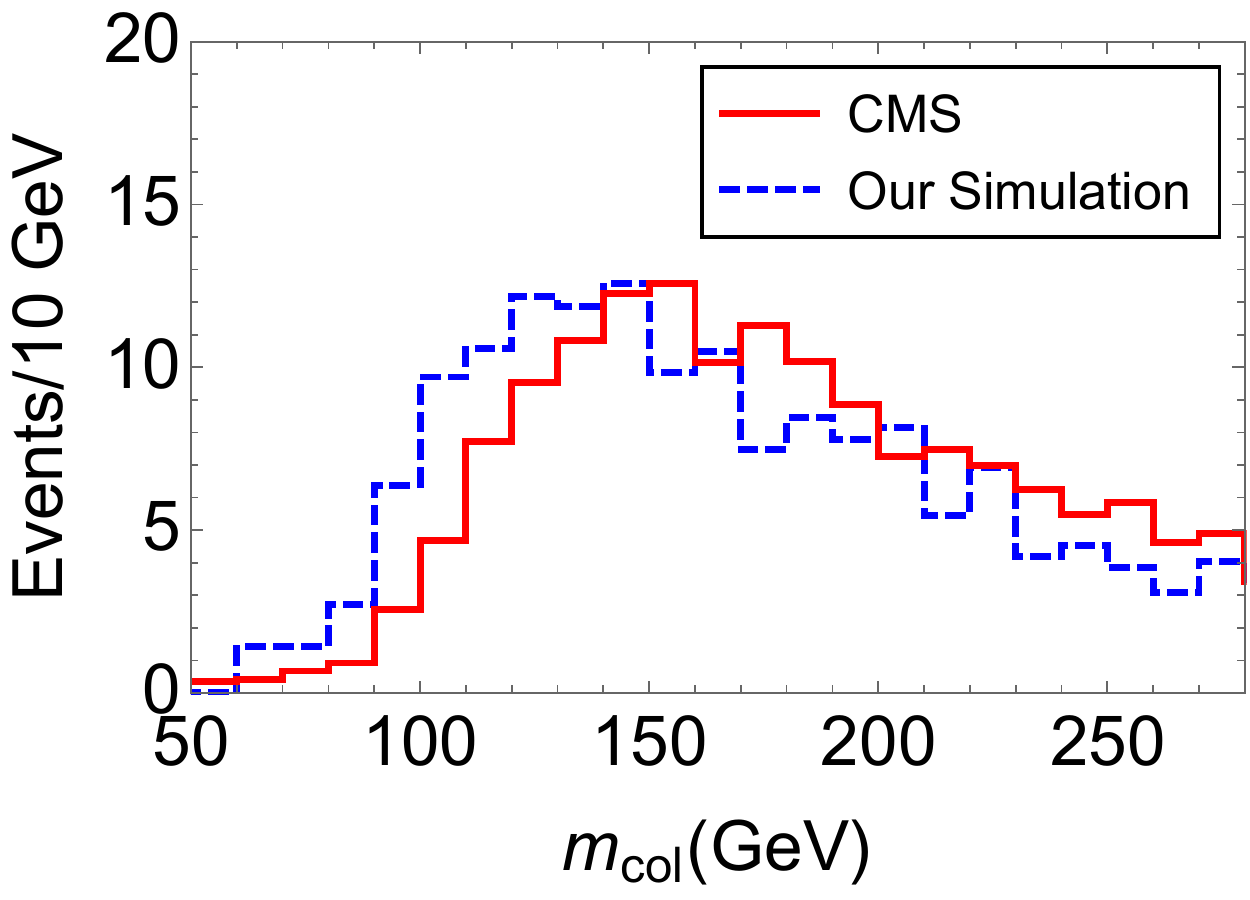}}
        \caption{The background for the 8 TeV LHC with 20 fb$^{-1}$ luminosity from our simulation compared to the CMS simulation~\cite{Khachatryan:2015kon}.}
         \label{fig:background}
\end{figure}

\begin{figure}
        \centering
        \includegraphics[width= 0.5\textwidth]{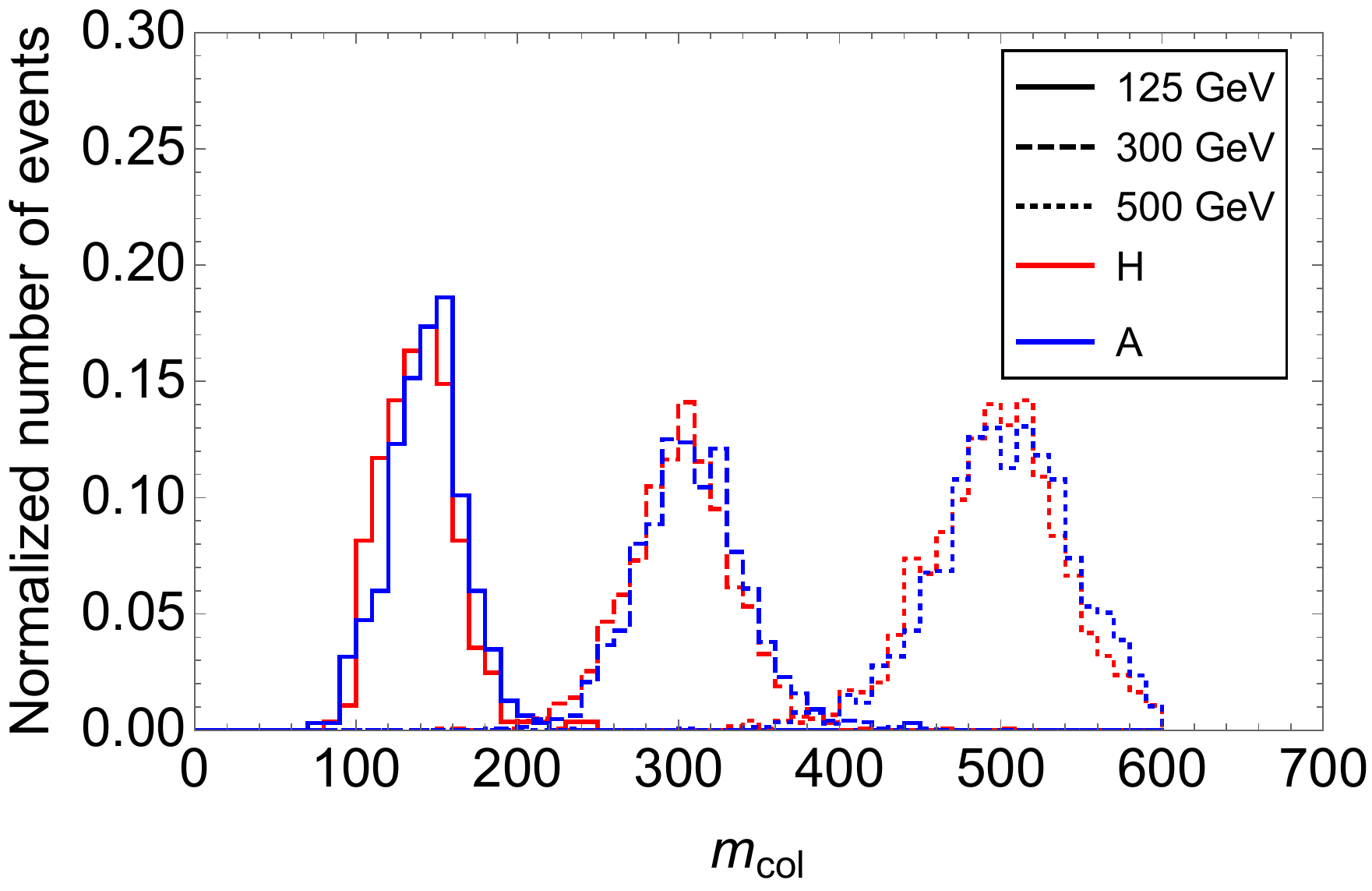}
        \caption{$m_{col}$ distributions for the scalar (red) and the pseudoscalar (blue) with $m_\phi=$ 125, 300 and 500 GeV.}
         \label{fig:singalplot}
\end{figure}

The main backgrounds for this search are $Z \rightarrow \tau\tau$, $WW$, and $ZZ$.\footnote{For the run-1 search, misidentified leptons have non negligible contributions to the background. Since we cannot simulate the misidentification accurately, we do not include them. We estimated that our conclusion does not change qualitatively even if this background is as large as the main one.} The signal and the background event samples are generated using Madgraph 5~\cite{Alwall:2014hca} followed by parton shower, hadronization and matching simulations in PYTHIA 6~\cite{Sjostrand:2006za}. Delphes 3~\cite{deFavereau:2013fsa} was used to simulate the detector environment. We include pileup in our simulation with 21 pileups per bunch for 8 TeV simulation and 40 pileups per bunch for 13 TeV run~\cite{Marshall:2014mza}.  The signal models are generated using Feynrules~\cite{Alloul:2013bka}. The comparisons between our simulation and the 8 TeV CMS simulation~\cite{Khachatryan:2015kon} are shown in Fig.~\ref{fig:background}, while the $m_\text{col}$ distribution for the signal is given in Fig.~\ref{fig:singalplot}.

\begin{figure}
        \centering
	\subfloat[$pp \rightarrow H \rightarrow \tau \mu$]{\includegraphics[width= 0.45\textwidth]{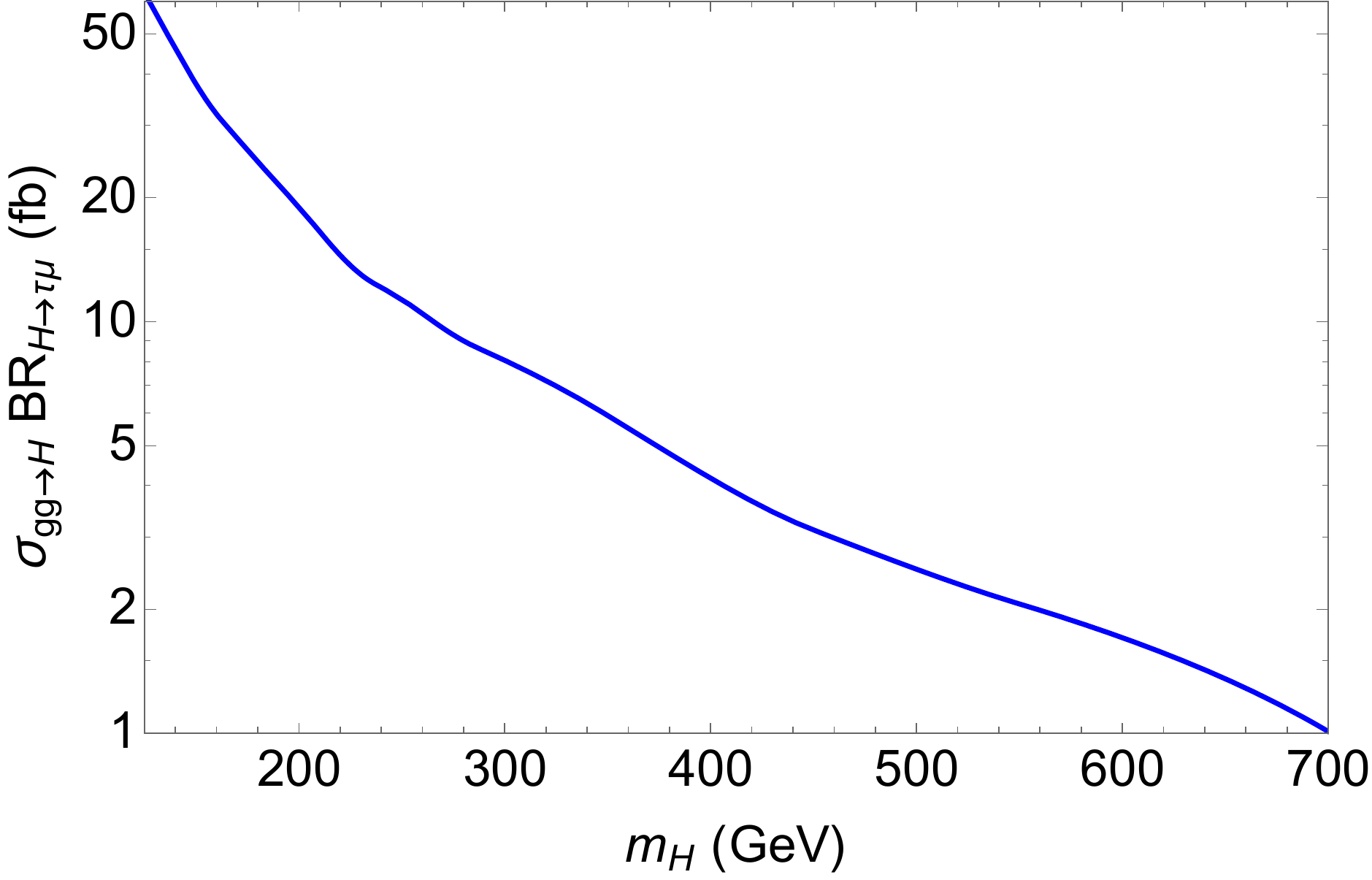}\label{fig:13TeVScalarBounds}} \hspace{5mm}
		\subfloat[$pp \rightarrow A \rightarrow \tau \mu$]{\includegraphics[width= 0.45\textwidth]{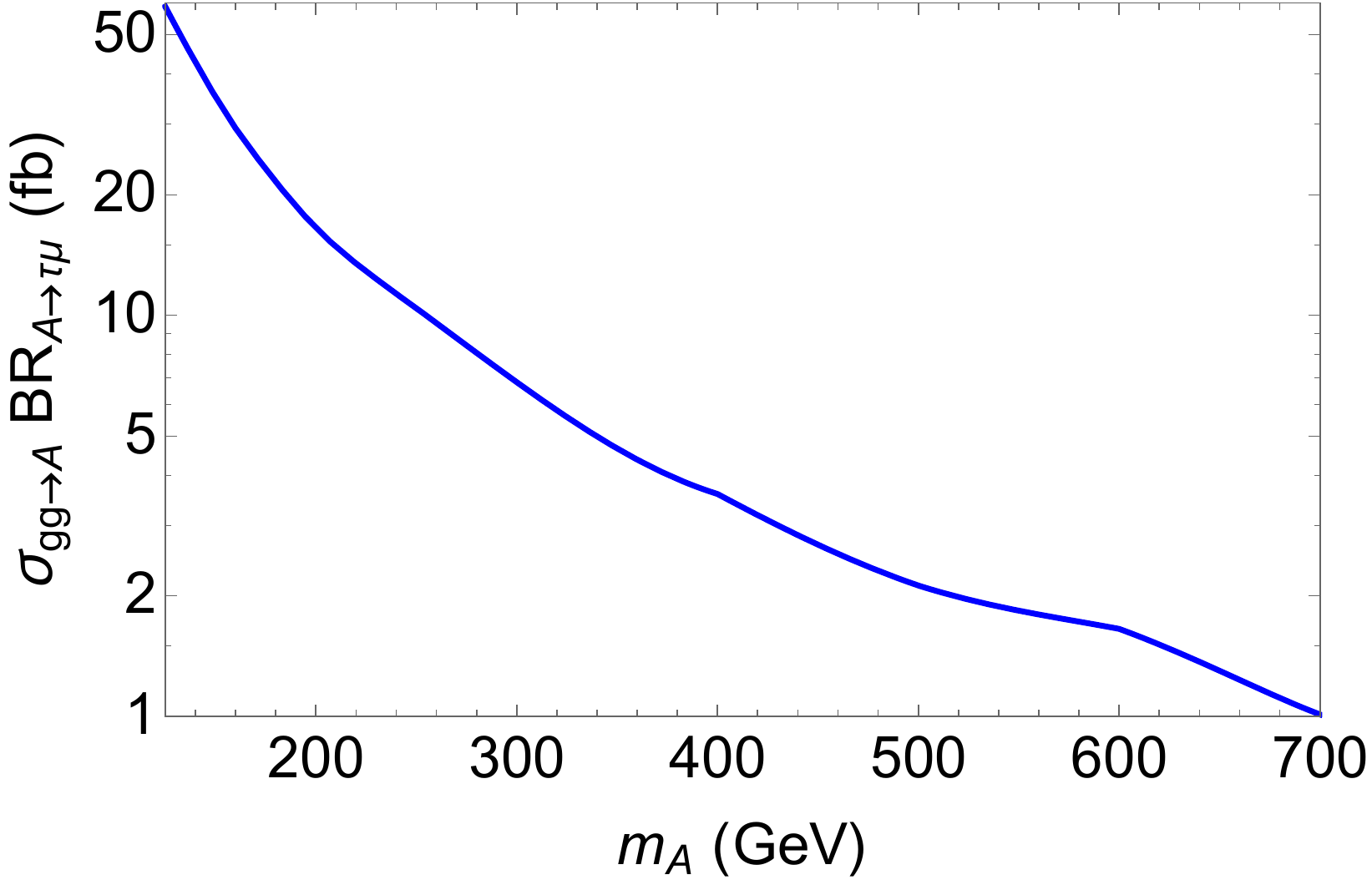}\label{fig:13TeVPseudoscalarBounds}}
		\caption{The estimated 95\% C.L. bounds for the scalar (left) and pseudoscalar (right) decaying into tau and muon at the 13 TeV LHC with 300 fb$^{-1}$ luminosity.} 
		\label{fig:sigmaBRbound}
\end{figure}

For the estimated 13 TeV reach, we define the signal regions to be $m_\phi - \Delta < m_\text{col} < m_\phi + \Delta$, where $m_\phi$ is the mass of the resonance we are interested in.
For each $m_\phi$, the value of $\Delta$ is varied to get the best estimated bound. The 95\% C.L. bound is found by solving for $N_\text{sig}$ that satisfies\footnote{We ignore systematic uncertainties in this work.} 
\begin{equation}
	\sum_{k=0}^{N_\text{back}} P\left( k; N_\text{back} + N_\text{sig} \right)  < 0.05,
\end{equation}
where $N_\text{back}$ is the number of estimated background events and the probability $P$ follows the Poisson distribution, $P\left(k,\lambda\right) = \frac{\lambda^k e^{-\lambda}}{k!}$. $N_\text{sig}$ is related to the signal cross section, $\sigma$, and the branching fraction BR$_{\phi\rightarrow \tau\mu}$, by $N_\text{sig} = \sigma\, \text{BR}_{\phi\rightarrow \tau\mu} \mathcal \, L \, \epsilon$, where $L$ is the luminosity and $\epsilon$ is the acceptance and efficiency of the detector estimated from the simulation. 
In our analysis below, we focus on the gluon-fusion production channel because it is the dominant production mechanism for both the heavy scalar and the pseudoscalar as can be seen in Fig.~\ref{fig:xsecA} and~\ref{fig:xsecH}. 

The estimated bounds for the 13 TeV LHC with 300 fb$^{-1}$ luminosity are given in Fig.~\ref{fig:sigmaBRbound}. In deriving these bounds, we only consider the 0-jet category as it contains signal predominantly produced by gluon-fusion~\cite{Khachatryan:2015kon}.

\subsection{Results}
\label{sec:result}
The constraints on LFV from the $\phi\to\tau\mu$ search ($\phi=H,A$) can roughly be categorized into two cases according to its production cross-section. If the production cross-section is large, the bounds are expected to be strong. On the contrary, if the cross-section is small, the bounds are expected to be weak.  For comparison, we also give the estimated bounds from $h\to\tau\mu$ search.

\subsubsection{Large cross-section case}\label{sec:large}

\begin{figure}
        \centering
        \includegraphics[width= 0.5\textwidth]{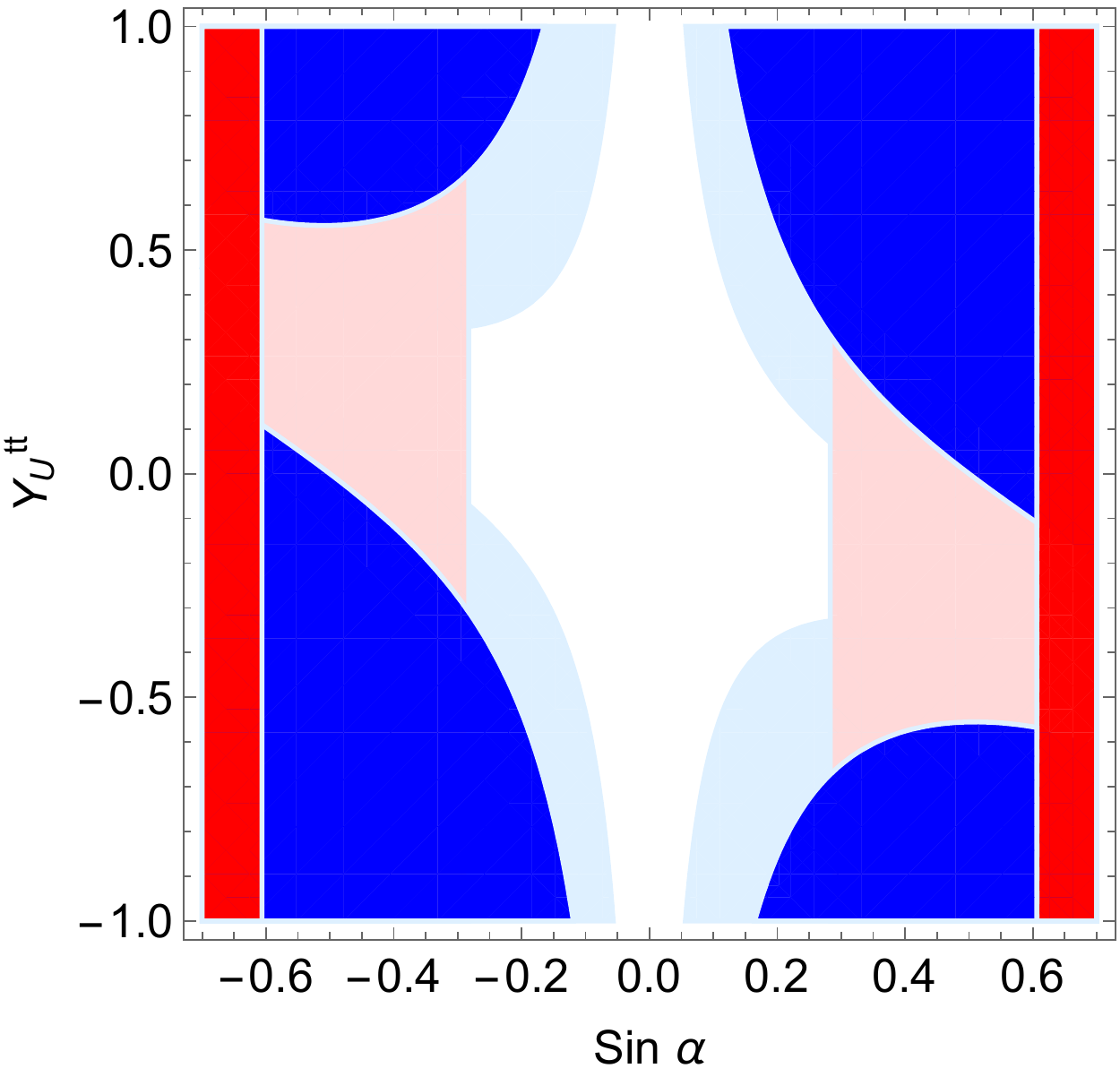}
        \caption{The excluded parameter space in the plane of $Y_U^{tt}$ and $\sin\alpha$. The dark blue and dark red regions are excluded by the LHC run-1 $ggh$ coupling and $WWh$ coupling measurements respectively. The light blue and light red regions are the predicted exclusion regions from the $ggh$ and the $WWh$ couplings measurements at the 14 TeV LHC with 300 $\text{fb}^{-1}$ luminosity.}
         \label{fig:etattsina}
\end{figure}

We begin with an optimistic scenario where the production cross sections of the heavy scalar and the pseudoscalar are large with a significant LFV branching fraction. This scenario can be achieved by taking $Y_U^{tt}$ large while keeping the other $Y$'s vanishing (except $Y_{\ell}^{\mu\tau}$ and $Y_{\ell}^{\tau\mu}$). 
However, $Y_U^{tt}$ is constrained by low energy experiment and hence cannot be arbitrary large. 
For example, the coupling $Y_U^{tt}$, in a combination with $Y_\ell^{\tau\mu}$ and $Y_\ell^{\mu\tau}$, can be constrained by the $\tau\to\mu\gamma$ measurement, see for example, Eq.~\eqref{eq:tgammaloop} and~\eqref{eq:tZloop}.
Moreover, Ref.~\cite{Mahmoudi:2009zx} analyzed constraints from $b\rightarrow s \gamma$ and $\Delta M_{B_d} $ and found that $|Y_U^{tt}| < 1$  for $m_{H^+} \lesssim 500$ GeV. 
Additionally, the LHC Higgs data provide another handle on $Y_U^{tt}$. 
A combined analysis of CMS and ATLAS from the LHC run-1 data found the gluon-fusion signal strength $\mu_{ggF} \equiv \frac{\sigma_{ggF}}{(\sigma_{ggF})_{SM}} = 1.03^{+0.16}_{-0.14}$~\cite{Khachatryan:2016vau}\footnote{This bound is set by assuming that the branching fraction to each channel is equal to the SM values. 
While this requirement is not strictly satisfied in our case, the variations of the light Higgs branching ratios to the observed final states are negligible. 
This is because the additional decay ($h\rightarrow g g$, $h \rightarrow \tau \mu$) only contributes a few percents of the total decay width.}. 
It is predicted that this signal strength can be measured with a precision of 6\% at the 14 TeV LHC with 300 fb$^{-1}$ luminosity~\cite{Dawson:2013bba}. 
The bound from the gluon-fusion signal strength can be seen in Fig.~\ref{fig:etattsina}. 
In the figure we also include the bounds on $\sin \alpha$ from $h \rightarrow WW^*$ measurements from the LHC run-1, $\frac{\sigma_{VBF}BR_{h\rightarrow WW}}{(\sigma_{VBF}BR_{h\rightarrow WW})_{SM}}= 0.84^{+0.4}_{-0.4}$~\cite{Khachatryan:2016vau}, together with the future estimates for the 14 TeV LHC with 300 fb$^{-1}$ luminosity~\cite{Dawson:2013bba}.

\begin{figure}
	\centering
	\subfloat[$y_{U,h}^{tt}$ -- $y_{h,\ell}^{\tau\mu}$ plane]{\includegraphics[width= 0.4\textwidth]{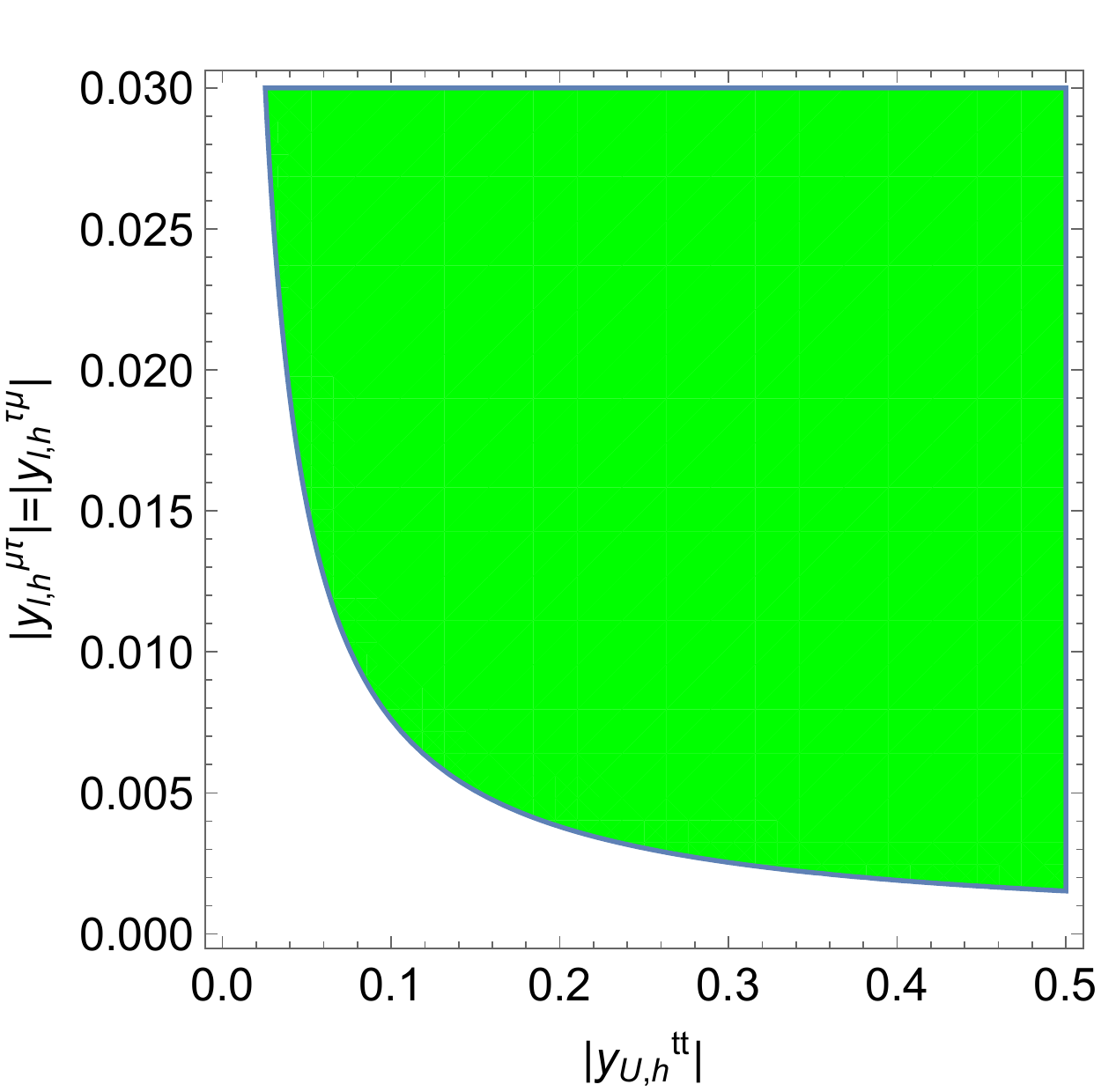}\label{fig:hymutausinytt}}
		\subfloat[$Y_U^{tt}$ - $Y_\ell^{\tau\mu}$ plane]{\includegraphics[width= 0.4\textwidth]{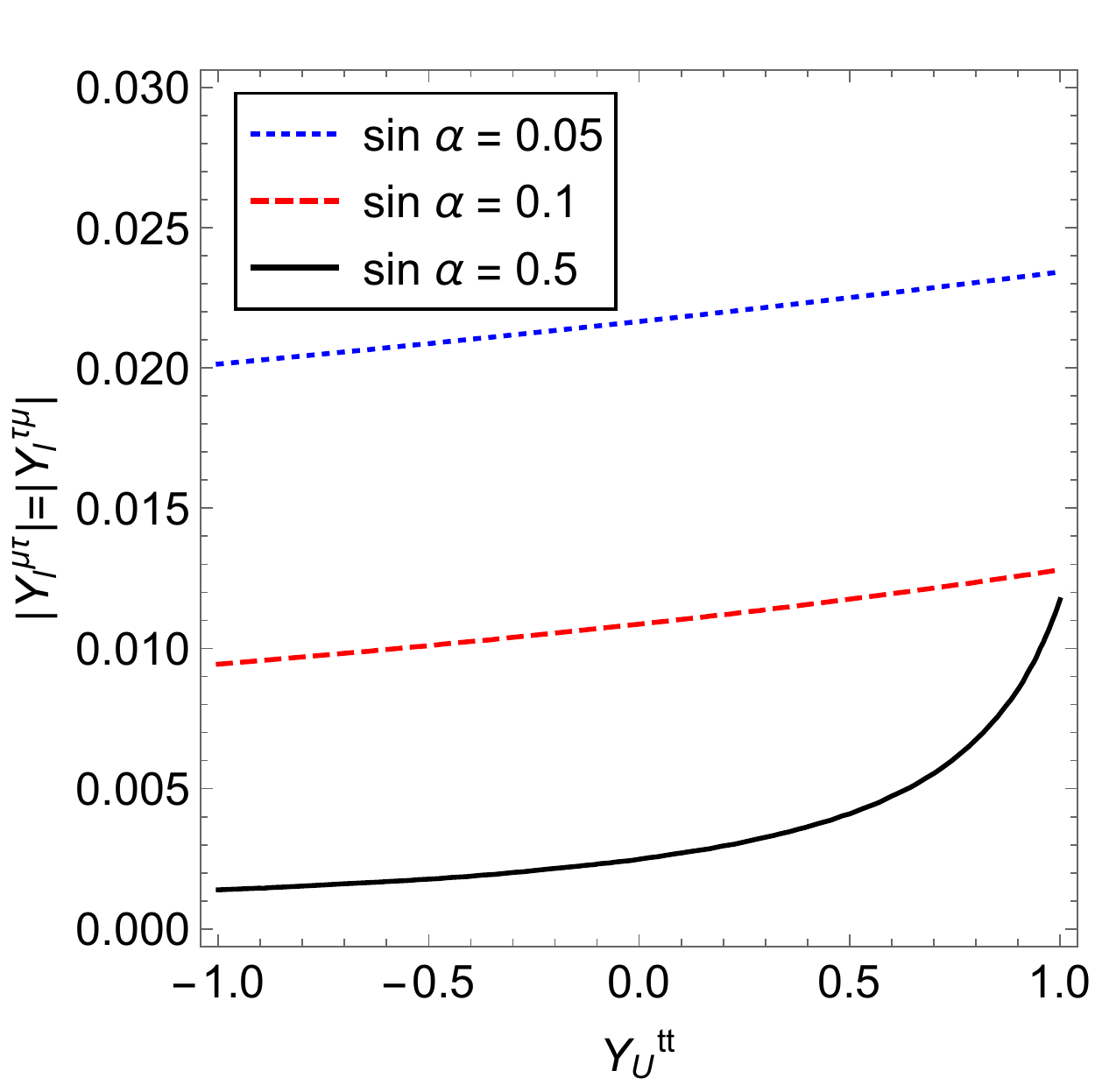}\label{fig:YtaumusinalphaYtt}}
	\caption{The estimated bounds on the $h$ LFV decays for the 13 TeV LHC with 300 fb$^{-1}$ luminosity.}
\end{figure}

The cross section of the light scalar, $h$, is determined by $y_{U,h}^{tt}$ which depends on the mixing angle $\alpha$ and $Y_U^{tt}$ as shown in Eq.~\eqref{eq:reducedcoupling}. Current data show that the branching fraction $BR_{h \rightarrow \tau \mu}$ is smaller than a few percents, hence it does not significantly affect the total decay width. Therefore the branching fraction $BR_{h \rightarrow \tau \mu}$ mostly depends on $y_{\ell,h}^{\tau\mu} = Y_{\ell}^{\tau\mu} \sin \alpha$ and $y_{\ell,h}^{\mu\tau}$, see Eq.~\eqref{eq:h2taumu}.
Hence the constraints from the $h \rightarrow \tau \mu$ decay can be displayed in the $y_{U,h}^{tt}$ -- $y_{\ell,h}^{\tau\mu}$ plane. Fig.~\ref{fig:hymutausinytt} shows our projection of such a bound for the 13 TeV LHC with 300 fb$^{-1}$ luminosity. In Fig.~\ref{fig:YtaumusinalphaYtt}, we take $\sin\alpha=0.05,0.1,0.5$ as our benchmark values. For small values of $\sin\alpha$, the bounds on $Y_\ell^{\tau\mu}$ does not significantly depend on $Y_U^{tt}$.

\begin{figure}[htpb]
\centering 
\subfloat[$\sin\alpha = 0.05$]{\includegraphics[width= 0.31\textwidth]{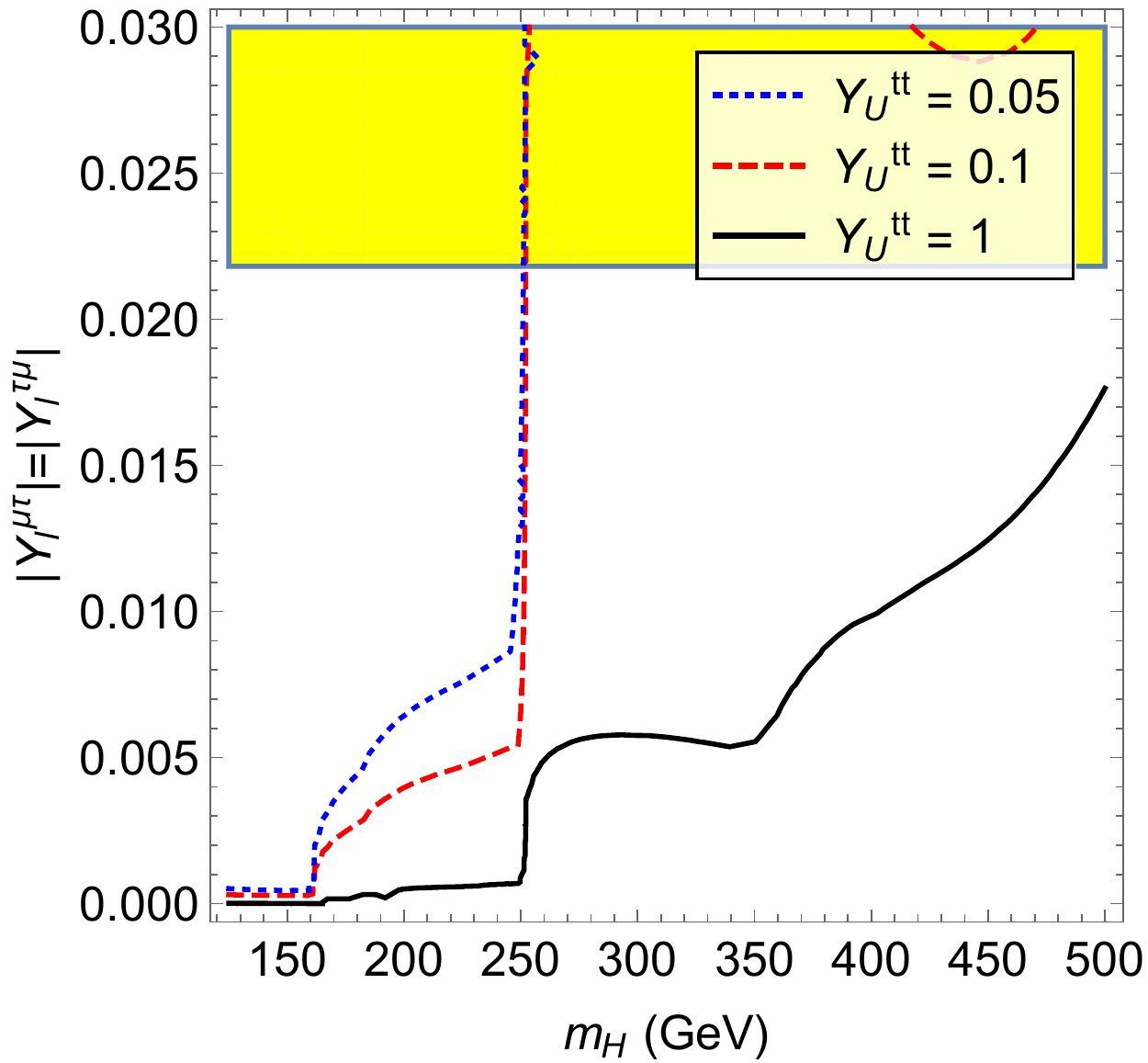}\label{fig:scalarboundsoptimistic005}} \hspace{2mm}
	\subfloat[$\sin\alpha = 0.1$]{\includegraphics[width= 0.31\textwidth]{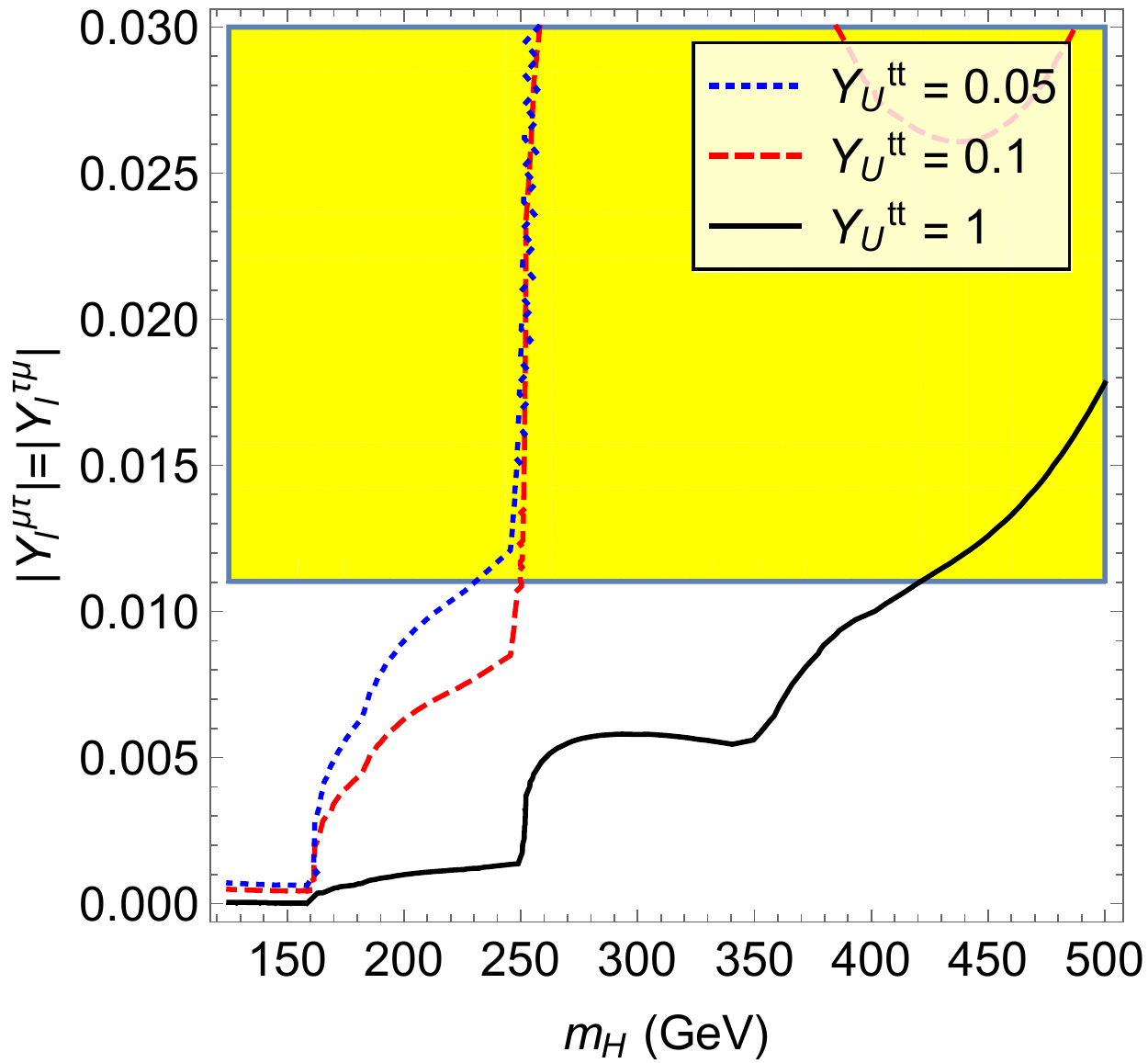}\label{fig:scalarboundsoptimistic01}} \hspace{2mm}
		\subfloat[$\sin\alpha = 0.5$]{\includegraphics[width= 0.31\textwidth]{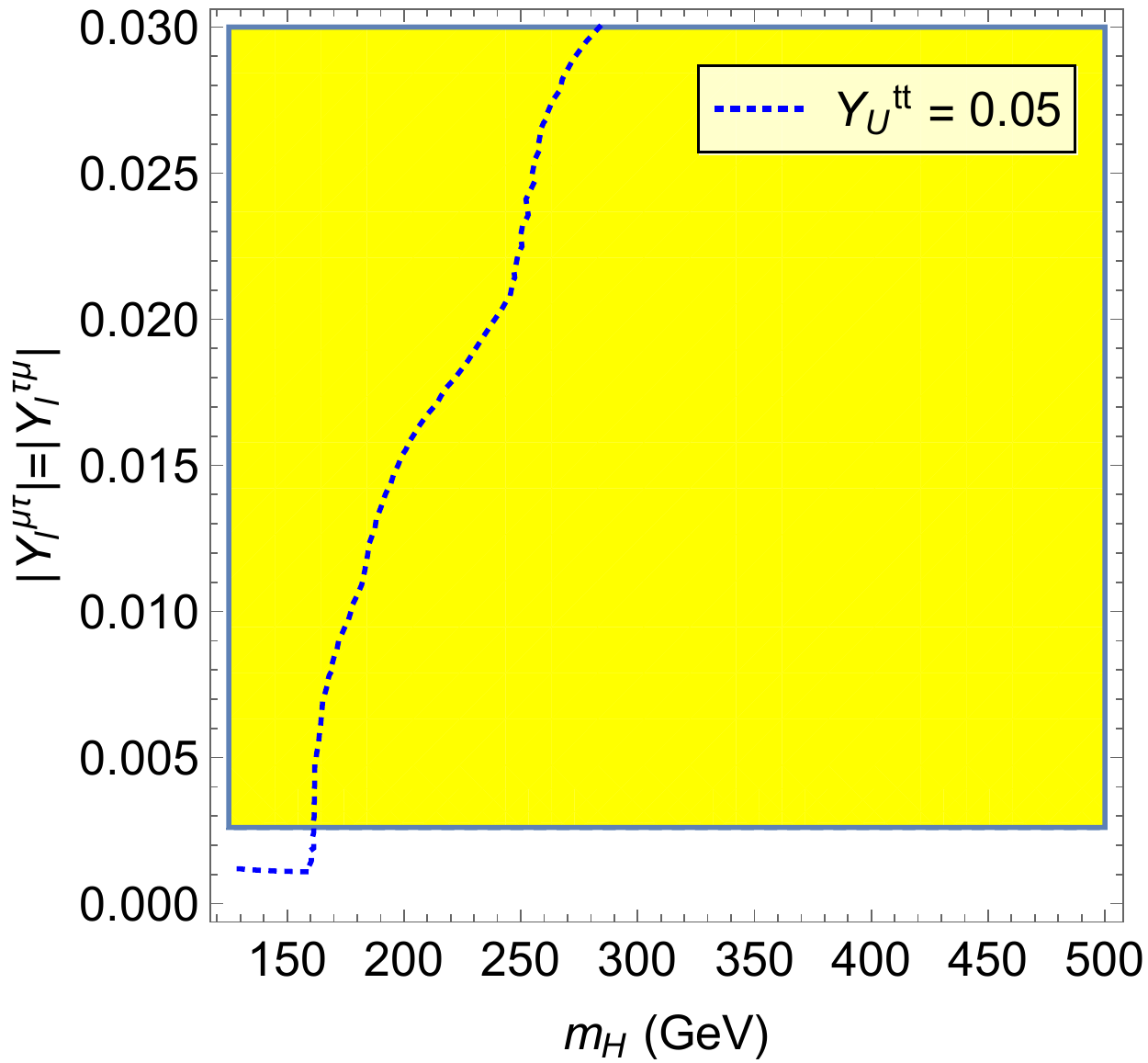}\label{fig:scalarboundsoptimistic05}} \\
					\subfloat[$\sin\alpha = -0.05$]{\includegraphics[width= 0.3\textwidth]{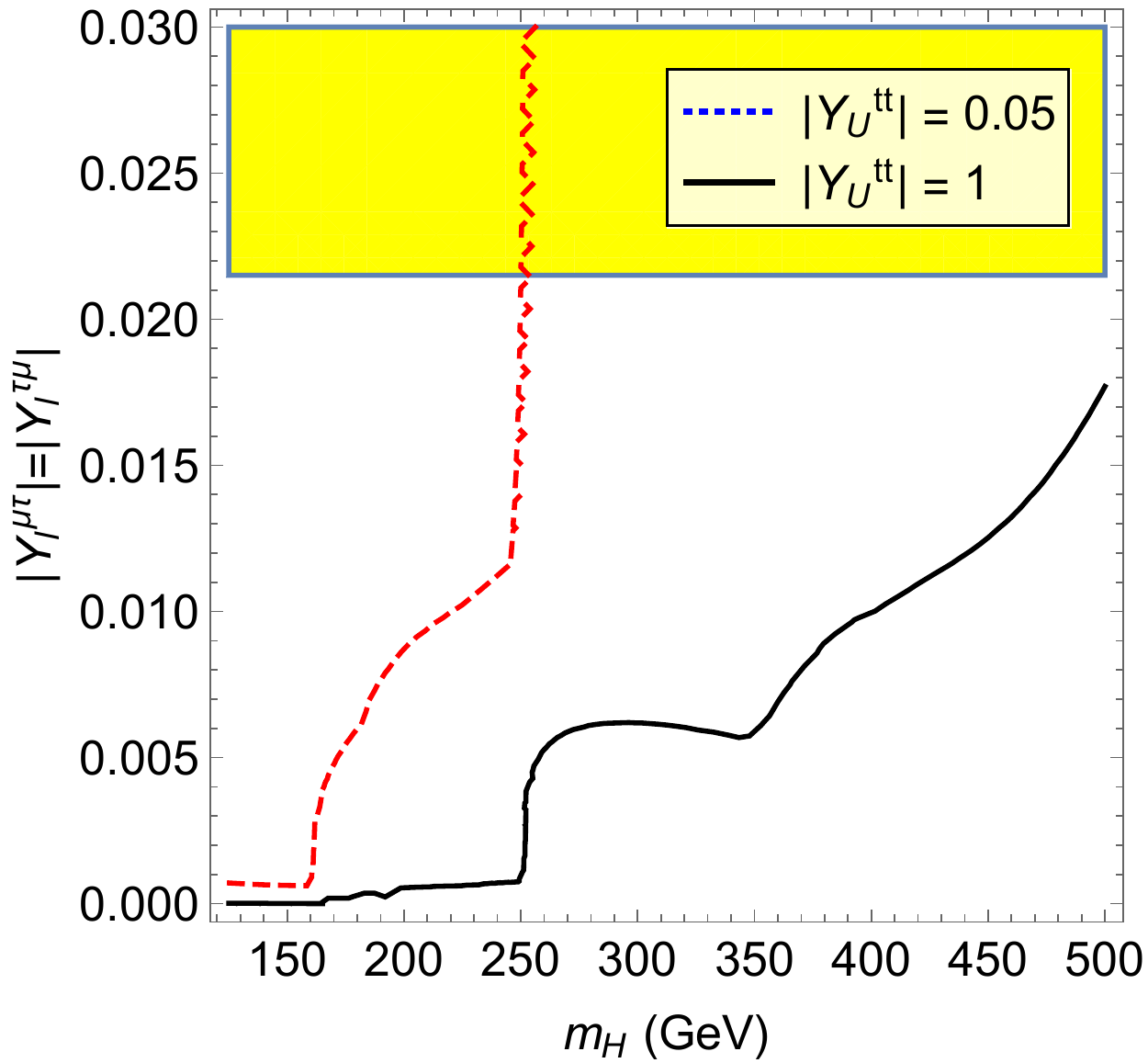}\label{fig:scalarboundsoptimisticminus005}} \hspace{5mm}
			\subfloat[$\sin\alpha = -0.1$]{\includegraphics[width= 0.3\textwidth]{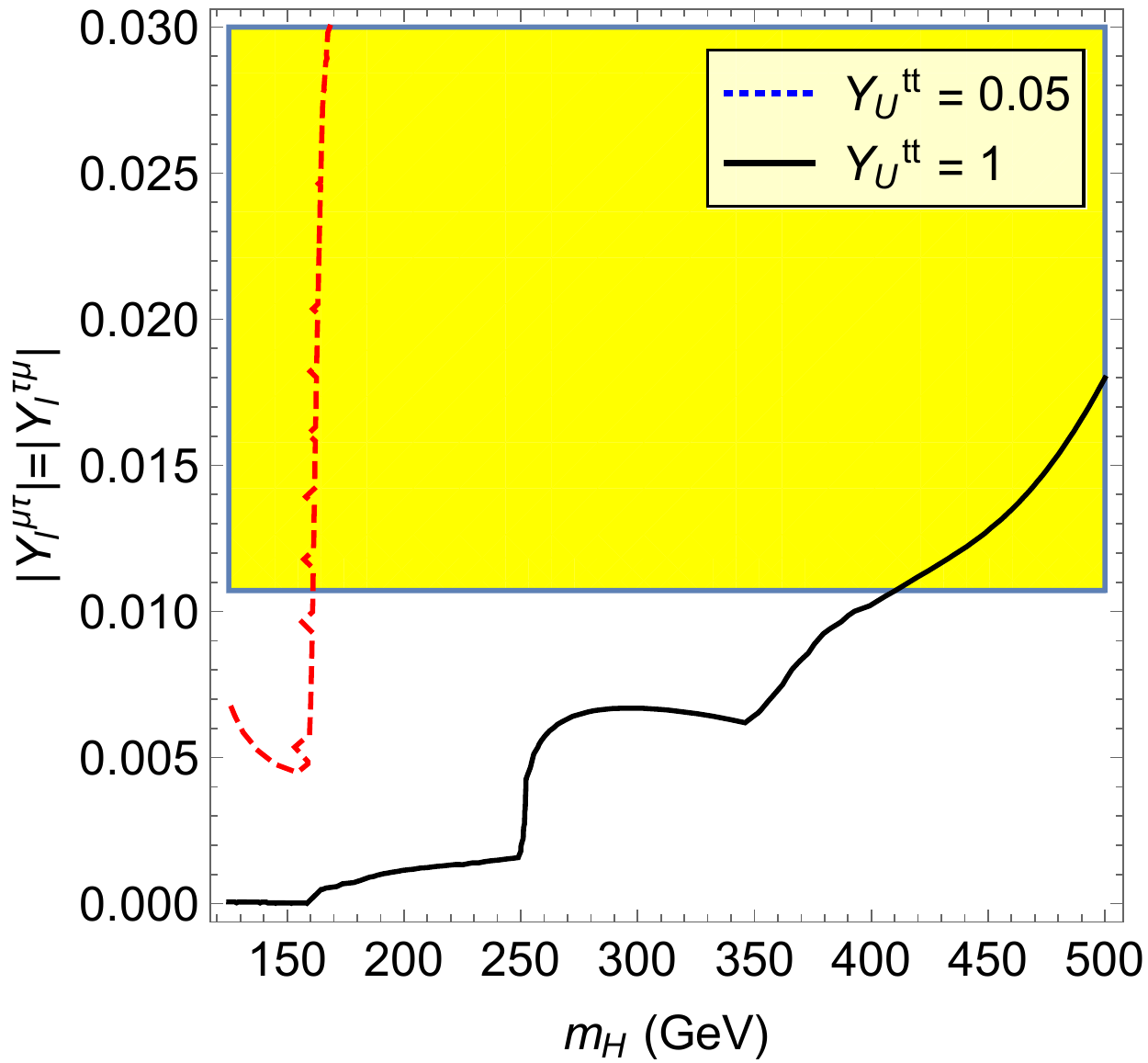}\label{fig:scalarboundsoptimisticminus01}} \hspace{5mm}
		\subfloat[$\sin\alpha = -0.5$]{\includegraphics[width= 0.3\textwidth]{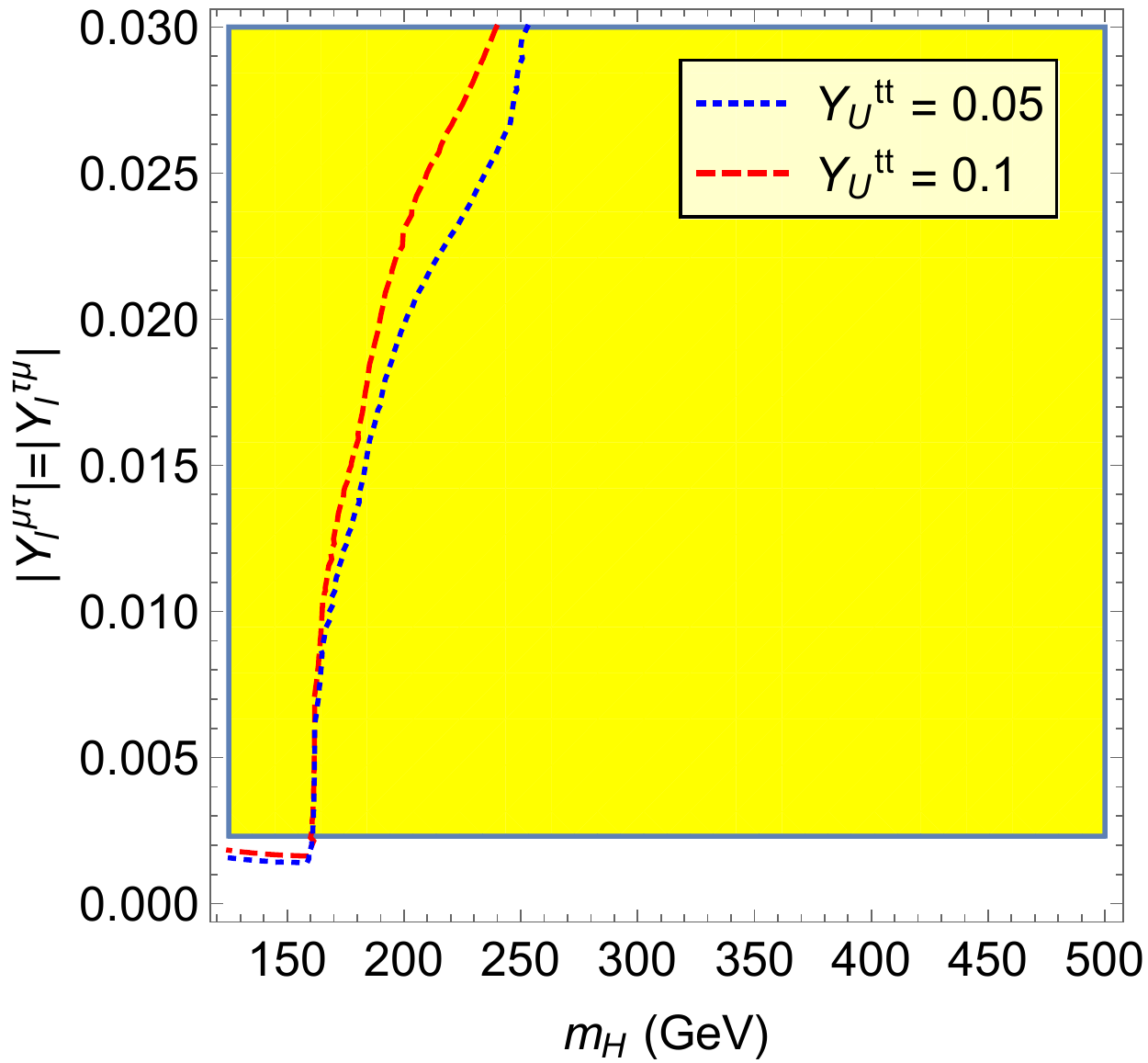}\label{fig:scalarboundsoptimisticminus05}}
		\caption{The estimated bounds for the heavy scalar LFV searches and their comparisons with the $h$ LFV searches. The yellow region is excluded by the $h$ LFV search. The region above the red, blue and black lines are excluded by the heavy scalar LFV search. Here we assume $\lambda_{hhH} = 1$, $m_H < 2 m_A$ and $m_H < 2 m_{H^+}$.} \label{fig:scalarboundsoptimistic}
\end{figure}

The production cross section of the heavy scalar depends on $m_H$ and $\alpha$ and $Y_U^{tt}$ through $y_{U,H}^{tt}$, while its LFV branching fraction also depends on $Y_{\ell}^{\tau\mu}$, $Y_{\ell}^{\mu\tau}$ and $\lambda_{hhH}$ (if the decay $H\to hh$ is open). Given the number of free parameters, we pick a set of benchmark points to show the LFV $H$ bounds. 
The estimated constraints in the $Y_\ell^{\tau\mu}$--$m_H$ plane are shown in Fig.~\ref{fig:scalarboundsoptimistic} for $\sin\alpha = \pm0.05, \pm0.1, \pm0.5$ and $Y_{U}^{tt}=0.05,0.1,1$.
Note that when a new decay channel of the $H$ opens, the $H$ LFV bound becomes weaker due to a smaller branching fraction for $H\to\tau\mu$. 
This is especially pronounced in the case of a small $Y_U^{tt}$ and a small mixing angle $\alpha$ where the production cross-section of the $H$ is small.
In the figure we also show the LFV $h$ bounds for comparison. 
Note also that for each value of $\sin\alpha$ in Fig.~\ref{fig:scalarboundsoptimistic}, the $h$ LFV bound is approximately the same for the values of $Y_U^{tt}$ under consideration.
For the cases of small mixing angle, $\sin\alpha=\pm0.05$ and $\pm0.1$, the $BR_{h\to\tau\mu}$ is small compared to $BR_{H\to\tau\mu}$. Hence the $H$ LFV bounds are generically stronger for a wide range of $m_H$, see Figs.~\ref{fig:scalarboundsoptimistic005},~\ref{fig:scalarboundsoptimistic01},~\ref{fig:scalarboundsoptimisticminus005} and~\ref{fig:scalarboundsoptimisticminus01}. 
Note that in Figs.~\ref{fig:scalarboundsoptimisticminus005} and~\ref{fig:scalarboundsoptimisticminus01}, there is no $H$ LFV bound for $Y_U^{tt}=0.05$ because the $H$ production cross-section is too small.
Also from the figure, it's clear that for the case $Y_U^{tt}=1$ the $H$ LFV bound is particularly strong because of a large production cross-section of the $H$.
On the other hand, for the case $\sin\alpha=\pm0.5$, only small value of $Y_U^{tt}$ is viable, see Fig.~\ref{fig:etattsina}.
Thus, we only consider the case $Y_U^{tt} = 0.05$ for $\sin\alpha=0.5$, Fig.~\ref{fig:scalarboundsoptimistic05} and $Y_U^{tt} = 0.05$ and $0.1$ for $\sin\alpha=-0.5$, Fig.~\ref{fig:scalarboundsoptimisticminus05}. In these cases the $H$ production cross-section is small while
the $h$ LFV bound is strong because of a large $BR_{h\to\tau\mu}$, see Fig.~\ref{fig:YtaumusinalphaYtt}.
As a result, we estimate the LFV $H$ searches only provide a better constraint for $m_H \lesssim 160$ GeV in the case of $\sin\alpha = \pm0.5$. 
It should be noted that as the mixing angle decreases, the LFV $h$ bound gets weaker. 
Hence the search for LFV $H$ decays become more and more relevant.

\begin{figure}
        \centering
\subfloat[$\sin\alpha = 0.05$]{\includegraphics[width= 0.31\textwidth]{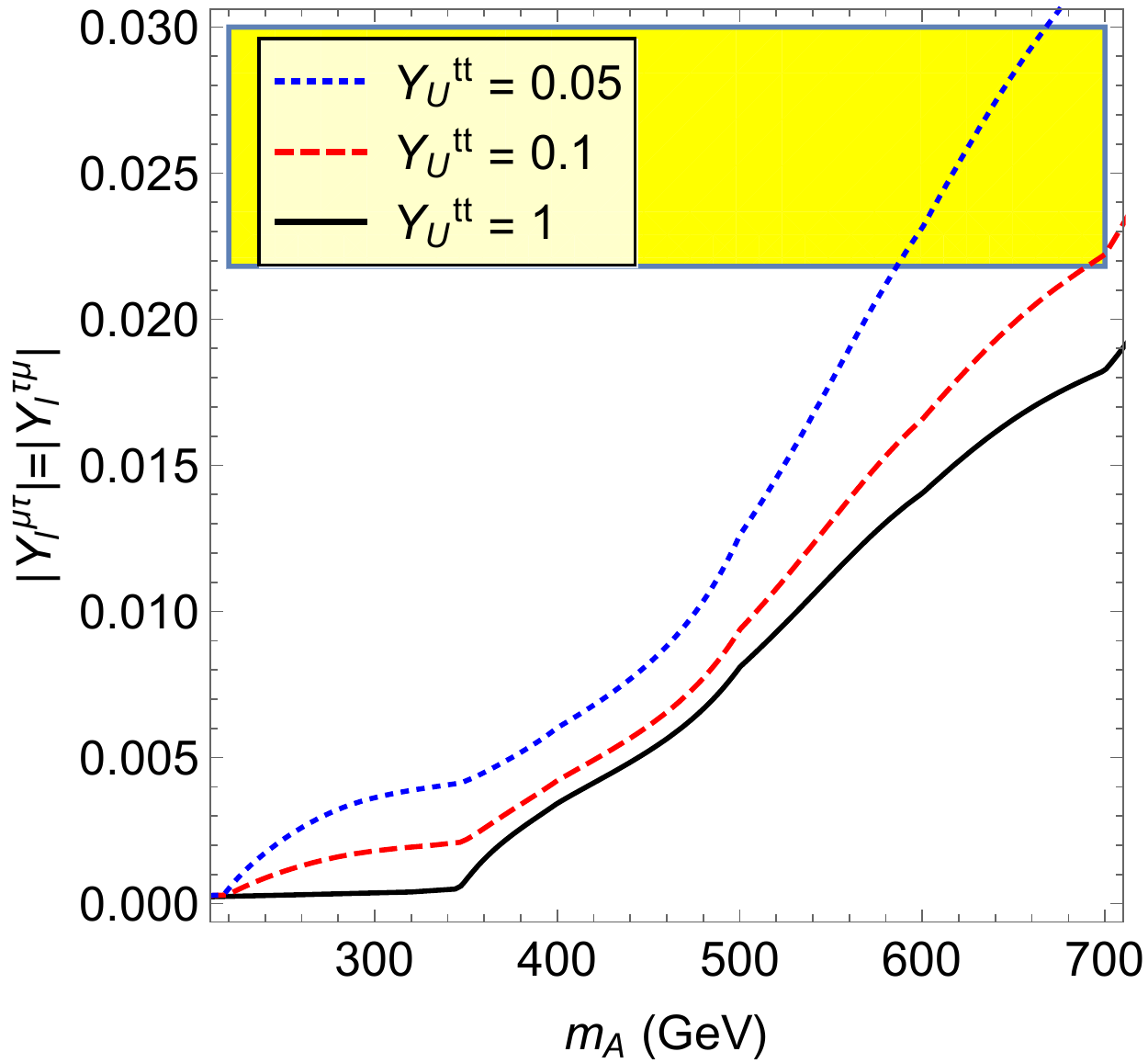}\label{fig:pseudoscalarboundsoptimistic01}} \hspace{2mm}
	\subfloat[$\sin\alpha = 0.1$]{\includegraphics[width= 0.31\textwidth]{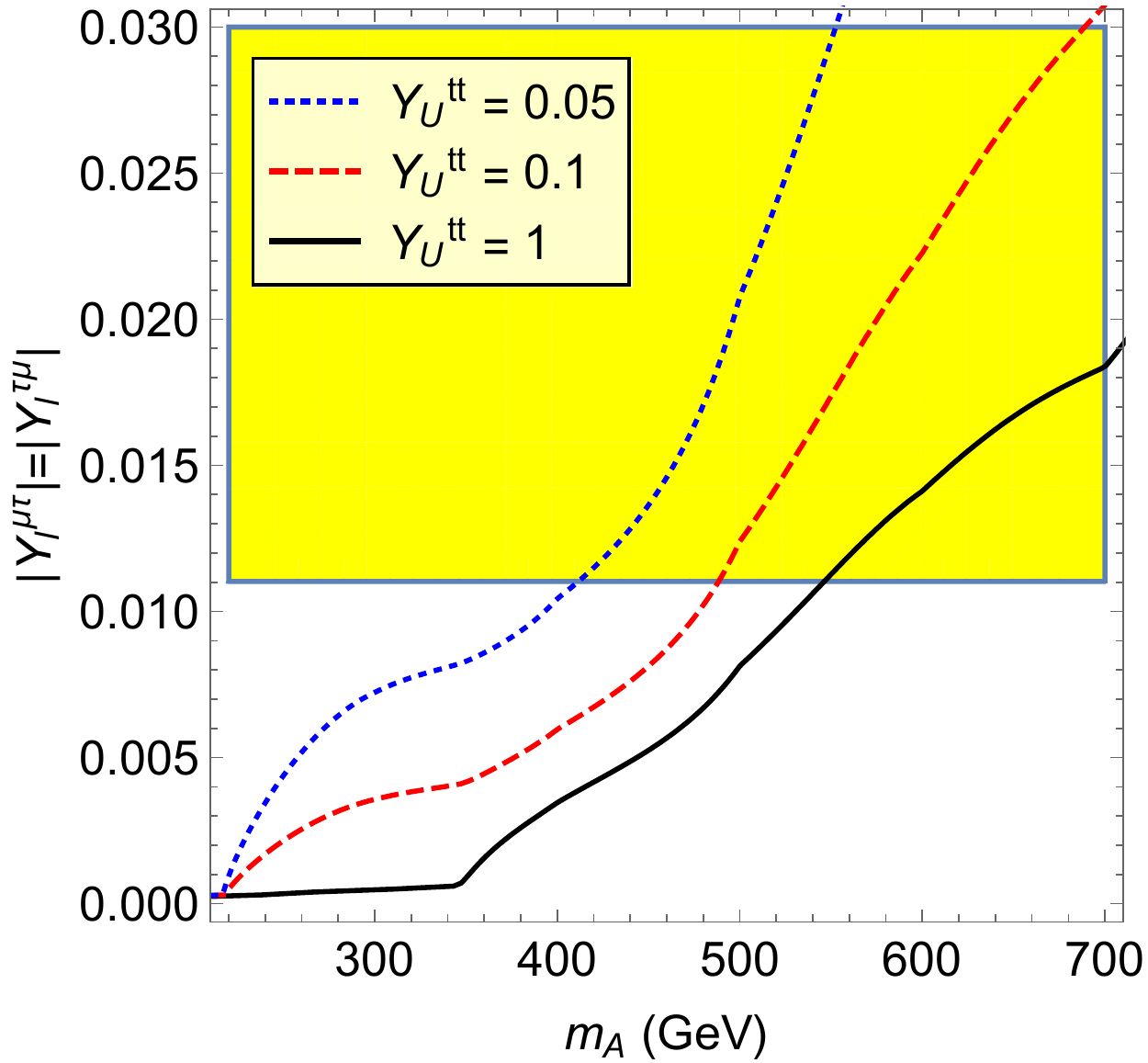}\label{fig:pseudoscalarboundsoptimistic01}} \hspace{2mm}
		\subfloat[$\sin\alpha = 0.5$]{\includegraphics[width= 0.31\textwidth]{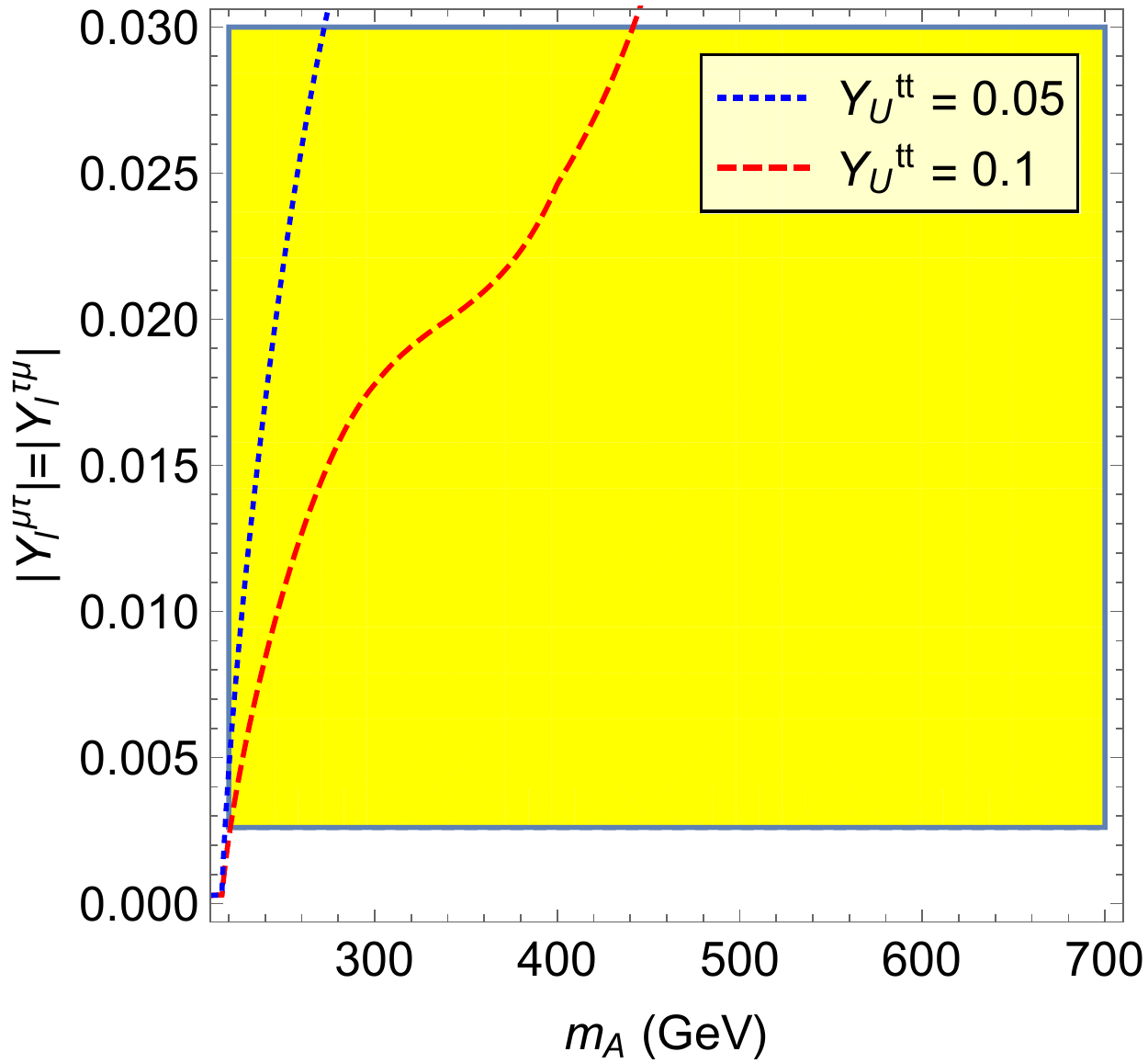}\label{fig:pseudoscalarboundsoptimistic05}} \\
					\subfloat[$\sin\alpha = 0.05$]{\includegraphics[width= 0.3\textwidth]{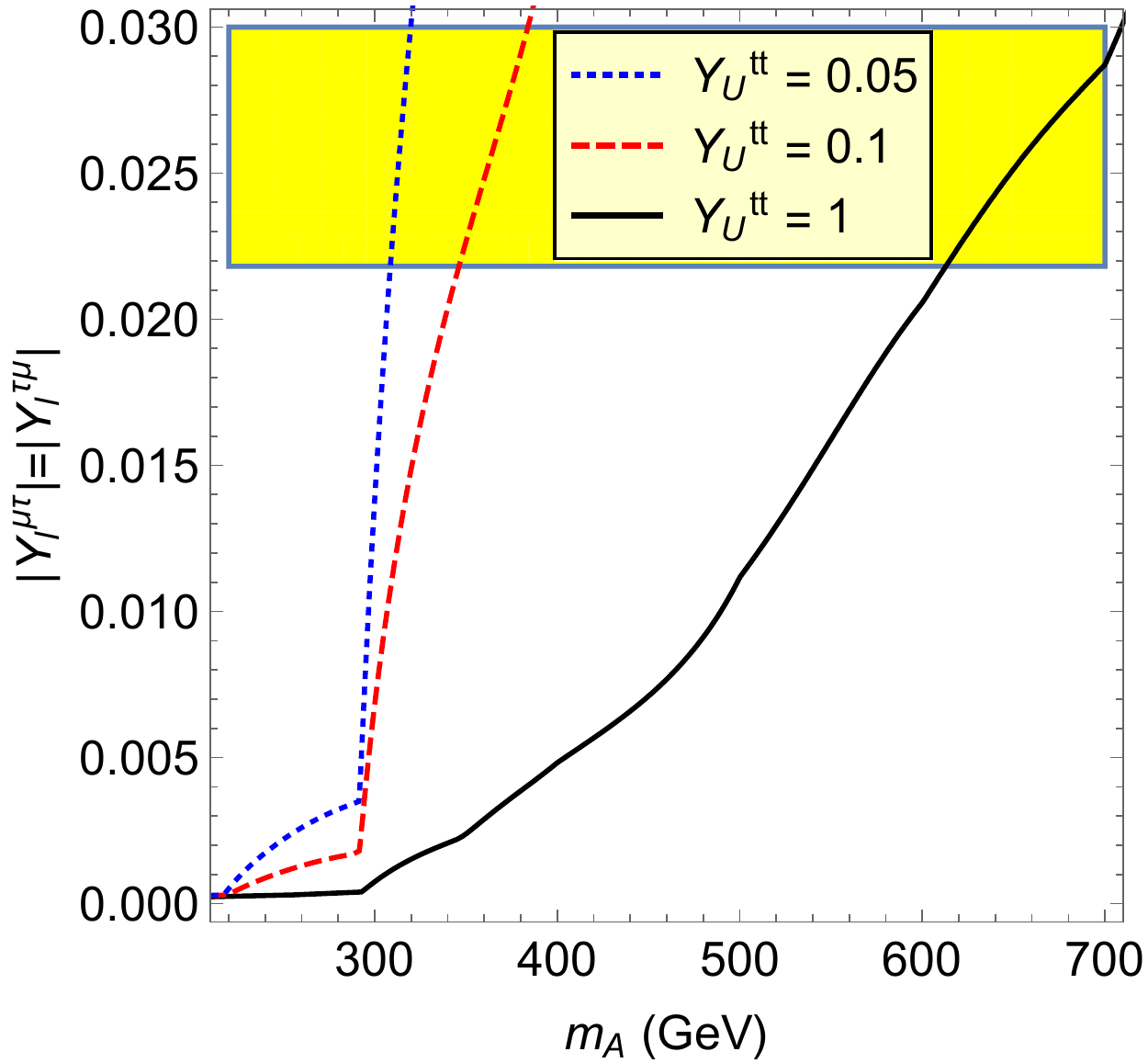}\label{fig:pseudoscalarboundsoptimistic01mH200}} \hspace{5mm}
			\subfloat[$\sin\alpha = 0.1$]{\includegraphics[width= 0.3\textwidth]{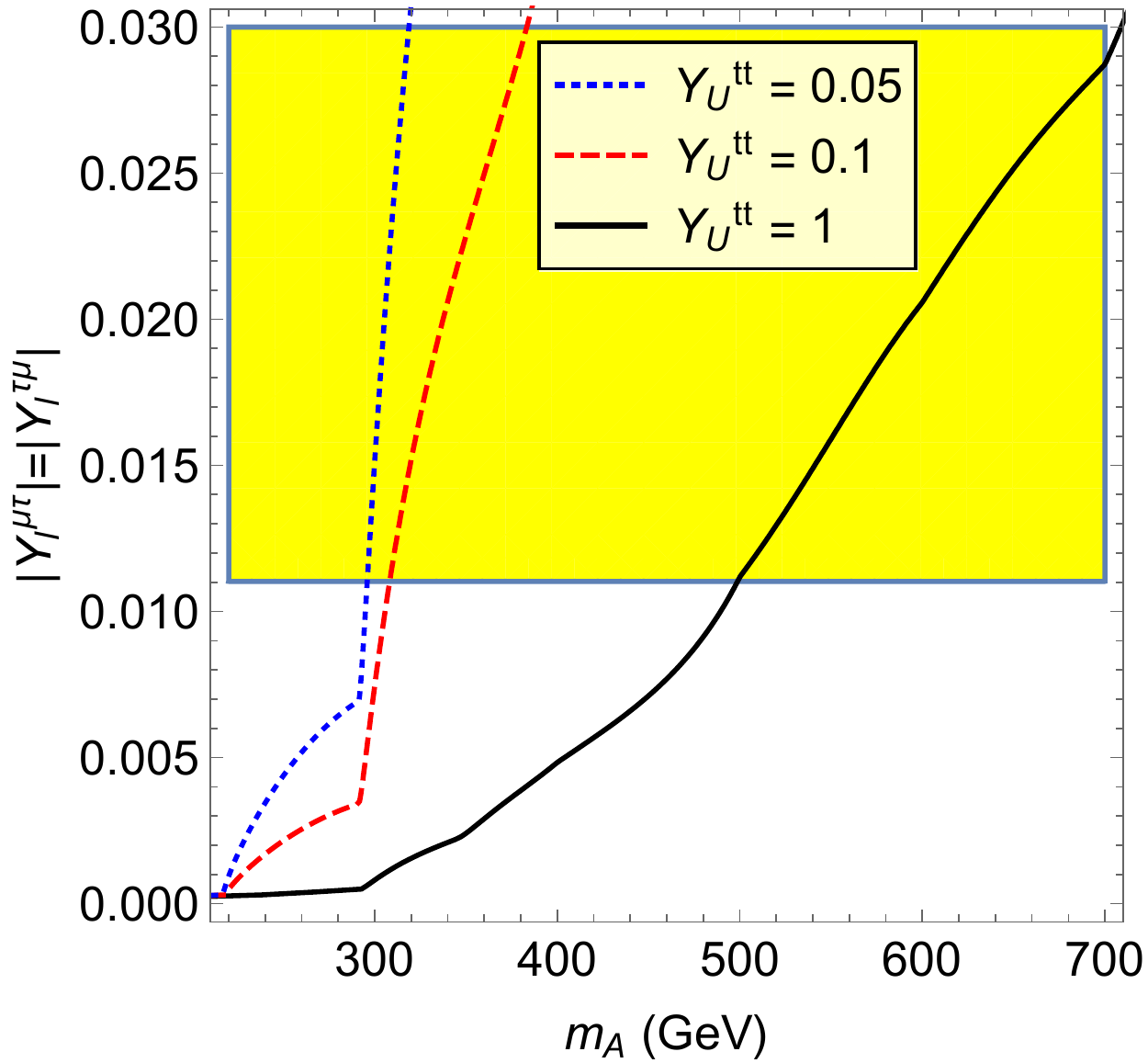}\label{fig:pseudoscalarboundsoptimistic01mH200}} \hspace{5mm}
		\subfloat[$\sin\alpha = 0.5$]{\includegraphics[width= 0.3\textwidth]{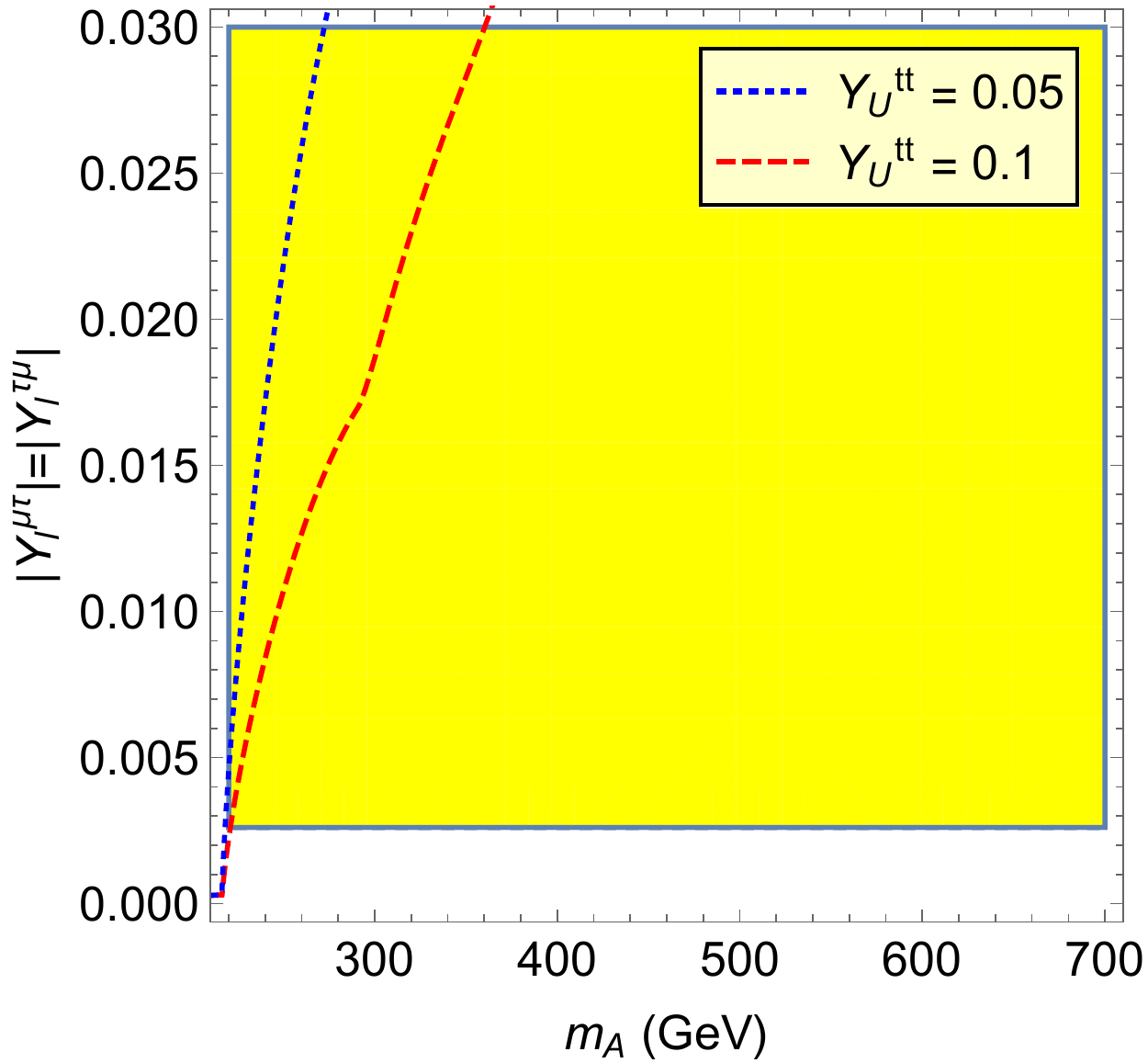}\label{fig:pseudoscalarboundsoptimistic05mH200}}
		\caption{The estimated bounds for the pseudoscalar LFV searches and their comparisons with the $h$ LFV searches for $\sin\alpha = 0.05, 0.1, 0.5$ and $Y_U^{tt}=0.05,0.1,1$. In the top pane, the decay $A\to HZ$ is closed ($m_A<m_H+m_Z$) while in the bottom pane the decay $A\to HZ$ may be open ($m_H = 200$ GeV). The yellow region is excluded by the $h$ LFV search. The region above the red, blue and black lines are excluded by the pseudoscalar LFV search} 
		\label{fig:pseudoscalaroptimistic}
\end{figure}

Having discussed the heavy scalar LFV bounds, now we turn our attention to the pseudoscalar LFV bounds.
The production cross section of $A$ depends only on $Y_U^{tt}$ and $m_A$ while the LFV branching ratio also depends $\sin\alpha$, $Y_\ell^{\tau\mu}$ and $Y_\ell^{\mu\tau}$. Similar to the case of the heavy scalar, the $A$ LFV branching fraction decreases as new decay channels ($A\to hZ$ and $A\to HZ$) open. 
The estimated LFV pseudoscalar bounds for various benchmark points are shown in Fig.~\ref{fig:pseudoscalaroptimistic}. 
Note the partial decay width of $A$ depends on $\sin^2\alpha$, thus we only consider the positive values of $\sin\alpha$.
Let's first consider the case where the decay channel $A\to HZ$ is closed (the top rows of Fig.~\ref{fig:pseudoscalaroptimistic}).
For the case of a small mixing angle, $\sin\alpha =0.05$ and $0.1$, the $BR_{h\to\tau\mu}$ is small, thus the bounds from the $A$ LFV search are stronger than the $h$ LFV bounds for a wide range of $m_A$. On the other hand, for a large mixing angle, ie. $\sin\alpha=0.5$, the viable value of $Y_U^{tt}$ is small. In this case the $BR_{h\to\tau\mu}$ is large while the production cross-section of $A$ is small. Thus the bound from the $h$ LFV search is much stronger than the $A$ LFV bound.
In the case that the decay channel $A \rightarrow H Z$ is open, the LFV branching fraction reduces significantly. 
Hence for low values of $Y_U^{tt}$, the LFV $A$ bounds is stronger than LFV $h$ bounds only for $m_A \lesssim m_H + m_Z$. 
For a bigger value of $Y_U^{tt}$, the $A$ production cross-section increases. 
As a result, the LFV $A$ bounds can be stronger for higher values of $m_A$. 
Finally, we note that in the $SO(3)$ limit where $m_H = m_A$, the pseudoscalar bounds is more constraining than the heavy scalar bounds.  

\subsubsection{Small cross section case: mixing}

\begin{figure}[h]
        \centering
        \includegraphics[width= 0.5\textwidth]{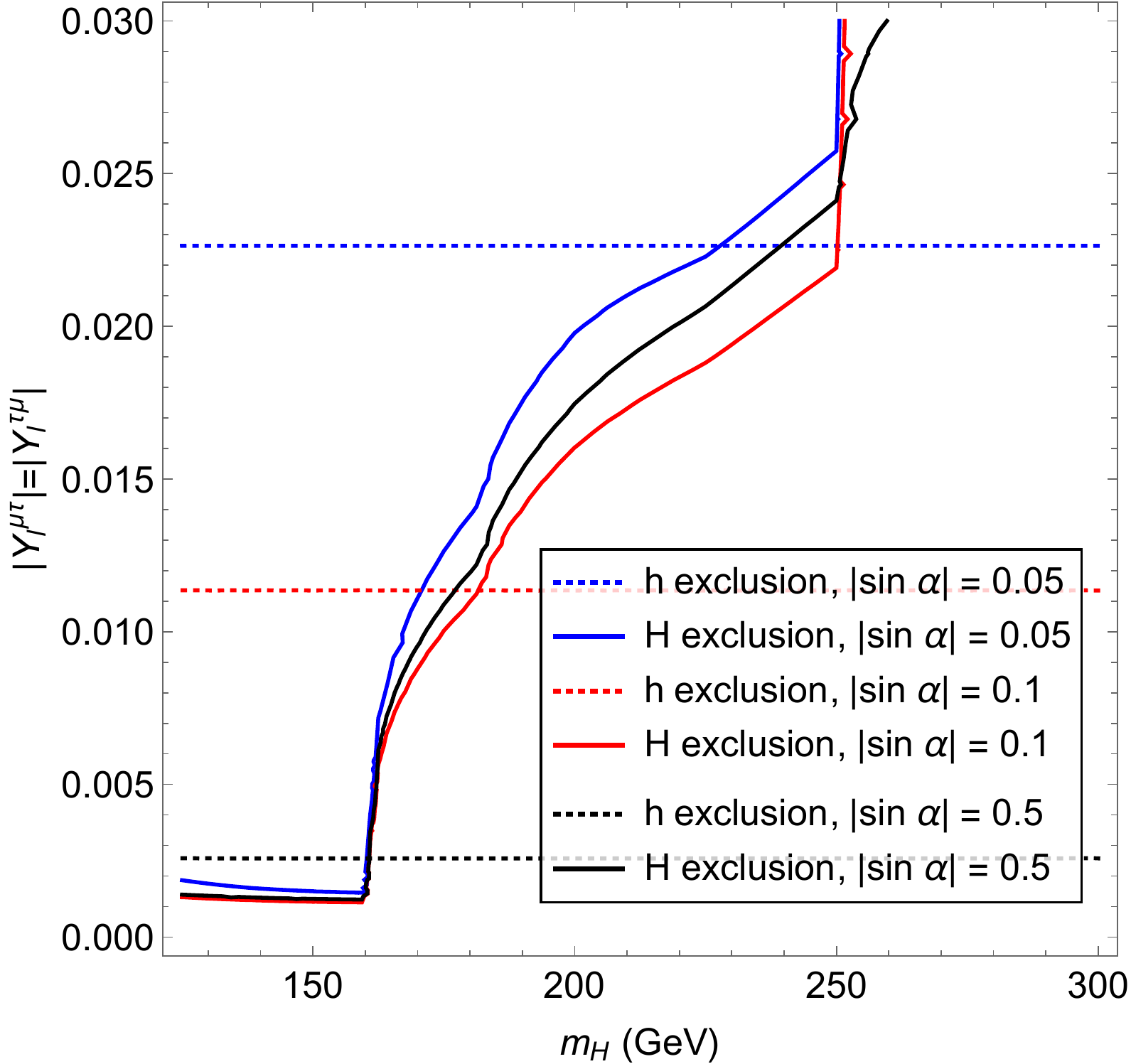}
        \caption{Projected 13 TeV with 300 $\text{fb}^{-1}$ luminosity LHC bounds on $Y_{\ell}^{\mu\tau}$ and $Y_{\ell}^{\tau\mu}$ as a function of the heavy scalar mass with $\sin\alpha=$ 0.05, 0.1 and 0.5 respectively. The region above each curve is excluded.  Here we assume $\lambda_{hhH} = 1$, $m_H < 2 m_A$ and $m_H < 2 m_{H^+}$.}
         \label{fig:pseudoscalarboundsmixing}
\end{figure}

For the rest of this section we'll consider the opposite limit where $Y_U^{tt}$ is vanishing. In this limit, the production cross-section of the heavy resonances, $H$ and $A$, are small. As a result, one would expect the bounds on LFV from $H$ and $A$ searches to get weaker. Nevertheless, the LFV bounds from the heavy resonances searches can still be more constraining than the bound from the 125 GeV search in some parts of parameter space.

We starts with the simplest case where all the $Y$'s are vanishing except for $Y_\ell^{\mu\tau}$ and $Y_\ell^{\tau\mu}$. 
In this scenario the pseudoscalar cannot be produced, hence it offers no LHC bounds.
The heavy scalar, $H$, on the other hand, can still be produced through the neutral scalars mixing.
Fig.~\ref{fig:pseudoscalarboundsmixing} shows the 13 TeV with 300 $\text{fb}^{-1}$ LHC bounds on the coupling $Y_\ell^{\mu\tau}$ and $Y_\ell^{\tau\mu}$ as a functions of $m_H$ with $\sin\alpha=0.05$, 0.1 and 0.5 respectively. For comparison, similar bounds from the 125 GeV search are also given. 
From the plot, we can see that in the case of $\sin\alpha \gtrsim 0.1$, the $H$ LFV search for $m_{H} \lesssim 160$ GeV yields stronger bounds than the $h$ LFV searches alone. 
In the case of $m_{H} \gtrsim 160$ GeV, the decay channel $H\to WW$ and $H\to ZZ$ are open, hence the bounds get weaken. For a lower value of $\sin\alpha$, the LFV $H$ search still provides a better bound until $m_H$ = 250 GeV where the decay channel to $hh$ opens.
 
In our analysis so far we have taken all the Yukawa couplings to vanish except $Y_\ell^{\mu\tau}$ and $Y_\ell^{\tau\mu}$. 
One could argue this is an optimistic scenario because the LFV branching faction of the heavy resonances are large, see for example Figs.~\ref{fig:brA} and~\ref{fig:brH}.
Introducing other non-zero Yukawa couplings, in principle, would dilute the LFV branching ratio and hence weakens our LFV bounds.   
Below, we relax this assumption and explore the consequences of allowing other Yukawa couplings to be non-vanishing. 

\subsubsection{Small cross section case: mixing and $Y_U^{cc}$}
\label{sec:smallYcc}

\begin{figure}
        \centering
        \includegraphics[width= 0.5\textwidth]{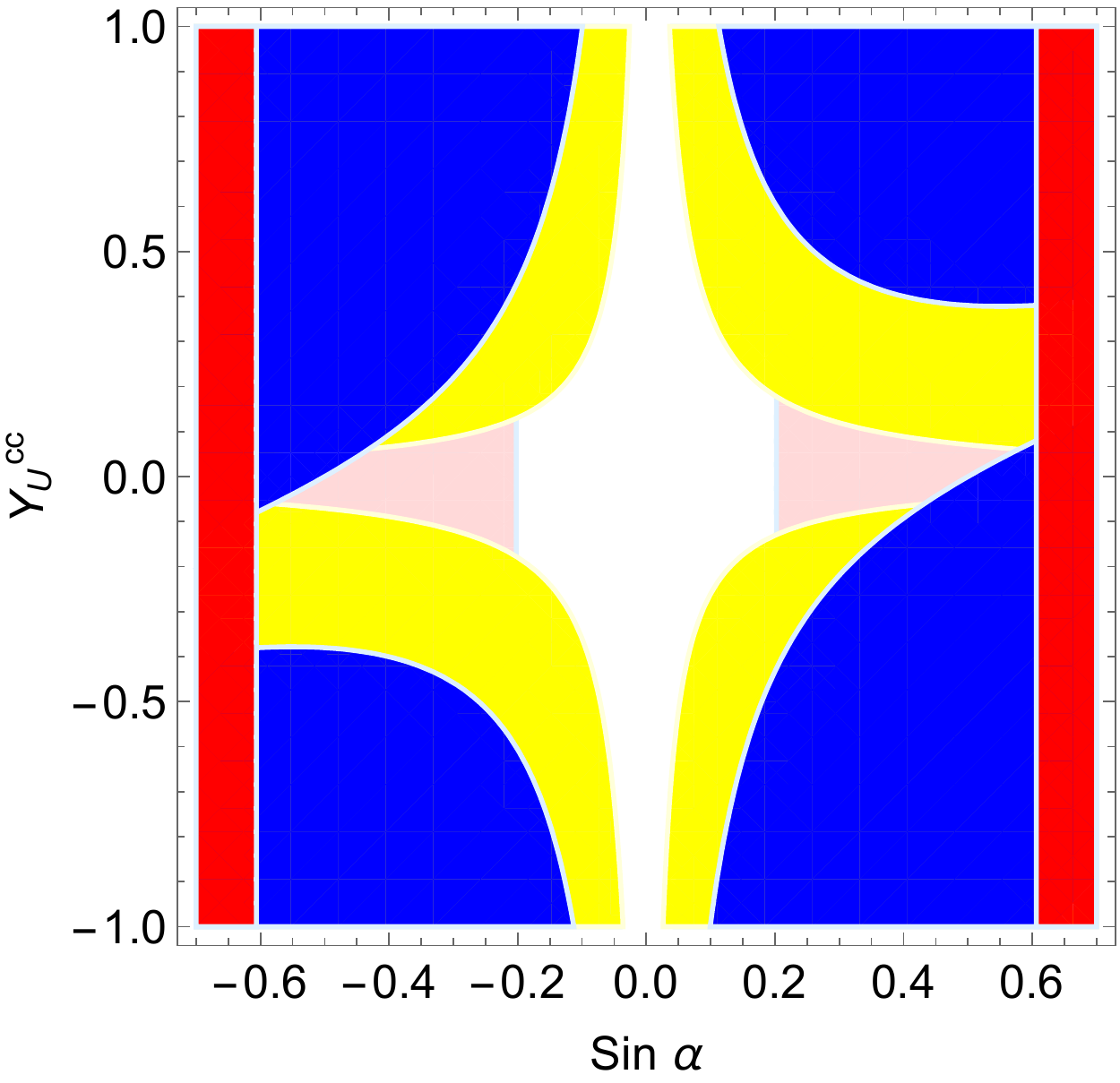}
        \caption{The excluded parameter space in the plane of $Y_U^{cc}$ and $\sin\alpha$. The dark blue and dark red regions are the excluded regions from the LHC run-1 $ggh$ coupling and $WWh$ coupling measurements respectively. The light red regions are the predicted exclusion regions from the $WWh$ coupling measurements at the 14 TeV LHC with 300 fb$^{-1}$ luminosity. The yellow region is the $cch$ coupling bounds taken from \cite{Perez:2015aoa}.}
         \label{fig:yccsina}
\end{figure}

We begin by introducing non-zero quark Yukawa couplings $Y_U$ and $Y_D$.
The case of non-zero $Y_U^{tt}$ has already been discussed in subsection~\ref{sec:large}, hence we will not consider it here. 
Similarly, we will not consider the off-diagonal elements of $Y_U$ and $Y_D$ since they are severely constrained by FCNC experiments. 
The couplings $Y_U^{uu}$, $Y_D^{dd}$ and $Y_D^{ss}$ are severely constrained, as discussed in Section \ref{subsection:taumum}. The coupling $Y_D^{bb}$ is also tightly constrained from the $h \rightarrow b\bar b$ search at the LHC. Therefore, this leaves the $Y_U^{cc}$ as the least constrained coupling in this scenario. Ref.~\cite{Perez:2015aoa} estimates the bounds on the light Higgs coupling to the charm quarks from a global fit of the LHC Higgs data and found that $y_{U,h}^{cc}\lesssim 0.045$. 
A non-zero value of $y_{U,h}^{cc}$ also contributes to the $h$ production cross-section through a charm loop and is also constrained by the Higgs measurements. The bounds on $Y_U^{cc}$ and $\sin\alpha$ from the above constraints are shown in Fig.~\ref{fig:yccsina}.

\begin{figure}
        \centering
        \includegraphics[width= 0.6\textwidth]{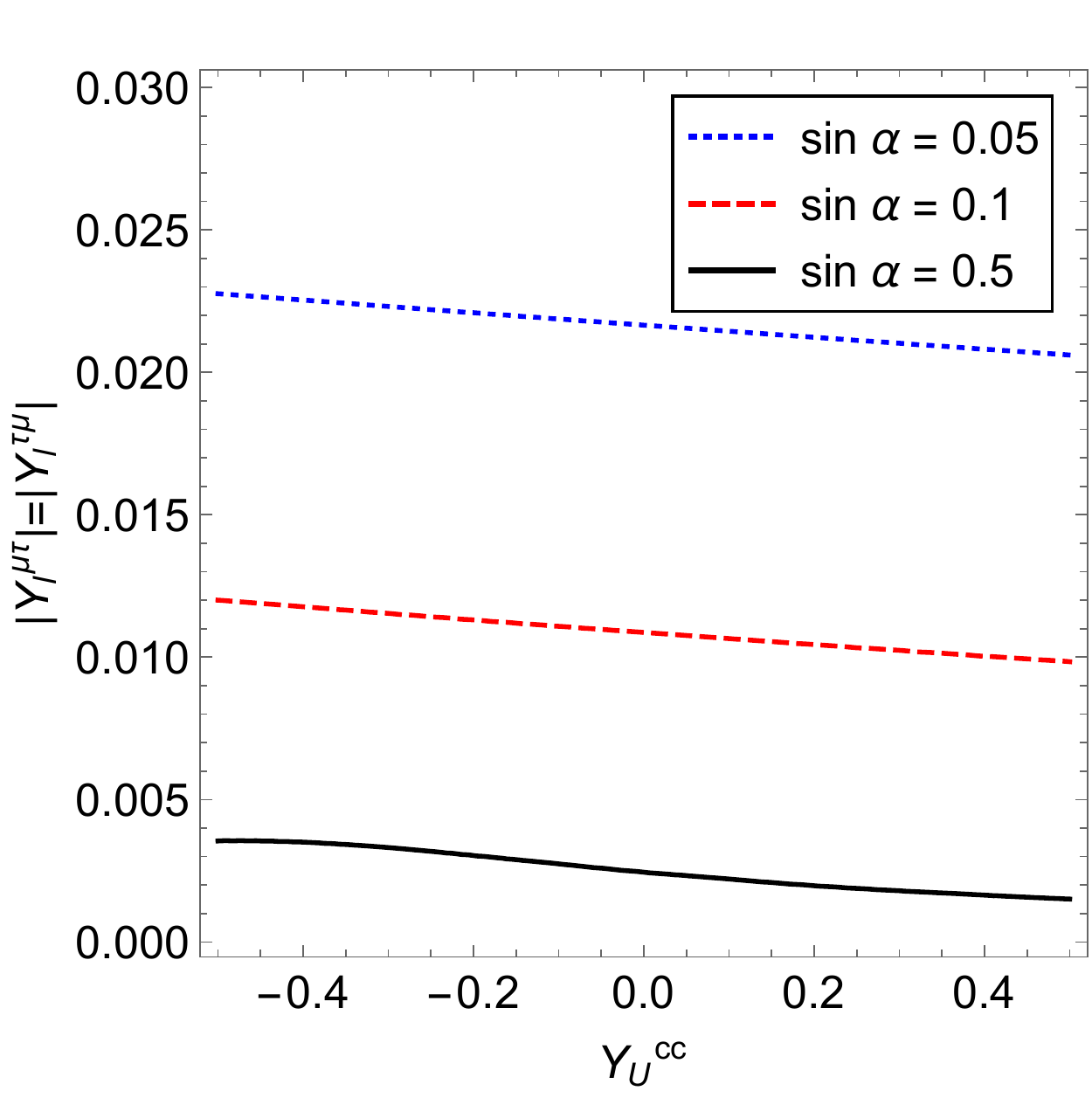}
        \caption{Estimated bounds on $Y_{\ell}^{\tau\mu}$ and $Y_{\ell}^{\mu\tau}$ as a function of $Y_{U}^{cc}$ for $\sin\alpha=0.05$, 0.1 and 0.5. The region above each curve is excluded.}
         \label{fig:ytaumuycc}
\end{figure}

As in the case of nonzero $Y_U^{tt}$, the LFV branching fraction of the light Higgs depends on $Y_\ell^{\tau\mu}$, $Y_\ell^{\mu\tau}$ and $\sin\alpha$ through $y_{\ell,h}^{\tau\mu}$ and $y_{\ell,h}^{\mu\tau}$. 
In this scenario, the charm-loop contribution to the $h$ production cross-section is enhanced by a non-zero $Y_U^{cc}$. However, in the case of $h$, the top-loop contribution dominates over the charm-loop contribution.
The LFV bounds from the light Higgs search are shown in Fig.~\ref{fig:ytaumuycc}. 
These bounds do not vary greatly with $Y_U^{cc}$, especially for small values of $\sin\alpha$, since in these cases the charm-loop contribution is not significantly enhanced. 

\begin{figure}[htpb]
\centering 
\subfloat[$\sin\alpha = 0.05$]{\includegraphics[width= 0.31\textwidth]{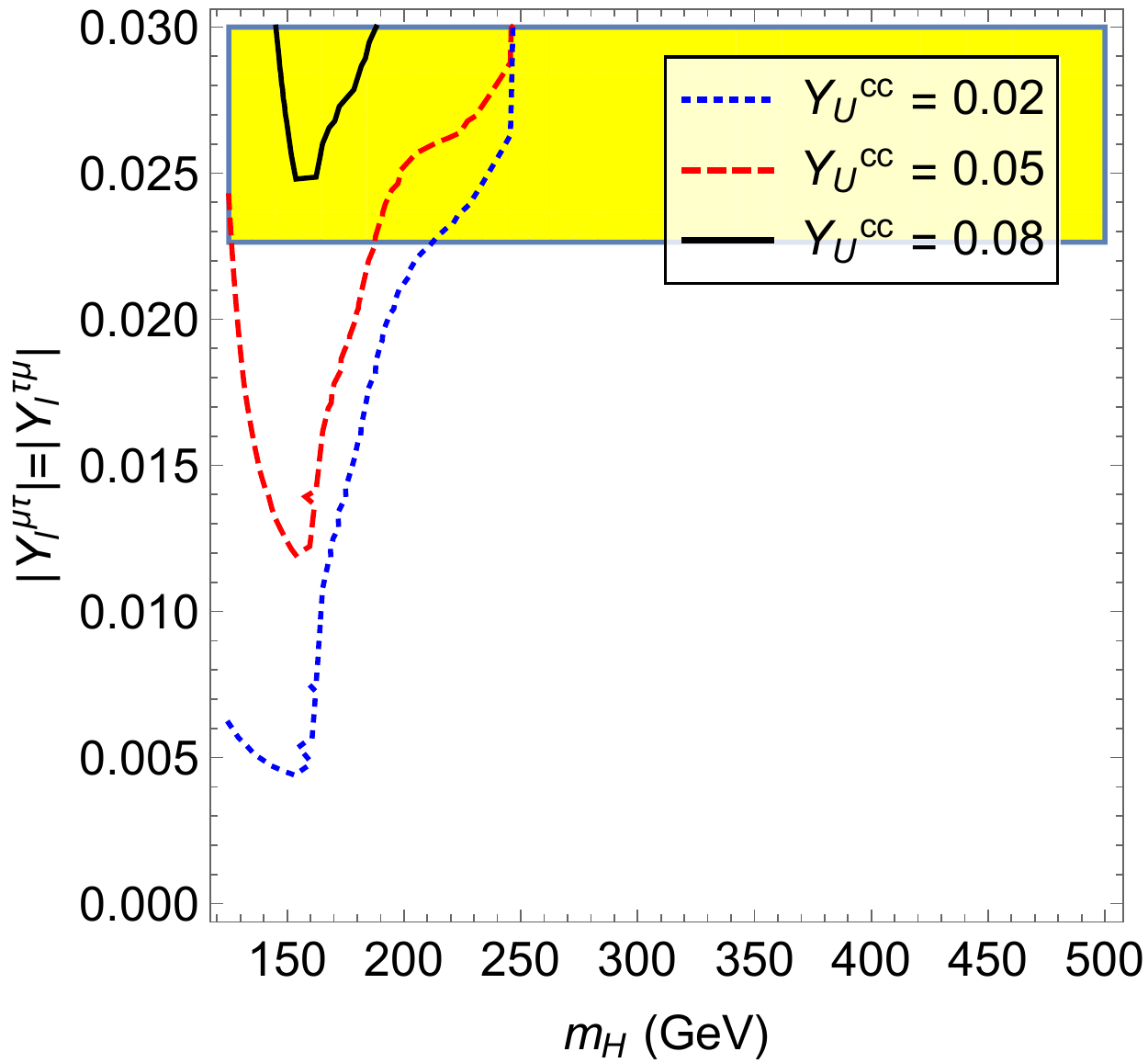}\label{fig:scalarboundsweakycc005}} \hspace{2mm}
	\subfloat[$\sin\alpha = 0.1$]{\includegraphics[width= 0.31\textwidth]{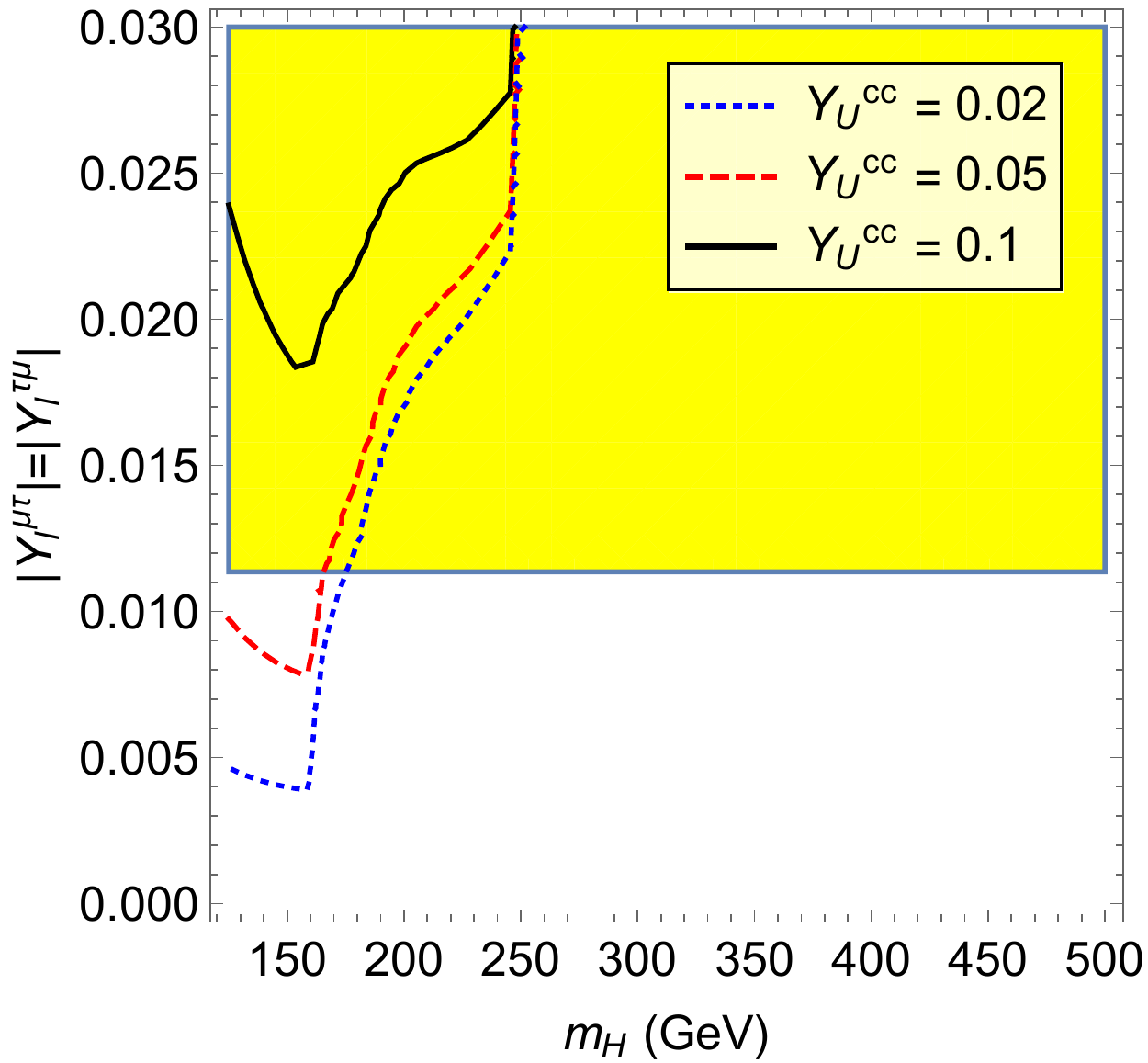}\label{fig:scalarboundsweakycc01}} \hspace{2mm}
		\subfloat[$\sin\alpha = 0.5$]{\includegraphics[width= 0.31\textwidth]{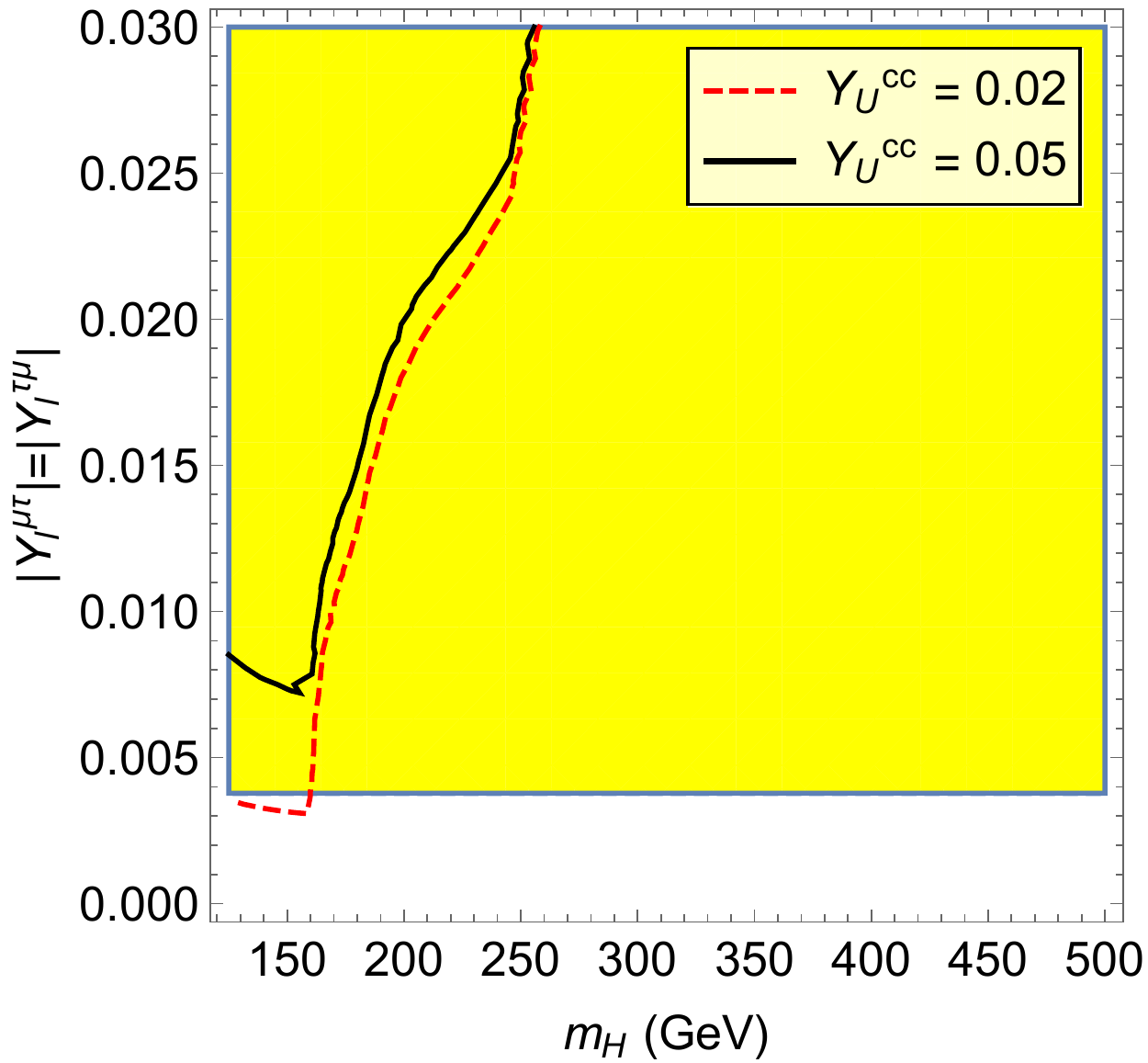}\label{fig:scalarboundsweakycc01}} \\
					\subfloat[$\sin\alpha = -0.05$]{\includegraphics[width= 0.3\textwidth]{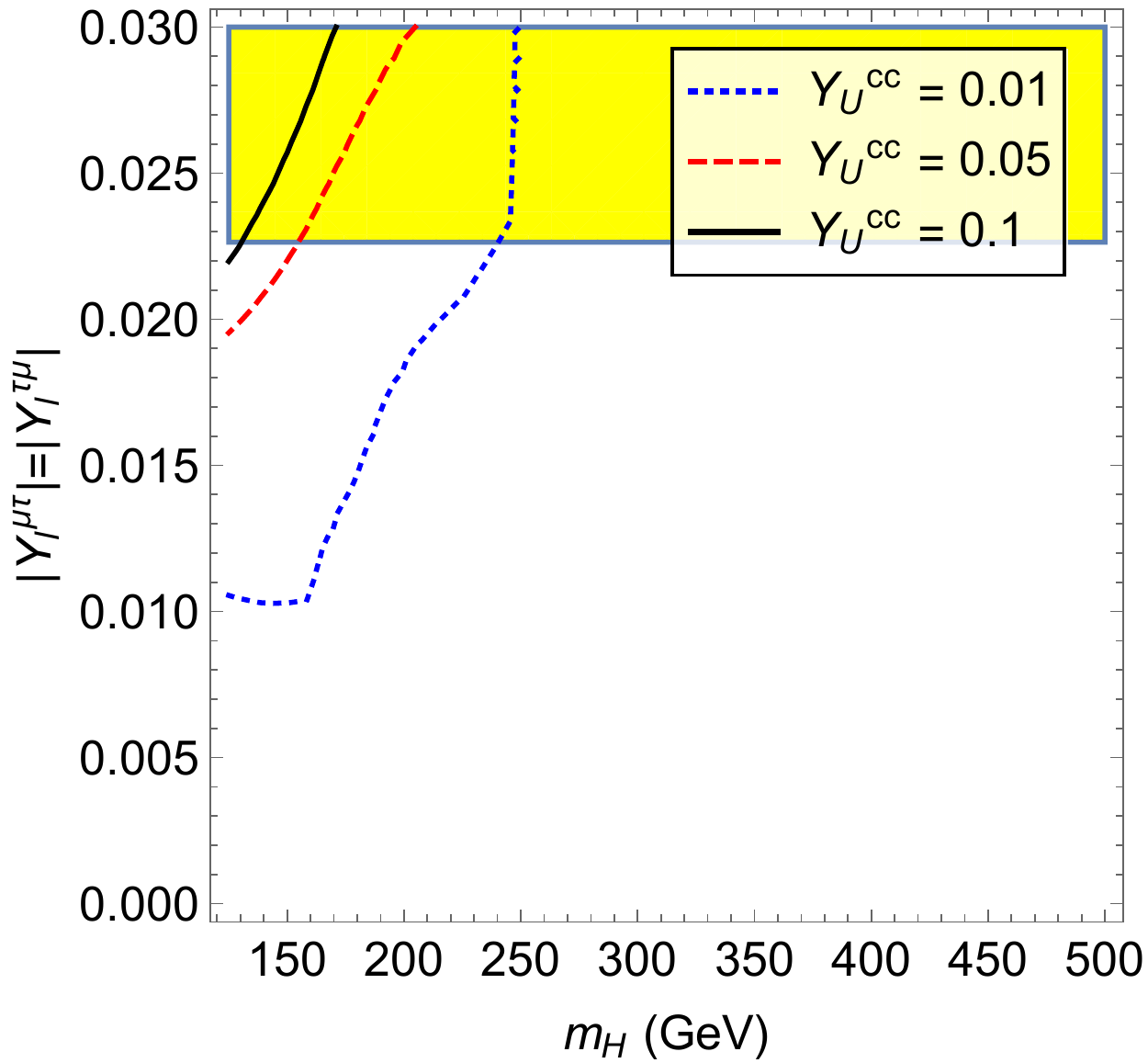}\label{fig:scalarboundsweakyccminus005}} \hspace{5mm}
			\subfloat[$\sin\alpha = -0.1$]{\includegraphics[width= 0.3\textwidth]{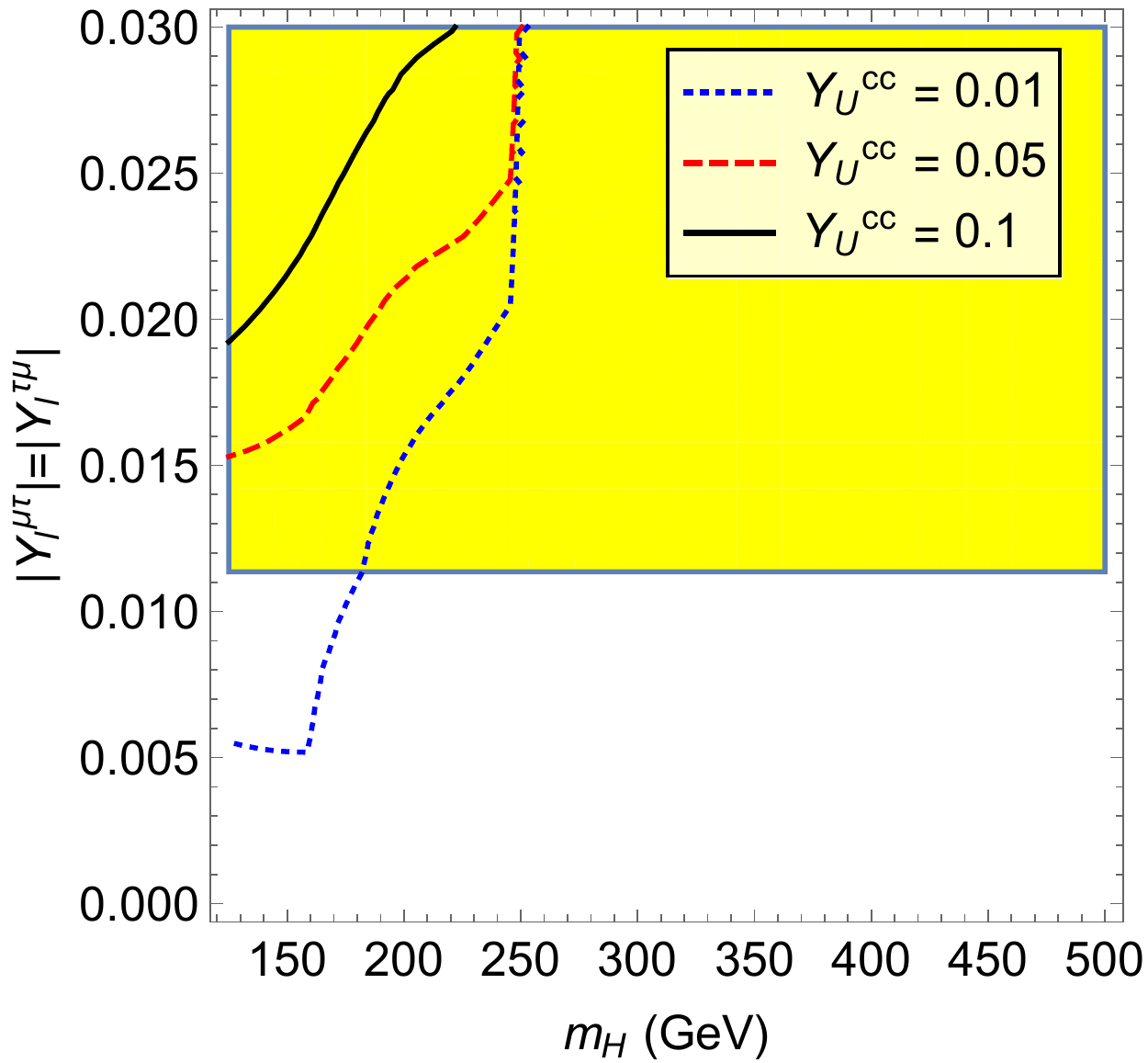}\label{fig:scalarboundsweakyccminus01}} \hspace{5mm}
		\subfloat[$\sin\alpha = -0.5$]{\includegraphics[width= 0.3\textwidth]{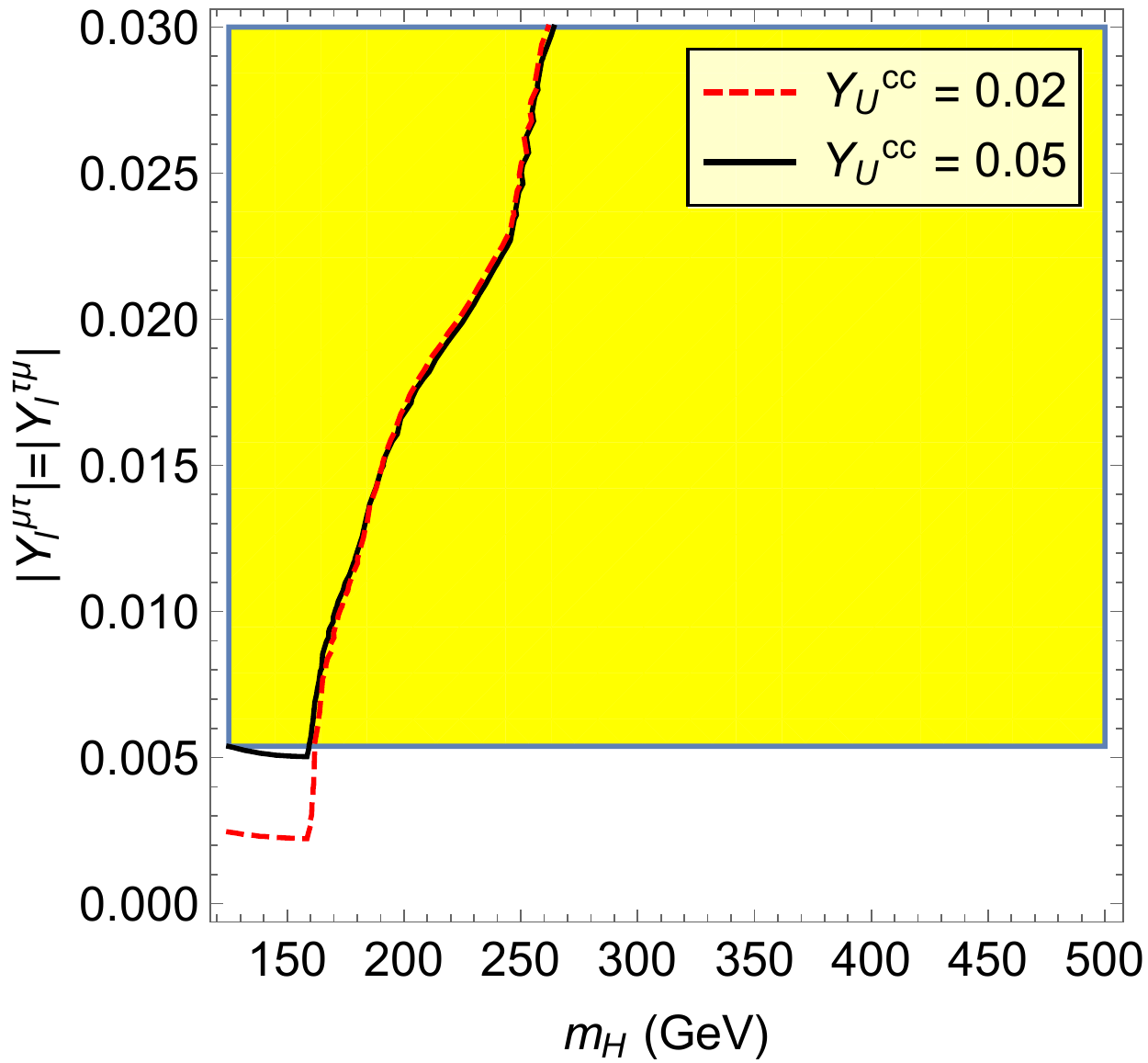}\label{fig:scalarboundsweakyccminus05}}
		\caption{The estimated bounds for the heavy scalar LFV searches and their comparisons with the $h$ LFV searches. The yellow region is excluded by the $h$ LFV search. The region above the red, blue and black lines are excluded by the heavy scalar LFV search. 
		Here we assume $\lambda_{hhH} = 1$, $m_H < 2 m_A$ and $m_H < 2 m_{H^+}$.} 
		\label{fig:scalarboundsweakycc}
\end{figure}

For the heavy scalar, $H$, its main production channel is through the gluon-fusion via the top-loop and the charm-loop. 
For a small mixing angle, $\sin\alpha=\pm0.05$ and $\pm0.1$, the top-loop contribution and the charm-loop contribution are comparable, while for a large mixing angle, $\sin\alpha=\pm0.5$, the top-loop contribution dominates over the charm-loop. 
Hence the $H$ cross-section does not vary significantly over the range of $Y_U^{cc}$ compatible with the LHC Higgs measurements (Fig.~\ref{fig:yccsina}).
On the other hand, the $H$ LFV branching ratio depends strongly on $Y_U^{cc}$. For $Y_u^{cc}\gtrsim0.01$ and $m_H\lesssim2m_W$, the main decay channel of the $H$ is $H\to c\bar c$. Thus in this case $BR_{H\to\tau\mu}\sim|Y_U^{cc}|^{-2}$.
As a result, the $H$ LFV bounds get weaker with increasing $Y_U^{cc}$ as can be seen in Fig.~\ref{fig:scalarboundsweakycc}. 
Note that when the $H\to WW$, $H\to ZZ$ and $H\to hh$ open, the $H$ LFV bounds get significantly weaker due to the large increase in the $H$ total decay width. 
For comparison, the $h$ LFV bounds is also given for each case of $\sin\alpha$ under consideration.
As can be seen from the figures, for $Y_U^{cc} \lesssim 0.05$, the heavy Higgs LFV search constrains more parameter space than the light Higgs LFV search for $m_H \lesssim 160-200$ GeV.

\begin{figure}
        \centering
\subfloat[$\sin\alpha = 0.05$]{\includegraphics[width= 0.31\textwidth]{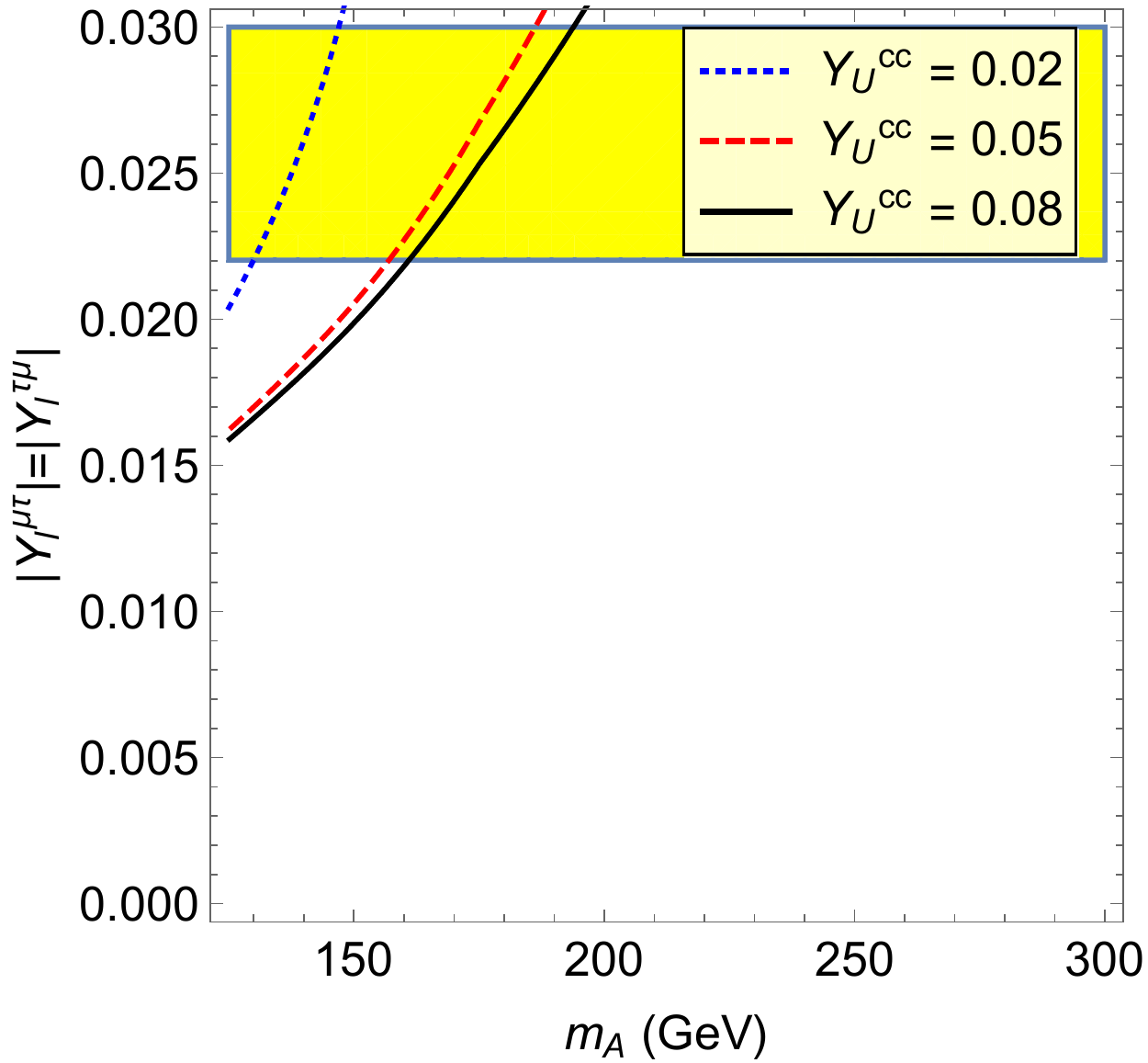}\label{fig:pseudoscalarboundsweakycc005}} \hspace{2mm}
	\subfloat[$\sin\alpha = 0.1$]{\includegraphics[width= 0.31\textwidth]{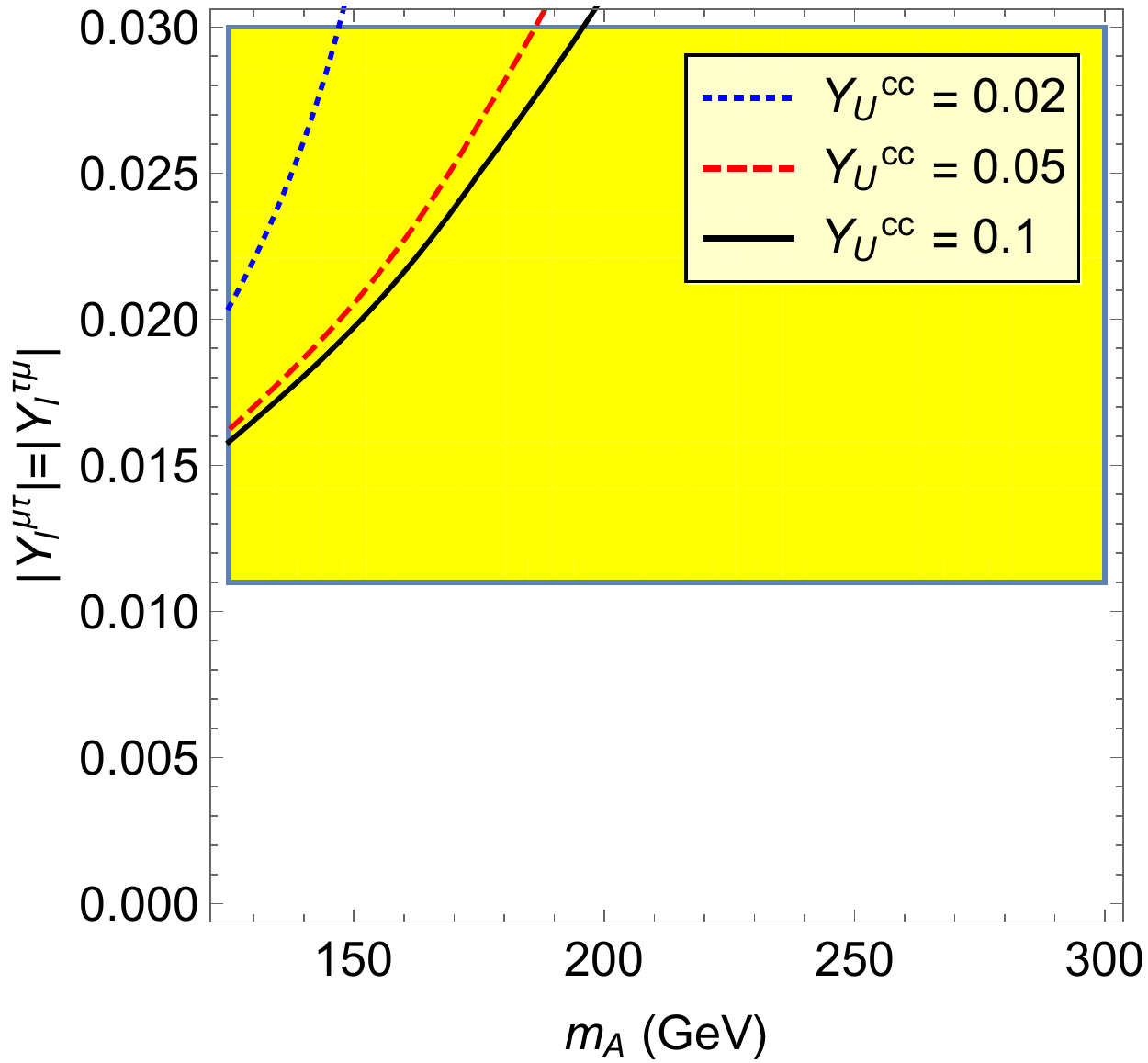}\label{fig:pseudoscalarboundsweakycc01}} \hspace{2mm}
		\subfloat[$\sin\alpha = 0.5$]{\includegraphics[width= 0.31\textwidth]{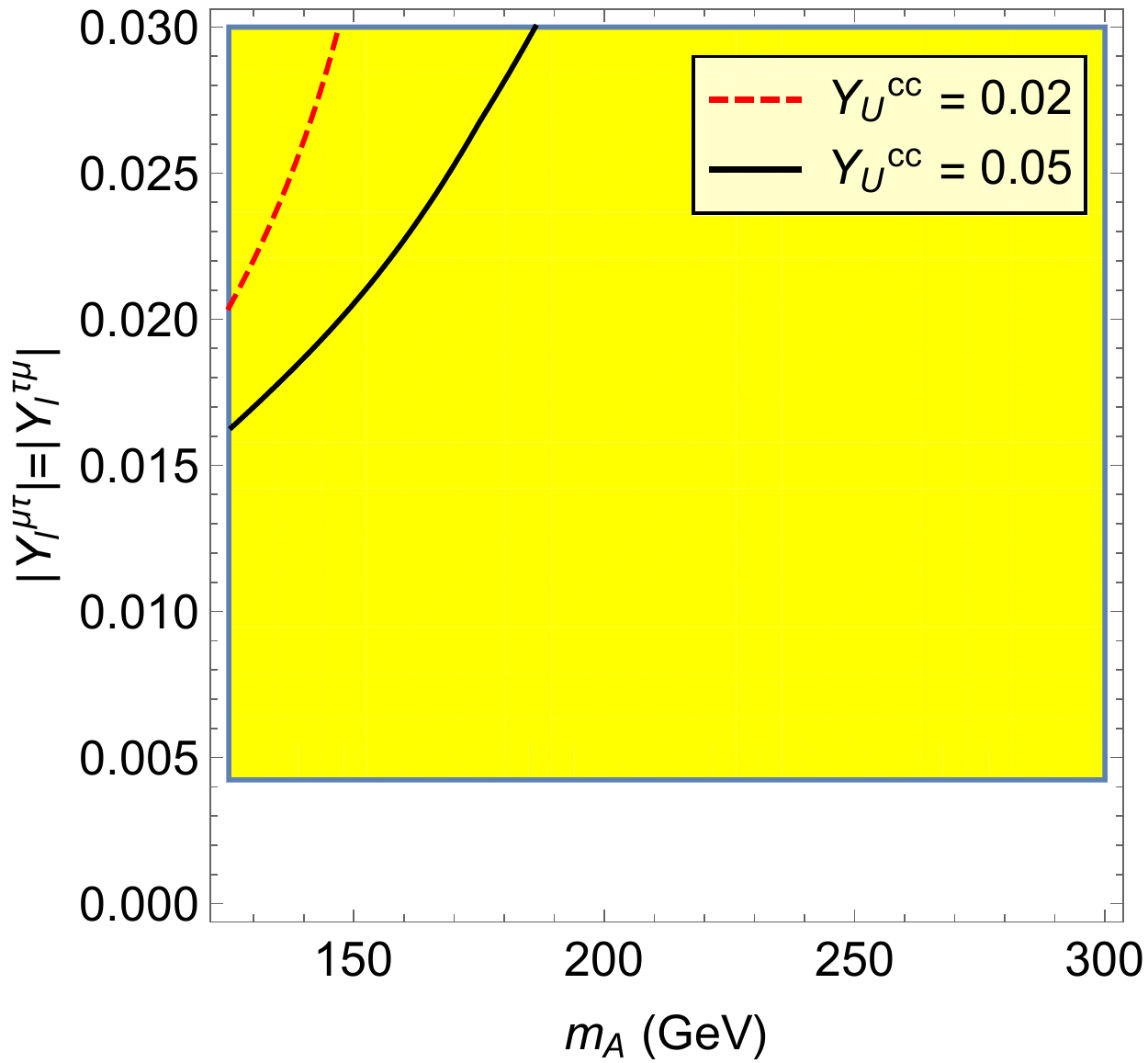}\label{fig:pseudoscalarboundsweakycc05}} 
		\caption{The estimated bounds for the pseudoscalar LFV searches and their comparisons with the $h$ LFV searches for $\sin\alpha = 0.05$, $\sin\alpha = 0.1$ and  $\sin\alpha = 0.5$. For all the figures, we take $m_A<m_H+m_Z$ so the decay channel $A\to HZ$ is closed. The yellow region is excluded by the $h$ LFV search. The region above the red, blue and black lines are excluded by the pseudoscalar LFV search.} 
		\label{fig:pseudoscalarboundsweakycc}
\end{figure}

For the pseudoscalar, its main production channel is through the gluon-fusion via a charm-loop. Thus the $A$ cross-section grows with $|Y_U^{cc}|^2$. Its LFV branching ratio, on the other hand, decreases with $|Y_U^{cc}|^2$ because of the increasing $A\to c\bar c$ partial width. However, the gain in production cross-section outweighs the drop in $BR_{A\to\tau\mu}$. Hence, the $A$ LFV bounds get more constraining for larger values of $Y_U^{cc}$.
Fig.~\ref{fig:pseudoscalarboundsweakycc} shows the estimated bounds for LFV pseudoscalar search for $\sin\alpha=0.05$, 0.1 and 0.5.
In obtaining these bounds, we assume the decay channel $A\to HZ$ is closed. If $A\to HZ$ is open, the bound would be even less constraining. For comparison, we also show
the corresponding LFV $h$ bounds. 
For each value of $\sin\alpha$, the $Y_U^{cc}$ values are chosen so that they are compatible with the LHC Higgs data (Fig.~\ref{fig:yccsina}).
From the plot, we can see that, unless the value of $\sin\alpha$ is small, the LFV $h$ search is more constraining than the LFV pseudoscalar search. Even for $\sin\alpha$ = 0.05, the LFV $A$ search is only more constraining for $m_A \lesssim 170$ GeV. Given this estimated results, we do not consider the case which the $A \rightarrow H Z$ decay channel is open. 
Finally we note that, contrary to the nonzero $Y_U^{tt}$ case, in this case the heavy scalar bounds is stronger than the pseudoscalar bounds in the $SO(3)$ limit where $m_A\simeq m_H$.

\subsubsection{Small cross section case: mixing and $Y_\ell^{\tau\tau}$}

\begin{figure}
        \centering
        \includegraphics[width= 0.6\textwidth]{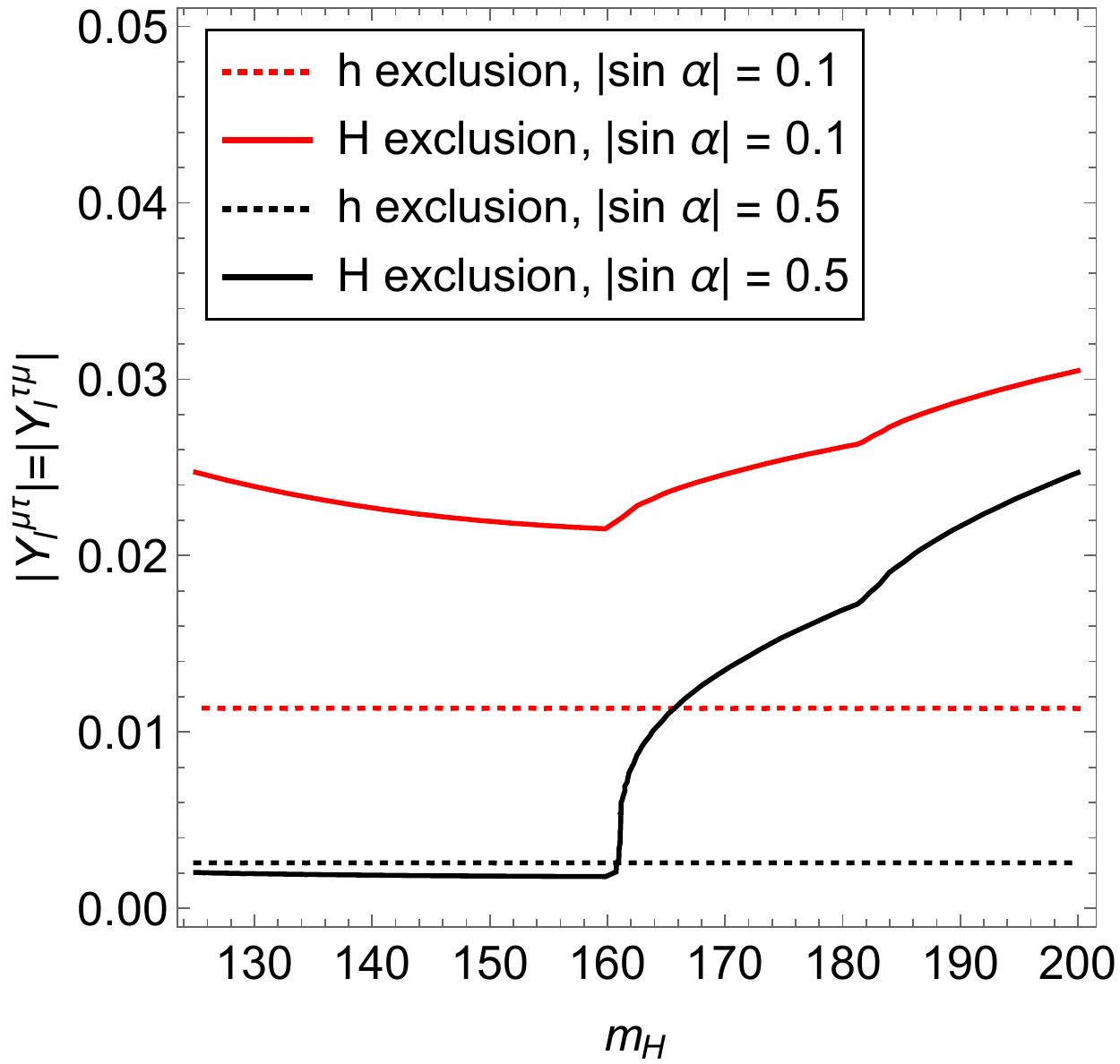}
        \caption{The estimated bounds from LFV decays of $h$ and $H$ in the case of nonzero $Y_\ell^{\tau\tau}$ and $Y_\ell^{\mu\tau}$ for 13 TeV 300 fb$^{-1}$ LHC. The value of  $Y_\ell^{\tau\tau}$ is taken to be the maximum value that still satisfies a 95\% C.L. of $\mu^{\tau\tau} = 1.11^{+0.24}_{-0.22}$~\cite{Khachatryan:2016vau}. The region above each curve is excluded.}
         \label{fig:weakytautau_ytaumumh}
\end{figure}

In the previous subsection, we reduce the LFV branching fractions by introducing $Y_U^{cc}$. In that scenario, the production cross-section in the gluon-fusion channel for the scalars are slightly boosted by charm loops. If instead one introduces a non-zero $Y_\ell^{\tau\tau}$, while keeping other $Y$'s zero (except $Y_\ell^{\mu\tau}$ and $Y_\ell^{\tau\mu}$), the production cross section of the scalars does not get affected\footnote{Non-zero value of $Y_\ell^{\mu\mu}$ and $Y_\ell^{ee}$ will have the same effect. However these couplings are constrained by $\tau\rightarrow \mu \ell^+ \ell^-$ and direct $h \rightarrow \ell^+\ell^-$ searches as discussed in Section \ref{sec:taumull}. }. However, in this scenario the pseudoscalar does not couple to quarks hence it will not get produced at the LHC. The value of $Y_\ell^{\tau\tau}$ is constrained by the measurement of $h\to\tau\tau$ decay which has the signal strength $\mu^{\tau\tau} = 1.11^{+0.24}_{-0.22}$~\cite{Khachatryan:2016vau}. Since we want to consider the most pessimistic scenario here, we take the maximum allowed value of $Y_\ell^{\tau\tau}$. The estimated bounds from the $h$ and the $H$ LFV searches are given in Fig.~\ref{fig:weakytautau_ytaumumh}. In this case a large value of mixing is needed to produce enough $H$. For a large, but still allowed, value of $\sin\alpha$, the $H$ LFV search provides better bounds than the $h$ search only in the low mass region, $m_H\lesssim160$ GeV.

\section{Conclusions and discussions}
\label{sec:conclusion}
The recently upgraded LHC offers an opportunity to probe physics at a higher energy scale and a promise of new physics discoveries. 
One of the generic phenomena in new physics scenarios is LFV. 
So far, the experimental efforts to probe LFV at the LHC has been focused on the 125 GeV scalar, $h$, decaying into tau and muon.  
However, since LFV requires new physics and new physics usually comes with extra particles, it could be beneficial to search for LFV in the decays of these new particles as well.
Thus in this paper we explore the possibility of utilizing these additional neutral particles in LFV search. 

In our work, we focus on the CP conserving Type-III 2HDM as a concrete setup for new physics with LFV. In this scenario, the LFV decay into tau and muon originates from the Yukawa couplings $Y_\ell^{\tau\mu}$ and $Y_\ell^{\mu\tau}$, see Eq.~\eqref{eq:yukbeforeewsb}.  These couplings are constrained by both the low energy measurements and the collider experiments. However, we found that the constraints from low energy measurement is typically weaker than the collider constraints, see for example Fig.~\ref{fig:ytaumuvssinalpha}.

The Yukawa couplings $Y_\ell^{\tau\mu}$ and $Y_\ell^{\mu\tau}$ correlate the $h\to\tau\mu$ decay with those of the heavy scalar ($H\to\tau\mu$) and the pseudoscalar ($A\to\tau\mu$).
We simulate the LHC reach for 13 TeV center of mass energy with 300 $\text{fb}^{-1}$ luminosity and find that the $H$ and $A$ search offer complimentary bounds on the LFV parameter space to the traditional $h$ search in many of our benchmark scenarios, see Sec.~\ref{sec:result}. 
Moreover, we have considered various scenarios of the Yukawa couplings and shown that even in the pessimistic case of a small production cross-section and a small LFV branching fraction, the heavy resonance searches are still well motivated.
Thus combining all three searches give the best possible bound on the LFV parameter space.

We assume in this work that CP is conserved, in particular, we take all the Yukawa couplings to be real. One could relax this assumption and consider complex Yukawas. In this case, one can study CP violation in the tau and muon decay.
In fact, CP violation in $h\to\tau\mu$ has been studied in Ref.~\cite{Kopp:2014rva}. However, one can expand such a study to include the $H$ and the $A$.
We leave this for a possible future project. 

\begin{acknowledgments}
The work of PU has been supported in part by 
the Thailand Research Fund under contract no.~TRG5880061, and the Faculty of Science, Srinakharinwirot University under grant no.~655/2559.

\end{acknowledgments}

\appendix
\section{Loop functions}

The loop induced production cross-sections and decays of the neutral Higgs bosons are given in terms of the loop functions~\cite{Djouadi:2005gj}
\begin{equation}
\begin{split}
	A_1^H(\tau) &= -\left[2 + 3\tau + 3\tau(2-\tau)f(\tau)\right],\\
	A_{1/2}^H(\tau) &= 2\tau\left[1+(1-\tau)f(\tau)\right],\\
	A_{1/2}^A(\tau) &= 2\tau f(\tau),
\end{split}
\label{eq:loopfunctions}
\end{equation}
where
\begin{equation}
	f(\tau) = \left\{\begin{aligned}&\left(\sin^{-1}\sqrt{1/\tau}\right)^2,\\ &-\frac14\left[\ln\left(\frac{1+\sqrt{1-\tau}}{1-\sqrt{1-\tau}}\right)-i\pi \right]^2,\end{aligned}
	\quad
	\begin{aligned}&\tau\ge1,\phantom{\frac14^2}\\ &\tau<1.\phantom{\frac14^2}\end{aligned}\right.
\end{equation}

\section{Wilson coefficients for $\tau \to \mu\gamma$}
\label{app:tautomugamma}
The two-loop contributions to the Wilson coefficients $c_L$ and $c_R$ had been worked out in Ref.~\cite{Chang:1993kw}. Below we translated their results into our notation.
The two-loop contributions can be split according to the particles running in the loop.
\begin{align}
	\label{eq:tgammaloop}\Delta c_L^{t\gamma} &= -6\kappa Q_t^2\frac{v}{m_t}\sum_\phi\Big(y^{\tau\mu\ast}_{\ell,\phi}\left[\text{Re}(y_{U,\phi}^{tt})f(z_{t\phi})-i\text{Im}(y_{U,\phi}^{tt})g(z_{t\phi})\right]\Big)\displaybreak[0]\\
	\Delta c_L^{W\gamma} &= \kappa\sum_\phi\left(y_{\ell,\phi}^{\tau\mu\ast}\delta_\phi\left[3f(z_{W\phi}) + \frac{23}{4}g(z_{W\phi}) +\frac34h(z_{W\phi}) + \frac{f(z_{W\phi})-g(z_{w\phi})}{2z_{W\phi}}\right]\right)\displaybreak[0]\\
	\label{eq:tZloop}\Delta c_L^{tZ} &= -6\kappa Q_t\frac{(1-4s_W^2)(1-4Q_ts_W^2)}{16s_W^2c_W^2}\frac{v}{m_t}\times\nn\\
	&\qquad\qquad\sum_\phi\left(y_{\ell,\phi}^{\tau\mu\ast}\left[\text{Re}(y_{U,\phi}^{tt})\tilde{f}(z_{t\phi},z_{tZ})-i\text{Im}(y_{U,\phi}^{tt})\tilde{g}(z_{t\phi},z_{tZ})\right]\right)\displaybreak[0]\\
	\Delta c_L^{WZ} &= \kappa\frac{1-4s_W^2}{4s_W^2}\sum_\phi\Bigg(y_{\ell,\phi}^{\tau\mu\ast}\delta_\phi\left[\frac{5-t_W^2}{2}\tilde{f}(z_{t\phi},z_{WZ}) + \frac{7-3t_W^2}{2}\tilde{g}(z_{th},z_{WZ})\right.\nn\\
	&\qquad\qquad\left.+\frac34g(z_{t\phi}) + \frac34h(z_{t\phi}) + \frac{1-t_W^2}{4z_{t\phi}}\left[\tilde{f}(z_{th},z_{WZ}) - \tilde{g}(z_{th},z_{WZ})\right] \right]\Bigg)\displaybreak[0]\\
	\Delta c_L^{CW} &= \frac{\kappa}{4s_W^2}\sum_\phi\Big(y_{\ell,\phi}^{\tau\mu\ast}\delta_\phi\left[D^{(3a)}(z_{W\phi}) + D^{(3b)}(z_{W\phi}) + D^{(3c)}(z_{W\phi})\right.\nn\\
	&\hspace{4cm}\left. + D^{(3d)}(z_{W\phi}) + D^{(3e)}(z_{W\phi})\right]\Big)\displaybreak[0]\\
	\Delta c_L^{CZ} &= \frac{\kappa}{4s_W^2}\sum_\phi y_{\ell,\phi}^{\tau\mu\ast}\delta_\phi\left[D^{(4a)}(z_{Z\phi}) + D^{(4b)}(z_{Z\phi}) + D^{(4c)}(z_{Z\phi})\right]
\end{align}
where $z_{xy}\equiv m_x^2/m_y^2$, ($\delta_h,\delta_H,\delta_A$) = ($\cos\alpha,\sin\alpha,1$),
\begin{equation}
	\kappa = \frac{\alpha}{16\pi}\frac{g^2}{m_W^2}\frac{v}{m_\tau} = \frac{\alpha}{2\sqrt{2}\pi}G_F\frac{v}{m_\tau},
\end{equation}
and the loop functions are
\begin{equation}
\begin{split}
	f(z) &= \frac{z}{2}\int_0^1dx\frac{1-2x(1-x)}{x(1-x)-z}\ln\frac{x(1-x)}{z},\\
	g(z) &= \frac{z}{2}\int_0^1dx\frac{1}{x(1-x)-z}\ln\frac{x(1-x)}{z},\\
	h(z) &= z^2\frac{\partial}{\partial z}\left(\frac{g(z)}{z^2}\right) = \frac{z}{2}\int_0^1\frac{dx}{z-x(1-x)}\left[1+\frac{z}{z-x(1-x)}\ln\frac{x(1-x)}{z}\right],\\
	\tilde{f}(x,y) &= \frac{xf(y)-yf(x)}{x-y} \\
	\tilde{g}(x,y) &= \frac{xg(y)-yg(x)}{x-y}
\end{split}
\end{equation}
Notice for $z<1/4$ the integral in the loop functions contain poles. In this case we follow Ref.~\cite{Chang:1993kw} and take the principle value of the integral. The dipole form factors are more involved.  They are given in terms of 
\begin{equation}
\begin{aligned}
	a_x &= x(x-1)\\
	b &= \frac{a_x}{z}
\end{aligned}
\qquad
\begin{aligned}
	A &= x + \frac{y}{z}\\
	B &= A - a_x\\
	B' &= A - a_y
\end{aligned}
\qquad
\begin{aligned}
	C &= \frac{A}{B}\ln\frac{A}{a_x} -1\\
	C' &= \frac{a_x}{B}\ln\frac{A}{a_x} -1\\
	C'' &= \frac{a_y}{B'}\ln\frac{A}{a_y} - 1
\end{aligned}
\end{equation}
where $x$ and $y$ are the Feynman parameters. The dipole form factors are
\begin{align}
	D^{(3a)}(z) &= -\frac12\int_0^1dx\,dy\frac{x}{B}\left(\frac{2C}{B}(3A-2xy)-\frac{3a_x-2xy}{a_x}\right)\displaybreak[0]\\
	D^{(3b)}(z) &= \int_0^1dx\,dy \frac{x}{B}\left[C'\left(\frac{3A-2xy}{B}+1+\frac{3B+ 3x(1-2y)}{2a_x}\right)+\frac{3A-2xy}{2a_x }\right]\displaybreak[0]\\
	D^{(3c)}(z) &= \int_0^1dx\,dy\frac{x^2y}{a_x(1-y-b)}\left[\frac{b}{1-y-b}\ln\frac{1-y}{b}-1\right]\displaybreak[0]\\
	D^{(3d)}(z) &= -\frac18\int_0^1dx\,dy \left[\frac{1}{zB}\left(1-\frac{2Ca_x}{B}\right) + \frac{x}{B}\left(1-\frac{2CA}{B}\right)\right]\displaybreak[0]\\
	D^{(3e)}(z) &= \frac18\int_0^1dx\,dy \frac{x}{a_x}\left[\frac{C'}{B^2}\big(xa_x(2x-1) + Bx(3x-1)-2B^2\big)-\left(2-\frac{x(2x-1)}{2B}\right)\right]\displaybreak[0]\\
	D^{(4a)}(z) &= 4s_W^2t_W^2\int_0^1dx\,dy \frac{2x}{a_x}\left[1+C'\left(1+\frac{x(1-x-y)}{2B}\right)\right]\displaybreak[0]\\
	D^{(4b)}(z) &= 4s_W^2t_W^2D^{(3c)}(z)\displaybreak[0]\\
	D^{(4c)}(z) &= 4s_W^2t_W^2\int_0^1dx\,dy \frac{1}{a_y}\left[y-x+C''\left(y-x+\frac{y^2(1-x-y)}{B'}\right)\right]
\end{align}

The 2-loop contributions to $c_R$ can be obtained from $\Delta c_L$ with the replacement $y_{f,\phi}^{ji\ast}\to y_{f,\phi}^{ij}$.

\section{Decay width of $\tau\rightarrow\mu M$} \label{app:taumum}
We follow Ref.\cite{Celis:2014asa} to calculate the decay width of $\tau\rightarrow\mu M$. The decay width of $\tau\rightarrow\mu\pi$ in the limit of $m_\mu = 0$ is given by
\begin{equation}
\Gamma_{\tau\rightarrow\mu\pi} = \frac{\left(m_\tau^2 - m_\pi^2 \right)^2 m_\pi^4 f_\pi^2 \left( (C^u_{PL} - C^d_{PL})^2 + (\text{L}\to\text{R})\right)}{256 \pi m_\tau},
\end{equation}
where we take $m_\tau = 1.777$ GeV, $m_\pi = 134.98$ MeV and $f_\pi = 130$ MeV.

The decay width of $\tau\rightarrow\mu\eta$ in the limit of $m_\mu = 0$ is given by
\begin{equation}
\Gamma_{\tau\rightarrow\mu\eta} = \frac{\left(m_\tau^2 - m_\eta^2 \right)^2 \left( \left(\left(C^u_{PL} + C^d_{PL}\right) h^q_\eta + \sqrt{2} C^s_{PL} h^s_\eta \right)^2 + (\text{L}\to\text{R}) \right)}{256 \pi m_\tau},
\end{equation}
where we take $m_\eta = 547.86$ MeV, $h^q_\eta = 0.001$ GeV$^3$ and  $h^s_\eta = -0.055$ GeV$^3$.

The decay width of $\tau\rightarrow\mu\eta'$ in the limit of $m_\mu = 0$ is given by
\begin{equation}
\Gamma_{\tau\rightarrow\mu\eta'} = \frac{\left(m_\tau^2 - m_{\eta'}^2 \right)^2 \left( \left(\left(C^u_{PL} + C^d_{PL}\right) h^q_{\eta'} + \sqrt{2} C^s_{PL} h^s_{\eta'} \right)^2 + (\text{L}\to\text{R}) \right)}{256 \pi m_\tau},
\end{equation}
where we take $m_{\eta'} = 957.78$ MeV, $h^q_{\eta'} = 0.001$ GeV$^3$ and  $h^s_{\eta'} = -0.068$ GeV$^3$.

In the equations above, we have 
\begin{equation}
C^q_{PL} = -C^q_{PR} = - \frac{Y_\ell^{\tau\mu}Y_{Q'}^{qq}}{m_q m_\tau m_A^2},
\end{equation}
where $q$ stands for the three light quarks: $u$, $d$ and $s$. For the quark masses we use $m_u = 2.2$ MeV, $m_d = 4.7$ MeV and $m_s = 96$ MeV.

The differential decay width of $\tau\rightarrow\mu\pi^+\pi^-$ as a function of the pion pair invariant mass squared, $s = (p_{\pi^+} + p_{\pi^-})^2$, is given by
\begin{equation} \label{eq:taumupipi}
\frac{d\Gamma_{\tau\rightarrow\mu\pi^+\pi^-}}{ds} = \frac{\sqrt{s-4m_\pi^2}\left(m_\tau-s\right)^2 }{512 \pi m_\tau \sqrt{s} } \left( \left| \left( C^u_{SL} + C^d_{SL} \right) \Gamma_\pi(s) + C^s_{SL} \Delta_\pi(s) \right|^2 + (\text{L}\to\text{R}) \right),
\end{equation}
where the hadronic form factors $\Gamma_\pi(s)$ and $\Delta_\pi(s)$ are taken from Ref. \cite{Celis:2013xja}. The coefficients $C^q_{SL}$ and $C^q_{RL}$ are given by
\begin{equation}
C^q_{SL} = C^q_{SR} = - \frac{Y_\ell^{\tau\mu}Y_{Q'}^{qq}}{m_q m_\tau}\left( \frac{\sin^2\alpha}{m_h^2} + \frac{\cos^2\alpha}{m_H^2}  \right).
\end{equation}
The decay width of $\tau\rightarrow\mu\pi^+\pi^-$ can be calculated by integrating Eq. \ref{eq:taumupipi} for $4m_\pi^2 \leq s \leq (m_\tau - m_\mu)^2$. The width of $\tau\rightarrow\mu\rho$ is calculated by integrating for Eq. \ref{eq:taumupipi} for 587 MeV $\leq \sqrt{s} \leq$ 962 MeV \cite{Miyazaki:2011xe}. In the equation above, we have ignored the contributions from the fermion masses (the first terms in Eq. \ref{eq:reducedcoupling}).

\bibliography{lit1} \bibliographystyle{unsrt}

\end{document}